\overfullrule=0pt
%
%
%
\def\unredoffs{} \def\redoffs{\voffset=-.31truein\hoffset=-.48truein}
\def\speclscape{}
%
%
%
%
%
\newbox\leftpage \newdimen\fullhsize \newdimen\hstitle \newdimen\hsbody
\tolerance=1000\hfuzz=2pt
\catcode`\@=11 
\ifx\hyperdef\UNd@FiNeD\def\hyperdef#1#2#3#4{#4}\def\hyperref#1#2#3#4{#4}\fi
\def\bigans{b }
\def\answ{b }
%
\ifx\answ\bigans\message{(This will come out unreduced.}
\magnification=1200\unredoffs\baselineskip=16pt plus 2pt minus 1pt
\hsbody=\hsize \hstitle=\hsize 
\else\message{(This will be reduced.} \let\l@r=L
\magnification=1000\baselineskip=16pt plus 2pt minus 1pt \vsize=7truein
\redoffs \hstitle=8truein\hsbody=4.75truein\fullhsize=10truein\hsize=\hsbody
\output={\ifnum\pageno=0 
  \shipout\vbox{\speclscape{\hsize\fullhsize\makeheadline}
    \hbox to \fullhsize{\hfill\pagebody\hfill}}\advancepageno
  \else
  \almostshipout{\leftline{\vbox{\pagebody\makefootline}}}\advancepageno
  \fi}
\def\almostshipout#1{\if L\l@r \count1=1 \message{[\the\count0.\the\count1]}
      \global\setbox\leftpage=#1 \global\let\l@r=R
 \else \count1=2
  \shipout\vbox{\speclscape{\hsize\fullhsize\makeheadline}
      \hbox to\fullhsize{\box\leftpage\hfil#1}}  \global\let\l@r=L\fi}
\fi
%
\newcount\yearltd\yearltd=\year\advance\yearltd by -2000

\def\Title#1#2{\nopagenumbers\abstractfont\hsize=\hstitle\rightline{#1}%
\vskip 1in\centerline{\titlefont #2}\abstractfont\vskip .5in\pageno=0}
\def\Date#1{\vfill\leftline{#1}\tenpoint\supereject\global\hsize=\hsbody%
\footline={\hss\tenrm\hyperdef\hypernoname{page}\folio\folio\hss}}%
%

\def\draftmode{\message{ DRAFTMODE }\def\draftdate{{\rm preliminary draft:
\number\month/\number\day/{0}\number\yearltd\ \ \hourmin}}%
\headline={\hfil\draftdate}\writelabels\baselineskip=20pt plus 2pt minus 2pt
 {\count255=\time\divide\count255 by 60 \xdef\hourmin{\number\count255}
  \multiply\count255 by-60\advance\count255 by\time
  \xdef\hourmin{\hourmin:\ifnum\count255<10 0\fi\the\count255}}}
\def\nolabels{\def\wrlabeL##1{}\def\eqlabeL##1{}\def\reflabeL##1{}}
\def\writelabels{\def\wrlabeL##1{\leavevmode\vadjust{\rlap{\smash%
{\line{{\escapechar=` \hfill\rlap{\sevenrm\hskip.03in\string##1}}}}}}}%
\def\eqlabeL##1{{\escapechar-1\rlap{\sevenrm\hskip.05in\string##1}}}%
\def\reflabeL##1{\noexpand\llap{\noexpand\sevenrm\string\string\string##1}}}
\nolabels
%
\global\newcount\secno \global\secno=0
\global\newcount\meqno \global\meqno=1
\def\s@csym{}
\def\newsec#1{\global\advance\secno by1%
{\toks0{#1}\message{(\the\secno. \the\toks0)}}%
\global\subsecno=0\eqnres@t\let\s@csym\secsym\xdef\secn@m{\the\secno}\noindent
{\bf\hyperdef\hypernoname{section}{\the\secno}{\the\secno.} #1}%
\writetoca{{\string\hyperref{}{section}{\the\secno}{\vskip2pt \bf \the\secno\quad}} {\bf #1}}%
\par\nobreak\medskip\nobreak}
\def\eqnres@t{\xdef\secsym{\the\secno.}\global\meqno=1\bigbreak\bigskip}
\def\sequentialequations{\def\eqnres@t{\bigbreak}}\xdef\secsym{}
\global\newcount\subsecno \global\subsecno=0
\def\subsec#1{\global\advance\subsecno by1%
{\toks0{#1}\message{(\s@csym\the\subsecno. \the\toks0)}}%
\ifnum\lastpenalty>9000\else\bigbreak\fi       \global\subsubsecno=0
\noindent{\it\hyperdef\hypernoname{subsection}{\secn@m.\the\subsecno}%
{\secn@m.\the\subsecno.} #1}\writetoca{\string\hskip1.45cm
{\string\hyperref{}{subsection}{\secn@m.\the\subsecno}{\secn@m.\the\subsecno.}}
{#1}}\par\nobreak\medskip\nobreak}
\def\appendix#1#2{\global\meqno=1\global\subsecno=0\xdef\secsym{\hbox{#1.}}%
\bigbreak\bigskip\noindent{\bf Appendix \hyperdef\hypernoname{appendix}{#1}%
{#1.} #2}{\toks0{(#1. #2)}\message{\the\toks0}}%
\xdef\s@csym{#1.}\xdef\secn@m{#1}%
\writetoca{{\string\hyperref{}{appendix}{#1}{\vskip2pt \bf {#1}\quad}} {\bf #2}}%
\par\nobreak\medskip\nobreak}
%
%
\def\checkm@de#1#2{\ifmmode{\def\f@rst##1{##1}\hyperdef\hypernoname{equation}%
{#1}{#2}}\else\hyperref{}{equation}{#1}{#2}\fi}
\def\eqnn#1{\DefWarn#1\xdef #1{(\noexpand\relax\noexpand\checkm@de%
{\s@csym\the\meqno}{\secsym\the\meqno})}%
\wrlabeL#1\writedef{#1\leftbracket#1}\global\advance\meqno by1}
\def\f@rst#1{\c@t#1a\em@ark}\def\c@t#1#2\em@ark{#1}
\def\eqna#1{\DefWarn#1\wrlabeL{#1$\{\}$}%
\xdef #1##1{(\noexpand\relax\noexpand\checkm@de%
{\s@csym\the\meqno\noexpand\f@rst{##1}}{\hbox{$\secsym\the\meqno##1$}})}
\writedef{#1\numbersign1\leftbracket#1{\numbersign1}}\global\advance\meqno by1}
\def\eqn#1#2{\DefWarn#1%
\xdef #1{(\noexpand\hyperref{}{equation}{\s@csym\the\meqno}%
{\secsym\the\meqno})}$$#2\eqno(\hyperdef\hypernoname{equation}%
{\s@csym\the\meqno}{\secsym\the\meqno})\eqlabeL#1$$%
\writedef{#1\leftbracket#1}\global\advance\meqno by1}
\def\xeqn{\expandafter\xe@n}\def\xe@n(#1){#1}
\def\xeqna#1{\expandafter\xe@n#1}
\def\eqns#1{(\e@ns #1{\hbox{}})}
\def\e@ns#1{\ifx\UNd@FiNeD#1\message{eqnlabel \string#1 is undefined.}%
\xdef#1{(?.?)}\fi{\let\hyperref=\relax\xdef\next{#1}}%
\ifx\next\em@rk\def\next{}\else%
\ifx\next#1\xeqn#1\else\def\n@xt{#1}\ifx\n@xt\next#1\else\xeqna#1\fi
\fi\let\next=\e@ns\fi\next}

\def\DefWarn#1{\ifx\UNd@FiNeD#1\else
\immediate\write16{*** WARNING: the label \string#1 is already defined ***}\fi}
%
\newskip\footskip\footskip14pt plus 1pt minus 1pt 
\def\footnotefont{\ninepoint}\def\f@t#1{\footnotefont #1\@foot}
\def\f@@t{\baselineskip\footskip\bgroup\footnotefont\aftergroup\@foot\let\next}
\setbox\strutbox=\hbox{\vrule height9.5pt depth4.5pt width0pt}
\global\newcount\ftno \global\ftno=0
\def\foot{\global\advance\ftno by1\def\foot@rg{\hyperref{}{footnote}%
{\the\ftno}{\the\ftno}\xdef\foot@rg{\noexpand\hyperdef\noexpand\hypernoname%
{footnote}{\the\ftno}{\the\ftno}}}\footnote{$^{\foot@rg}$}}
%
\newwrite\ftfile
\def\footend{\def\foot{\global\advance\ftno by1\chardef\wfile=\ftfile
\hyperref{}{footnote}{\the\ftno}{$^{\the\ftno}$}%
\ifnum\ftno=1\immediate\openout\ftfile=\jobname.fts\fi%
\immediate\write\ftfile{\noexpand\smallskip%
\noexpand\item{\noexpand\hyperdef\noexpand\hypernoname{footnote}
{\the\ftno}{f\the\ftno}:\ }\pctsign}\findarg}%
\def\footatend{\vfill\eject\immediate\closeout\ftfile{\parindent=20pt
\centerline{\bf Footnotes}\nobreak\bigskip\input \jobname.fts }}}
\def\footatend{}
%
%
\global\newcount\refno \global\refno=1
\newwrite\rfile
\def\ref{[\hyperref{}{reference}{\the\refno}{\the\refno}]\nref}
\def\nref#1{\DefWarn#1%
\xdef#1{[\noexpand\hyperref{}{reference}{\the\refno}{\the\refno}]}%
\writedef{#1\leftbracket#1}%
\ifnum\refno=1\immediate\openout\rfile=\jobname.refs\fi
\chardef\wfile=\rfile\immediate\write\rfile{\noexpand\item{[\noexpand\hyperdef%
\noexpand\hypernoname{reference}{\the\refno}{\the\refno}]\ }%
\reflabeL{#1\hskip.31in}\pctsign}\global\advance\refno by1\findarg}
\def\findarg#1#{\begingroup\obeylines\newlinechar=`\^^M\pass@rg}
{\obeylines\gdef\pass@rg#1{\writ@line\relax #1^^M\hbox{}^^M}%
\gdef\writ@line#1^^M{\expandafter\toks0\expandafter{\striprel@x #1}%
\edef\next{\the\toks0}\ifx\next\em@rk\let\next=\endgroup\else\ifx\next\empty%
\else\immediate\write\wfile{\the\toks0}\fi\let\next=\writ@line\fi\next\relax}}
\def\striprel@x#1{} \def\em@rk{\hbox{}}
\def\lref{\begingroup\obeylines\lr@f}
\def\lr@f#1#2{\DefWarn#1\gdef#1{\let#1=\UNd@FiNeD\ref#1{#2}}\endgroup\unskip}
\def\semi{;\hfil\break}
\def\addref#1{\immediate\write\rfile{\noexpand\item{}#1}} 
\def\listrefs{\footatend\vfill\supereject\immediate\closeout\rfile\writestoppt
\baselineskip=\footskip\centerline{{\bf References}}\bigskip{\parindent=20pt%
\frenchspacing\escapechar=` \input \jobname.refs\vfill\eject}\nonfrenchspacing}
\def\startrefs#1{\immediate\openout\rfile=\jobname.refs\refno=#1}
\def\xref{\expandafter\xr@f}\def\xr@f[#1]{#1}
\def\refs#1{\count255=1[\r@fs #1{\hbox{}}]}
\def\r@fs#1{\ifx\UNd@FiNeD#1\message{reflabel \string#1 is undefined.}%
\nref#1{need to supply reference \string#1.}\fi%
\vphantom{\hphantom{#1}}{\let\hyperref=\relax\xdef\next{#1}}%
\ifx\next\em@rk\def\next{}%
\else\ifx\next#1\ifodd\count255\relax\xref#1\count255=0\fi%
\else#1\count255=1\fi\let\next=\r@fs\fi\next}
%

%
\newwrite\ffile\global\newcount\figno \global\figno=1
\def\fig{fig.~\hyperref{}{figure}{\the\figno}{\the\figno}\nfig}
\def\nfig#1{\DefWarn#1%
\xdef#1{fig.~\noexpand\hyperref{}{figure}{\the\figno}{\the\figno}}%
\writedef{#1\leftbracket fig.\noexpand~\xfig#1}%
\ifnum\figno=1\immediate\openout\ffile=\jobname.figs\fi\chardef\wfile=\ffile%
{\let\hyperref=\relax
\immediate\write\ffile{\noexpand\medskip\noexpand\item{Fig.\ %
\noexpand\hyperdef\noexpand\hypernoname{figure}{\the\figno}{\the\figno}. }
\reflabeL{#1\hskip.55in}\pctsign}}\global\advance\figno by1\findarg}
\def\listfigs{\vfill\eject\immediate\closeout\ffile{\parindent40pt
\baselineskip14pt\centerline{{\bf Figure Captions}}\nobreak\medskip
\escapechar=` \input \jobname.figs\vfill\eject}}
\def\xfig{\expandafter\xf@g}\def\xf@g fig.\penalty\@M\ {}
\def\figs#1{figs.~\f@gs #1{\hbox{}}}
\def\f@gs#1{{\let\hyperref=\relax\xdef\next{#1}}\ifx\next\em@rk\def\next{}\else
\ifx\next#1\xfig #1\else#1\fi\let\next=\f@gs\fi\next}
\def\figin{\epsfcheck\figin}\def\figins{\epsfcheck\figins}
\def\epsfcheck{\ifx\epsfbox\UNd@FiNeD
\message{(NO epsf.tex, FIGURES WILL BE IGNORED)}
\gdef\figin##1{\vskip2in}\gdef\figins##1{\hskip.5in}
\else\message{(FIGURES WILL BE INCLUDED)}%
\gdef\figin##1{##1}\gdef\figins##1{##1}\fi}
\def\DefWarn#1{}
\def\figinsert{\goodbreak\midinsert}
\def\ifig#1#2#3{\DefWarn#1\xdef#1{Fig.~\noexpand\hyperref{}{figure}%
{\the\figno}{\the\figno}}\writedef{#1\leftbracket fig.\noexpand~\xfig#1}%
\figinsert\figin{\centerline{#3}}\medskip\centerline{\vbox{\baselineskip12pt
\advance\hsize by -1truein\noindent\wrlabeL{#1=#1}\footnotefont%
{\bf Fig.~\hyperdef\hypernoname{figure}{\the\figno}{\the\figno}:} #2}}
\bigskip\endinsert\global\advance\figno by1}
\newwrite\lfile
{\escapechar-1\xdef\pctsign{\string\%}\xdef\leftbracket{\string\{}
\xdef\rightbracket{\string\}}\xdef\numbersign{\string\#}}
\def\writedefs{\immediate\openout\lfile=\jobname.defs \def\writedef##1{%
{\let\hyperref=\relax\let\hyperdef=\relax\let\hypernoname=\relax
 \immediate\write\lfile{\string\def\string##1\rightbracket}}}}%
\def\writestop{\def\writestoppt{\immediate\write\lfile{\string\pageno
 \the\pageno\string\startrefs\leftbracket\the\refno\rightbracket
 \string\def\string\secsym\leftbracket\secsym\rightbracket
 \string\secno\the\secno\string\meqno\the\meqno}\immediate\closeout\lfile}}
\def\writestoppt{}\def\writedef#1{}
\def\seclab#1{\DefWarn#1%
\xdef #1{\noexpand\hyperref{}{section}{\the\secno}{\the\secno}}%
\writedef{#1\leftbracket#1}\wrlabeL{#1=#1}}
\def\subseclab#1{\DefWarn#1%
\xdef #1{\noexpand\hyperref{}{subsection}{\secn@m.\the\subsecno}%
{\secn@m.\the\subsecno}}\writedef{#1\leftbracket#1}\wrlabeL{#1=#1}}
\def\applab#1{\DefWarn#1%
\xdef #1{\noexpand\hyperref{}{appendix}{\secn@m}{\secn@m}}%
\writedef{#1\leftbracket#1}\wrlabeL{#1=#1}}
\newwrite\tfile \def\writetoca#1{}
\def\leaderfill{\leaders\hbox to 1em{\hss.\hss}\hfill}
\def\writetoc{\immediate\openout\tfile=\jobname.toc
   \def\writetoca##1{{\edef\next{\write\tfile{\noindent ##1
   \string\leaderfill {\string\hyperref{}{page}{\noexpand\number\pageno}%
                       {\noexpand\number\pageno}} \par}}\next}}}
\newread\ch@ckfile
\def\listtoc{\immediate\closeout\tfile\immediate\openin\ch@ckfile=\jobname.toc
\ifeof\ch@ckfile\message{no file \jobname.toc, no table of contents this pass}%
\else\closein\ch@ckfile\centerline{\bf Contents}\nobreak\medskip%
{\baselineskip=16pt\footnotefont\parskip=0pt\catcode`\@=11\input\jobname.toc
\catcode`\@=12\bigbreak\bigskip}\fi}
\catcode`\@=12 
%
\edef\tfontsize{\ifx\answ\bigans scaled\magstep3\else scaled\magstep4\fi}
\font\titlerm=cmr10 \tfontsize \font\titlerms=cmr7 \tfontsize
\font\titlermss=cmr5 \tfontsize \font\titlei=cmmi10 \tfontsize
\font\titleis=cmmi7 \tfontsize \font\titleiss=cmmi5 \tfontsize
\font\titlesy=cmsy10 \tfontsize \font\titlesys=cmsy7 \tfontsize
\font\titlesyss=cmsy5 \tfontsize \font\titleit=cmti10 \tfontsize
\skewchar\titlei='177 \skewchar\titleis='177 \skewchar\titleiss='177
\skewchar\titlesy='60 \skewchar\titlesys='60 \skewchar\titlesyss='60
\def\titlefont{\def\rm{\fam0\titlerm}
\textfont0=\titlerm \scriptfont0=\titlerms \scriptscriptfont0=\titlermss
\textfont1=\titlei \scriptfont1=\titleis \scriptscriptfont1=\titleiss
\textfont2=\titlesy \scriptfont2=\titlesys \scriptscriptfont2=\titlesyss
\textfont\itfam=\titleit \def\it{\fam\itfam\titleit}\rm}
 \ifx\answ\bigans\else scaled\magstep1\fi
\ifx\answ\bigans\def\abstractfont{\tenpoint}\else
\font\absit=cmti10 scaled \magstep1
\font\abssl=cmsl10 scaled \magstep1
\font\absrm=cmr10 scaled\magstep1 \font\absrms=cmr7 scaled\magstep1
\font\absrmss=cmr5 scaled\magstep1 \font\absi=cmmi10 scaled\magstep1
\font\absis=cmmi7 scaled\magstep1 \font\absiss=cmmi5 scaled\magstep1
\font\abssy=cmsy10 scaled\magstep1 \font\abssys=cmsy7 scaled\magstep1
\font\abssyss=cmsy5 scaled\magstep1 \font\absbf=cmbx10 scaled\magstep1
\skewchar\absi='177 \skewchar\absis='177 \skewchar\absiss='177
\skewchar\abssy='60 \skewchar\abssys='60 \skewchar\abssyss='60
\def\abstractfont{\def\rm{\fam0\absrm}
\textfont0=\absrm \scriptfont0=\absrms \scriptscriptfont0=\absrmss
\textfont1=\absi \scriptfont1=\absis \scriptscriptfont1=\absiss
\textfont2=\abssy \scriptfont2=\abssys \scriptscriptfont2=\abssyss
\textfont\itfam=\absit \def\it{\fam\itfam\absit}\def\footnotefont{\tenpoint}%
\textfont\slfam=\abssl \def\sl{\fam\slfam\abssl}%
\textfont\bffam=\absbf \def\bf{\fam\bffam\absbf}\rm}\fi
\def\tenpoint{\def\rm{\fam0\tenrm}
\textfont0=\tenrm \scriptfont0=\sevenrm \scriptscriptfont0=\fiverm
\textfont1=\teni  \scriptfont1=\seveni  \scriptscriptfont1=\fivei
\textfont2=\tensy \scriptfont2=\sevensy \scriptscriptfont2=\fivesy
\textfont\itfam=\tenit \def\it{\fam\itfam\tenit}\def\footnotefont{\ninepoint}%
\textfont\bffam=\tenbf \def\bf{\fam\bffam\tenbf}\def\sl{\fam\slfam\tensl}\rm}
\font\ninerm=cmr9 \font\sixrm=cmr6 \font\ninei=cmmi9 \font\sixi=cmmi6
\font\ninesy=cmsy9 \font\sixsy=cmsy6 \font\ninebf=cmbx9
\font\nineit=cmti9 \font\ninesl=cmsl9 \skewchar\ninei='177
\skewchar\sixi='177 \skewchar\ninesy='60 \skewchar\sixsy='60
\def\ninepoint{\def\rm{\fam0\ninerm}
\textfont0=\ninerm \scriptfont0=\sixrm \scriptscriptfont0=\fiverm
\textfont1=\ninei \scriptfont1=\sixi \scriptscriptfont1=\fivei
\textfont2=\ninesy \scriptfont2=\sixsy \scriptscriptfont2=\fivesy
\textfont\itfam=\ninei \def\it{\fam\itfam\nineit}\def\sl{\fam\slfam\ninesl}%
\textfont\bffam=\ninebf \def\bf{\fam\bffam\ninebf}\rm}
%
%

\hyphenation{anom-aly anom-alies coun-ter-term coun-ter-terms}
\def\inv{^{\raise.15ex\hbox{${\scriptscriptstyle -}$}\kern-.05em 1}}

\def\Dsl{\,\raise.15ex\hbox{/}\mkern-13.5mu D} 
\def\dsl{\raise.15ex\hbox{/}\kern-.57em\partial}

 \def\Tr{{\rm Tr}}
\def\lspace{\ifx\answ\bigans{}\else\qquad\fi}
\def\lbspace{\ifx\answ\bigans{}\else\hskip-.2in\fi} 
\def\boxeqn#1{\vcenter{\vbox{\hrule\hbox{\vrule\kern3pt\vbox{\kern3pt
	\hbox{${\displaystyle #1}$}\kern3pt}\kern3pt\vrule}\hrule}}}
\def\mbox#1#2{\vcenter{\hrule \hbox{\vrule height#2in
		\kern#1in \vrule} \hrule}}  
%

\def\vev#1{\langle #1 \rangle}

\def\darr#1{\raise1.5ex\hbox{$\leftrightarrow$}\mkern-16.5mu #1}

\def\roughly#1{\raise.3ex\hbox{$#1$\kern-.75em\lower1ex\hbox{$\sim$}}}

\global\newcount\subsubsecno \global\subsubsecno=0
\def\subsubsec#1{\global\advance\subsubsecno by1%
{\toks0{#1}\message{(\the\secno\the\subsecno\the\subsubsecno. \the\toks0)}}%
\ifnum\lastpenalty>9000\else\bigbreak\fi
\noindent{\it\hyperdef\hypernoname{subsubsection}{\the\secno.\the\subsecno\the\subsubsecno}%
{\the\secno.\the\subsecno.\the\subsubsecno.} #1}
\par\nobreak\medskip\nobreak}
\def\boxit#1{\vbox{\hrule\hbox{\vrule\kern8pt
\vbox{\hbox{\kern8pt}\hbox{\vbox{#1}}\hbox{\kern8pt}}
\kern8pt\vrule}\hrule}}
\def\mathboxit#1{\vbox{\hrule\hbox{\vrule\kern8pt\vbox{\kern8pt
\hbox{$\displaystyle #1$}\kern8pt}\kern8pt\vrule}\hrule}}
\def\slashchar#1{\setbox0=\hbox{$#1$}           
   \dimen0=\wd0                                 
   \setbox1=\hbox{/} \dimen1=\wd1               
   \ifdim\dimen0>\dimen1                        
      \rlap{\hbox to \dimen0{\hfil/\hfil}}      
      #1                                        
   \else                                        
      \rlap{\hbox to \dimen1{\hfil$#1$\hfil}}   
      /                                         
   \fi}
\def\sqr#1#2{{\vcenter{\vbox{\hrule height.#2pt
         \hbox{\vrule width.#2pt height#1pt \kern#1pt
            \vrule width.#2pt}
         \hrule height.#2pt}}}}


\input epsf.sty

\def\si{{\sigma}}

\def\p{{\partial}}

\def\({\left(}
\def\){\right)}

\def\ap{{\alpha'}}

\lref\PARTONE{
  C.R.~Mafra, O.~Schlotterer and S.~Stieberger,
``Complete N-Point Superstring Disk Amplitude. I. Pure Spinor Computation,''
  AEI--2011--34, MPP-2011--47.}

\def\crr{\noalign{\vskip5pt}}

\def\comment#1{{}}
\def\ss#1{{\scriptstyle{#1}}}

\def\ap{\alpha'}

\def\cf{cf.\ }
\def\ie{i.e.\ }
\def\eg{e.g.\ }
\def\eqq{eq.\ }
\def\eqqs{eqs.\ }

\def\al{\alpha}

\def\si{\sigma}

\def\bet{\beta}




\input epsf
\def\figin{\epsfcheck\figin}\def\figins{\epsfcheck\figins}
\def\epsfcheck{\ifx\epsfbox\UnDeFiNeD
\message{(NO epsf.tex, FIGURES WILL BE IGNORED)}
\gdef\figin##1{\vskip2in}\gdef\figins##1{\hskip.5in}
\else\message{(FIGURES WILL BE INCLUDED)}%
\gdef\figin##1{##1}\gdef\figins##1{##1}\fi}
\def\DefWarn#1{}
\def\figinsert{\goodbreak\midinsert}
\def\ifig#1#2#3{\DefWarn#1\xdef#1{Fig.~\the\figno}
\writedef{#1\leftbracket fig.\noexpand~\the\figno}%
\figinsert\figin{\centerline{#3}}\medskip\centerline{\vbox{\baselineskip12pt
\advance\hsize by -1truein\noindent\footnotefont\centerline{{\bf
Fig.~\the\figno}\ #2}}}
\bigskip\endinsert\global\advance\figno by1}

\def\appA{A}
\def\appB{B}
\def\appC{C}

\def\wtilde{\widetilde}

\def\h {{1\over 2}}

\def\ov {\overline}
\def\o {\over}
\def\fc#1#2{{#1 \o #2}}

\def\IZ{ {\bf Z}}\def\IR{ {\bf R}}


\def\br{\hfill\break}

\def\mod {{\rm mod}}
\def\lf {\left}
\def\ri {\right}
\def\ra {\rightarrow}
\def\lra {\longrightarrow}

\def\p {\partial}

 \def\Pc{{\cal P}}
 
 \def\Oc {{\cal O}}
 
\def\Mc {{\cal M}} \def\Ac {{\cal A}}
\def\Pc {{\cal P}}

\def\Kc {{\cal K}}

\Title{\vbox{\rightline{AEI--2011--35}\rightline{MPP--2011--65}
\vskip-0.5cm}}{
	       	\vbox{
		\centerline{Complete $N$--Point Superstring Disk Amplitude}\vskip6pt
		\centerline{II. Amplitude and Hypergeometric Function Structure}
		}
	}
\centerline{Carlos R. Mafra$^{a,b,}$\foot{e-mail: crmafra@aei.mpg.de}, 
Oliver Schlotterer$^{b,c,}$\foot{e-mail: olivers@mppmu.mpg.de}, 
and Stephan Stieberger$^{b,c,}$\foot{e-mail: stephan.stieberger@mpp.mpg.de}}

\bigskip\medskip
\centerline{\it $^a$ Max--Planck--Institut f\"ur Gravitationsphysik} 
\centerline{\it Albert--Einstein--Institut, 14476 Potsdam, Germany}
\vskip4pt
\centerline{\it $^b$ Kavli Institute for Theoretical Physics} 
\centerline{\it University of California, Santa Barbara, CA 93106, USA}
\vskip4pt
\centerline{\it $^c$ Max--Planck--Institut f\"ur Physik}
\centerline{\it Werner--Heisenberg--Institut, 80805 M\"unchen, Germany}
\bigskip\medskip\medskip
\noindent
Using the pure spinor formalism in part I \PARTONE\ we compute the complete 
tree--level amplitude of 
$N$ massless open strings  and find a striking simple and compact form in terms of minimal building 
blocks: the full $N$--point amplitude is expressed by a sum over  $(N-3)!$ Yang--Mills partial subamplitudes each multiplying
a multiple Gaussian hypergeometric function. While the former 
capture the space--time kinematics of the amplitude the latter encode the string effects.
This result disguises a lot of structure linking aspects of gauge amplitudes as color and kinematics with properties of generalized Euler integrals.
In this part II the structure of the multiple hypergeometric functions is analyzed in detail:
their relations to monodromy equations, their minimal basis structure, and methods to determine their poles and transcendentality properties are proposed.
Finally, a Gr\"obner basis analysis provides independent sets of rational functions in the Euler integrals.

\Date{}
\noindent
\goodbreak

\lref\ElvangWD{
H.~Elvang, D.Z.~Freedman, M.~Kiermaier,
``Solution to the Ward Identities for Superamplitudes,''
JHEP {\bf 1010}, 103 (2010).
[arXiv:0911.3169 [hep-th]].
}

\lref\GrisaruVM{
  M.T.~Grisaru, H.N.~Pendleton, P.~van Nieuwenhuizen,
``Supergravity and the S Matrix,''
Phys.\ Rev.\  {\bf D15}, 996 (1977);\br
M.T.~Grisaru, H.N.~Pendleton,
``Some Properties of Scattering Amplitudes in Supersymmetric Theories,''
Nucl.\ Phys.\  {\bf B124}, 81 (1977).
}

\lref\ParkePN{
S.J.~Parke, T.R.~Taylor,
``Perturbative QCD Utilizing Extended Supersymmetry,''
Phys.\ Lett.\  {\bf B157}, 81 (1985), [Erratum-ibid. 174B, 465 (1986)];\br
Z.~Kunszt,
``Combined Use of the Calkul Method and N=1 Supersymmetry to Calculate QCD Six Parton Processes,''
Nucl.\ Phys.\  {\bf B271}, 333 (1986).
}

\lref\LANCEeff{
  L.J.~Dixon,
``Calculating scattering amplitudes efficiently,''
[hep-ph/9601359].
}

\lref\MANGO{
  M.L.~Mangano, S.J.~Parke,
``Multiparton amplitudes in gauge theories,''
Phys.\ Rept.\  {\bf 200}, 301-367 (1991).
[hep-th/0509223].
}

\lref\Frampton{
P. Frampton,
``Dual Resonance Models,'' Frontiers in Physics, Benjamin 1974.}

\lref\HopkinsonER{
  J.F.L.~Hopkinson, E.~Plahte,
``Infinite series representation of the n-point function in the generalized veneziano model,''
Phys.\ Lett.\  {\bf B28}, 489-492 (1969).
}

\lref\MafraJQ{
  C.R.~Mafra, O.~Schlotterer, S.~Stieberger and D.~Tsimpis,
``A recursive formula for N-point SYM tree amplitudes,''
  arXiv:1012.3981 [hep-th], to appear in Phys.\ Rev.\  D.
}

\lref\mhv{
  S.J.~Parke and T.R.~Taylor,
  ``An Amplitude for $n$ Gluon Scattering,''
  Phys.\ Rev.\ Lett.\  {\bf 56}, 2459 (1986).
}

\lref\BerendsME{
  F.A.~Berends, W.T.~Giele,
``Recursive Calculations for Processes with n Gluons,''
Nucl.\ Phys.\  {\bf B306}, 759 (1988).
}

\lref\DixonIK{
  L.J.~Dixon, J.M.~Henn, J.~Plefka, T.~Schuster,
``All tree-level amplitudes in massless QCD,''
JHEP {\bf 1101}, 035 (2011).
[arXiv:1010.3991 [hep-ph]].
}

\lref\BernDW{
  Z.~Bern, L.J.~Dixon and D.A.~Kosower,
``On-Shell Methods in Perturbative QCD,''
  Annals Phys.\  {\bf 322}, 1587 (2007)
  [arXiv:0704.2798 [hep-ph]];\br
J.J.M.~Carrasco, H.~Johansson,
``Generic multiloop methods and application to N=4 super-Yang-Mills,''
[arXiv:1103.3298 [hep-th]];\br
Z.~Bern, Y.-t.~Huang,
``Basics of Generalized Unitarity,''
[arXiv:1103.1869 [hep-th]].
}

\lref\BernFY{
Z.~Bern, J.J.M.~Carrasco, H.~Johansson,
``The Structure of Multiloop Amplitudes in Gauge and Gravity Theories,''
Nucl.\ Phys.\ Proc.\ Suppl.\  {\bf 205-206}, 54-60 (2010).
[arXiv:1007.4297 [hep-th]];\br
Z.~Bern, J.J.M. Carrasco, L.J.~Dixon, H.~Johansson, R.~Roiban,
 ``Amplitudes and Ultraviolet Behavior of N=8 Supergravity,''
[arXiv:1103.1848 [hep-th]].
}

\lref\KawaiXQ{
  H.~Kawai, D.C.~Lewellen and S.H.H.~Tye,
``A Relation Between Tree Amplitudes Of Closed And Open Strings,''
  Nucl.\ Phys.\  B {\bf 269}, 1 (1986).
}

\lref\BCJ{
  Z.~Bern, J.J.M.~Carrasco and H.~Johansson,
  ``New Relations for Gauge-Theory Amplitudes,''
  Phys.\ Rev.\  D {\bf 78}, 085011 (2008)
  [arXiv:0805.3993 [hep-ph]].
}

\lref\mafrabcj{
  C.R.~Mafra,
``Simplifying the Tree-level Superstring Massless Five-point Amplitude,''
  JHEP {\bf 1001}, 007 (2010)
  [arXiv:0909.5206 [hep-th]].
}

\lref\FTAmps{
  C.R.~Mafra,
  ``Towards Field Theory Amplitudes From the Cohomology of Pure Spinor
  Superspace,''
  JHEP {\bf 1011}, 096 (2010)
  [arXiv:1007.3639 [hep-th]].
}

\lref\BjerrumBohrRD{
  N.E.J.~Bjerrum-Bohr, P.H.~Damgaard and P.~Vanhove,
  ``Minimal Basis for Gauge Theory Amplitudes,''
  Phys.\ Rev.\ Lett.\  {\bf 103}, 161602 (2009)
  [arXiv:0907.1425 [hep-th]].
}

\lref\STi{
  S.~Stieberger and T.R.~Taylor,
``Amplitude for N-gluon superstring scattering,''
  Phys.\ Rev.\ Lett.\  {\bf 97}, 211601 (2006)
  [arXiv:hep-th/0607184].
  }
\lref\STii{S.~Stieberger and T.R.~Taylor,  
``Multi-gluon scattering in open superstring theory,''
  Phys.\ Rev.\  D {\bf 74}, 126007 (2006)
  [arXiv:hep-th/0609175].
}

\lref\stieS{
  S.~Stieberger and T.R.~Taylor,
  ``Supersymmetry Relations and MHV Amplitudes in Superstring Theory,''
  Nucl.\ Phys.\  B {\bf 793}, 83 (2008)
  [arXiv:0708.0574 [hep-th]].
}

\lref\StiebergerAM{
  S.~Stieberger and T.R.~Taylor,
 ``Complete Six--Gluon Disk Amplitude in Superstring Theory,''
  Nucl.\ Phys.\  B {\bf 801}, 128 (2008)
  [arXiv:0711.4354 [hep-th]].
}

\lref\StieOpr{
  D.~Oprisa and S.~Stieberger,
  ``Six gluon open superstring disk amplitude, multiple hypergeometric  series
  and Euler-Zagier sums,''
  arXiv:hep-th/0509042.
}

\lref\StiebergerHQ{
  S.~Stieberger,
  ``Open \& Closed vs. Pure Open String Disk Amplitudes,''
  arXiv:0907.2211 [hep-th].
}

\lref\Jaxo{D.~Binosi and L.~Theussl,
``JaxoDraw: A graphical user interface for drawing Feynman diagrams,''
  Comput.\ Phys.\ Commun.\  {\bf 161}, 76 (2004)
  [arXiv:hep-ph/0309015].
}

\lref\Medinas{
  R.~Medina, F.T.~Brandt and F.R.~Machado,
  ``The open superstring 5-point amplitude revisited,''
  JHEP {\bf 0207}, 071 (2002)
  [arXiv:hep-th/0208121]
\semi
  L.A.~Barreiro and R.~Medina,
  ``5-field terms in the open superstring effective action,''
  JHEP {\bf 0503}, 055 (2005)
  [arXiv:hep-th/0503182].
}

\lref\BernIA{
  Z.~Bern, T.~Dennen,
``A Color Dual Form for Gauge-Theory Amplitudes,''\br
[arXiv:1103.0312 [hep-th]].
}

\lref\MSST{
C.R.~Mafra, O.~Schlotterer, S.~Stieberger, D.~Tsimpis,
 ``Six Open String Disk Amplitude in Pure Spinor Superspace,''
Nucl.\ Phys.\  {\bf B846}, 359-393 (2011).
[arXiv:1011.0994 [hep-th]].
}

\lref\KITP{
C.R.~Mafra, O.~Schlotterer, S.~Stieberger,
``Explicit BCJ Numerators from Pure Spinors,''
[arXiv:1104.5224 [hep-th]].
}

\lref\BeisertJX{
  N.~Beisert, H.~Elvang, D.Z.~Freedman, M.~Kiermaier, A.~Morales, S.~Stieberger,
``$E_{7,7}$ constraints on counterterms in N=8 supergravity,''
Phys.\ Lett.\  {\bf B694}, 265-271 (2010).
[arXiv:1009.1643 [hep-th]].
}

\lref\ACHT{J.~Br\"odel, L.J.~Dixon,
``$R^4$ counterterm and $E_{7,7}$ symmetry in maximal supergravity,''
JHEP {\bf 1005}, 003 (2010).
[arXiv:0911.5704 [hep-th]];\br
H.~Elvang, D.Z.~Freedman, M.~Kiermaier,
``A simple approach to counterterms in N=8 supergravity,''
JHEP {\bf 1011}, 016 (2010).
[arXiv:1003.5018 [hep-th]];\br
H.~Elvang, M.~Kiermaier,
``Stringy KLT relations, global symmetries, and $E_{7,7}$ violation,''
JHEP {\bf 1010}, 108 (2010).
[arXiv:1007.4813 [hep-th]]
}

\lref\BOHR{N.E.J.~Bjerrum-Bohr, P.H.~Damgaard, T.~Sondergaard, P.~Vanhove,
``The Momentum Kernel of Gauge and Gravity Theories,''
JHEP {\bf 1101}, 001 (2011).
[arXiv:1010.3933 [hep-th]].
}

\lref\GRAV{S.~Stieberger,
``Constraints on Tree-Level Higher Order Gravitational Couplings in Superstring Theory,''
Phys.\ Rev.\ Lett.\  {\bf 106}, 111601 (2011).
[arXiv:0910.0180 [hep-th]].
}

\lref\progress{C.R.~Mafra, O.~Schlotterer, and S.~Stieberger,
``Complete  $N$--Graviton Tree--Level Amplitude,'' MPP--2011--66, to appear.}

\lref\BCFW{F.~Cachazo, P.~Svrcek, E.~Witten,
``MHV vertices and tree amplitudes in gauge theory,''
JHEP {\bf 0409}, 006 (2004).
[hep-th/0403047].\br
R.~Britto, F.~Cachazo, B.~Feng, E.~Witten,
 ``Direct proof of tree-level recursion relation in Yang-Mills theory,''
Phys.\ Rev.\ Lett.\  {\bf 94}, 181602 (2005);
[hep-th/0501052].
}

\lref\FOURLIT{M.B.~Green, J.H.~Schwarz,
``Supersymmetrical Dual String Theory. 2. Vertices and Trees,''
Nucl.\ Phys.\  {\bf B198}, 252-268 (1982);\br
J.H.~Schwarz,
``Superstring Theory,''
Phys.\ Rept.\  {\bf 89}, 223-322 (1982);\br
A.A.~Tseytlin,
``Vector Field Effective Action in the Open Superstring Theory,''
Nucl.\ Phys.\  {\bf B276}, 391 (1986).
}

\lref\psf{
  N.~Berkovits,
  ``Super-Poincare covariant quantization of the superstring,''
  JHEP {\bf 0004}, 018 (2000)
  [arXiv:hep-th/0001035].
}
\lref\PSS{
  C.R.~Mafra,
  ``PSS: A FORM Program to Evaluate Pure Spinor Superspace Expressions,''
[arXiv:1007.4999 [hep-th]].
}

\lref\Sturmfels{
B. Sturmfels,
``Algorithms in Invariant Theory,'' Second edition, 2008, Springer, Wien.}

\lref\Cox{
D.A. Cox, J.B. Little, D. O'Shea, 
``Ideals, Varieties, and Algorithms,'' Third Edition, 2007, Springer, Berlin.}

\lref\LANCE{
  L.J.~Dixon,
``Scattering amplitudes: the most perfect microscopic structures in the universe,''
[arXiv:1105.0771 [hep-th]].
}

\lref\Srivastava{
H.M. Srivastava and P.W. Karlsson
``Multiple Gaussian hypergeometric series,''
Chichester, West Sussex, 1985.}

\listtoc
\writetoc
\break

\newsec{Introduction}

During the last years remarkable progress has been accumulated in our understanding and 
in our ability to compute scattering amplitudes, both for theoretical and 
phenomenological purposes, \cf ref.~\BernDW\ for a recent account.
Striking relations have emerged and simple structures have been discovered 
leading to a beautiful harmony between seemingly different structures and aspects of gauge and gravity scattering amplitudes \cf ref.~\LANCE.
As an example we mention the duality between color and kinematics, which exhibits a new
structure in gauge theory \BCJ. This property allows to rearrange the kinematical factors
in the amplitude such, that the form of the amplitude becomes rather simple. 
Moreover, recently it has been shown \BernIA, that the duality between color and kinematics  
allows to essentially interchange the role of color and kinematics in the full 
color decomposition of the amplitude. Many of the nice properties encountered in gauge amplitudes
take over to graviton scattering.

The properties of scattering amplitudes in both gauge and gravity theories suggest a deeper 
understanding from string theory, \cf ref.~\BernFY\ for a recent review.  
In fact, many striking field theory relations such as Bern--Carrasco--Johansson (BCJ) or 
Kawai--Lewellen--Tye (KLT) relations can be easily derived from and understood in string theory 
by tracing these identities back to the monodromy properties of the string world--sheet 
\refs{\KawaiXQ\StiebergerHQ-\BjerrumBohrRD}.
Furthermore, recently it has been shown, how superstring amplitudes can be used to 
efficiently provide numerators satisfying the color identities \KITP.
We shall demonstrate in this work, that the complete result for the $N$--point superstring 
amplitudes displays properties and symmetries inherent in field--theory and reveals structures relevant to field--theory.
Moreover, we find a beautiful harmony of the string amplitudes with strong interrelations 
between field--theory and string theory.

When computing amplitudes it is highly desirable to obtain results which are both simple and compact.
In \PARTONE\ we show how the pure spinor formalism \psf\
can be used to accomplish this
for the complete $N$--point superstring disk amplitude, which is given by
\eqn\Simple{
\Ac(1,\ldots,N)=\sum_{\si\in S_{N-3}} \Ac_{YM}(1,2_\si,\ldots,(N-2)_\si,N-1,N)\ F^\si(\ap)\ ,}
where $\Ac_{YM}$ represent 
$(N-3)!$ color ordered Yang--Mills (YM) subamplitudes, $F^\si(\ap)$ are generalized
Euler integrals encoding the full $\ap$--dependence of the string amplitude and $i_\si=\si(i)$.
The intriguing result \Simple\ disguises a lot of structure linking aspects of gauge amplitudes
as color and kinematics with properties of generalized Euler integrals. 
Both the Yang-Mills subamplitudes $\Ac_{YM}$ and the
hypergeometric integrals $F$ are reduced to a minimal basis of $(N-3)!$ elements each.
Relations among the  integrals $F$ and relations among the string-- or field--theory subamplitudes 
are found to be in one--to--one correspondence, hinting a duality between
color and kinematics at the level of the full fledged superstring amplitude.

The pure spinor formalism proved to be crucial to arrive at the compact expression \Simple. It 
provides a manifestly spacetime supersymmetric approach to superstring theory which can still 
be quantized covariantly \psf. Correlation functions of the worldsheet CFT in the pure spinor 
formalism can be efficiently organized in terms of so-called BRST building blocks \refs{\MSST, \MafraJQ}. 
These are composite superfields which transform covariantly under the BRST operator and have the 
right symmetry properties to allow for an interpretation in terms of diagrams made of cubic 
vertices \FTAmps. As shown in \PARTONE, manipulations of the BRST-covariant building blocks and the 
hypergeometric integrals reduce the number of distinct integrals in the $N$--point disk
amplitude down to $(N-3)!$. At the same time, field theory subamplitudes 
${\cal A}_{YM}(1,2_\si,\ldots,(N-2)_\si,N-1,N)$ build up as the associated kinematic factors.

So far, $N$--point superstring disk amplitudes have been computed up to seven open strings, 
\ie $N\leq 7$.
The scattering amplitude of four open superstrings has been known for a long time \FOURLIT.
Five--point superstring disk amplitudes have been computed in the RNS formalism in 
refs.~\refs{\Medinas,\StieOpr}, while in the pure spinor formalism in refs.~\refs{\FTAmps,\mafrabcj}.
Furthermore, six open string amplitudes have been computed in refs.~\refs{\StieOpr,\STi\STii\stieS-\StiebergerAM} 
in the RNS formalism, while in pure spinor superspace in refs.~\MSST.
Finally, seven open string amplitudes with MHV helicity configurations have been computed in the RNS formalism in \stieS.
However, the result \Simple\ represents the first superstring disk 
amplitude beyond $N\geq 7$ including the complete kinematics. In addition, in contrast to the
previous results, \eqq \Simple\ yields also very compact expressions for arbitrary $N$ and 
independent on the chosen helicity configuration and the space--time dimension.

The organization of the present work is as follows. In section 2 we discuss and explore the 
result \Simple\ to reveal the various structures shared by this result.
We find a complementarity between the system of equations derived by the monodromy relations 
(giving rise to relations between subamplitudes of different color ordering for the same kinematics) 
and the system of equations derived from partial fraction decomposition or partial integrations 
(giving rise to relations between functions referring to different kinematics for the same color ordering).
We display the full color decomposition of the full string amplitude and comment on a possible 
string manifestation of the recently anticipated swapping symmetry between color and kinematics in the color decomposition of the full amplitude \BernIA. 
In section 3 the module of  multiple hypergeometric functions is analyzed in detail. 
We present a method to determine the leading poles  of Euler integrals. Partial fraction 
expansion of these integrals can be made according to their leading
pole structure. Furthermore, 
a Gr\"obner basis analysis provides an independent set of rational functions or monomials 
for the Euler integrals without poles. Any  partial fraction decomposition of finite 
Euler integrals can be expressed in terms of this basis.
In section 4 we have some concluding remarks and comment on applications and implications 
of our result in view of effective D--brane action, recursion relation and graviton scattering amplitudes.
In appendix A we propose a method to analyze the transcendentality properties  of Euler integrals.  In appendix B for the six open superstring amplitude 
we present the extended  set of functions  and its relation to the minimal basis set.
Finally, in appendix C we present  $\ap$--expansions
of the basis functions $F^\si$ for $N\geq 7$.


\newsec{The structure of the $N$--point superstring disk amplitude}

The complete superstring $N$--point disk subamplitude is given by \PARTONE
\eqn\SimpleN{
\Ac(1,\ldots,N)=\sum_{\si\in S_{N-3}} \Ac_{YM}(1,2_\si,\ldots,(N-2)_\si,N-1,N)\ 
F_{(1,\ldots,N)}^\si(\ap)\ .}
In \eqq \SimpleN, 
$F^\si\equiv F_{(1,\ldots,N)}^\si$ denotes the set of $(N-3)!$ integrals which will be explicitly given
in subsection 2.3. The labels $(1,\ldots,N)$ in $F_{(1,\ldots,N)}^\si$ 
are related to the integration region of the integrals and are responsible for dictating which color--ordering of the 
superstring subamplitude is being computed.
The result \SimpleN\ is valid in any space--time dimension $D$, for 
any compactification and any amount of supersymmetry. Furthermore, the expression 
\SimpleN\ does not make any reference to any kinematical or helicity choices.
In the following we explore the  result \SimpleN\ to illuminate the role of color and kinematics.

\subsec{Basis representations: kinematics vs. color}

In field--theory there are in total $(N-3)!$ independent YM color--ordered subamplitudes $\Ac_{YM}$ \BCJ,
see refs.\ \refs{\StiebergerHQ,\BjerrumBohrRD} for a string theory derivation of this result.
Hence, in field--theory any subamplitude $\Ac_{YM}(1_\Pi,\ldots,N_\Pi)$,
with $\Pi\in S_N$, can be expressed as 
\eqn\DECO{
\Ac_{YM}(1_\Pi,\ldots,N_\Pi)=\sum_{\si\in S_{N-3}} K_\Pi^\si\ \Ac_{YM,\si}\ ,}
with $i_\Pi=\Pi(i)$, some universal and state--independent kinematic coefficients
$K_\Pi^\si$ generically depending on the kinematic invariants, \cf \eqq  (2.7) for a straightforward
derivation. Besides, we introduced the abbreviation:
\eqn\DEFI{
\Ac_{YM,\si}:=\Ac_{YM}(1,2_\si,\ldots,(N-2)_\si,N-1,N)\ .}
One crucial property of \SimpleN\ is the fact that the superstring $N$--point (sub)amplitude is
decomposed in terms of a $(N-3)!$ basis of Yang--Mills color ordered amplitudes $\Ac_{YM,\si}$, i.e. the whole superstring amplitude can be decomposed w.r.t. the 
kinematics described by the set of $\Ac_{YM,\si},\ \si\in S_{N-3}$.
Hence, by these results it is obvious, that in the sum of \SimpleN\ only $(N-3)!$ terms 
and as many different multiple hypergeometric functions can appear since any additional
kinematical term could be eliminated by redefining the functions $F^\si$ thanks to the 
amplitude relations \DECO.

Moreover, the string subamplitudes \SimpleN\ 
solve the system of relations given by 
\eqn\DUAL{\eqalign{
\Ac(1,2,\ldots,N)&+e^{i\pi s_{12}}\ \Ac(2,1,3,\ldots,N-1,N)+
e^{i\pi( s_{12}+ s_{13})}\ \Ac(2,3,1,\ldots,N-1,N)\cr
&+\ldots+e^{i\pi(s_{12}+s_{13}+\ldots+ s_{1N-1})}\ \Ac(2,3,\ldots,N-1,1,N)=0}}
and permutations thereof. Throughout this work, we will be mostly using dimensionless 
Mandelstam invariants $s_{ij}=\ap(k_i+k_j)^2$. The set of identities \DUAL\ 
has been derived from the monodromy properties of the disk world--sheet 
\refs{\StiebergerHQ,\BjerrumBohrRD}.

Furthermore, since there exists a basis of $(N-3)!$ YM building blocks allowing for the decomposition
\DECO, we may express any string subamplitude by {\it one} specific set of YM amplitudes 
$\Ac_{YM,\si}$ referring \eg to the string amplitude \SimpleN:
\eqn\furthernice{
\Ac(1_\Pi,\ldots,N_\Pi)=\sum_{\si\in S_{N-3}}\ \Ac_{YM,\si}\ F^\si_\Pi(\ap)\ ,}
with $\Pi\in S_N$.
Inserting the set \furthernice\ into the monodromy relations
yields  a set of relations for the functions $F_\Pi^\si$ for each given $\si\in S_{N-3}$.
{\it E.g.} \DUAL\ gives the following set of identities: 
\eqn\DUALF{\eqalign{
F^\si_{(1,\ldots,N)}&+e^{i\pi s_{12}}\ F^\si_{(2,1,3,\ldots,N-1,N)}+
e^{i\pi( s_{12}+ s_{13})}\ F^\si_{(2,3,1,\ldots,N-1,N)}\cr
&+\ldots+e^{i\pi(s_{12}+s_{13}+\ldots+ s_{1N-1})}\ F^\si_{(2,3,\ldots,N-1,1,N)}=0\ \ \ ,\ \ \ 
\si\in S_{N-3}\ .}}
Hence, for a given $\si\in S_{N-3}$ corresponding to the given YM amplitude 
$\Ac_{YM,\si}$ the set of functions $F_\Pi^\si,\ \Pi\in S_N$ 
enjoys the monodromy relations. 
As a consequence for each permutation $\si\in S_{N-3}$ or YM basis amplitude $\Ac_{YM,\si}$ 
there are $(N-3)!$ different functions $F^\si_\Pi$ all related through the 
equations \DUALF\ and permutations thereof.

Note that the $\ap\ra0$ limit of \eqq \furthernice\ reproduces explicit expressions of the kinematic coefficients 
$K_\Pi^\si$ introduced in \DECO\ (which were already given in \BCJ\ for $N$--point field theory amplitudes):
\eqn\CARLOS{
K_\Pi^\si=\lf.F^\si_\Pi(\ap)\ \ri|_{\ap=0}\ .}
This relation enables to compute the matrix elements $K_\Pi^\si$ {\it directly} by means of extracting
the  field--theory limit  of the string world--sheet integrals $F^\si_\Pi(\ap)$ 
(by the method described in section 3.3) rather than by solving the monodromy relations \DUAL.

Further insights can be gained when looking at different representations for the same
amplitude \SimpleN:
\eqn\OTHER{
\Ac(1,\ldots,N)=\sum_{\pi\in S_{N-3}} \Ac_{YM,\pi}\ F^\pi_{(1,\ldots,N)}(\ap)\ ,}
with some permutations $\pi\in S_{N-3}$ singling out a basis of $(N-3)!$ independent basis amplitudes $A_{YM,\pi}$. More precisely, in contrast to the set $\Ac_{YM,\si}$ in \DEFI, the new set $\Ac_{YM,\pi}$ in \OTHER\ represents a more general basis of $(N-3)!$ independent
subamplitudes $\Ac_{YM,\pi}$, where three legs $i,j,k$ (possibly other than $1,N-1,N$) 
are fixed and the remaining ones are permuted by $\pi \in S_{N-3}$.

By applying the decomposition \DECO\ and comparing the two expressions 
\OTHER\ and \SimpleN\ we find the relation between the set of $(N-3)!$ new and old 
independent basis functions $F^\pi_{(1,\ldots,N)}$ and $F^\si_{(1,\ldots,N)}$:
\eqn\FINDREL{
F^\si_{(1,\ldots,N)}=\sum_{\pi\in S_{N-3}} (K^{-1})_\pi^\si\ F^\pi_{(1,\ldots,N)}\ \ \ ,\ \ \ 
\si\in S_{N-3}\ .}
In this case the matrix $K$ becomes a quadratic $(N-3)!\times(N-3)!$ matrix, \cf
subsection 2.3 for explicit examples.
Hence, for a given {\it fixed} color 
ordering $(1,\ldots,N)$ any function $F^\si$ may be expressed in terms of a basis
of $(N-3)!$ functions $F^\pi$ referring to the {\it same} color ordering.
With \FINDREL\ sets of systems of equations involving 
the kinematics functions $F^\pi$ (of the {\it same} color ordering) can be generated. 
According to \CARLOS\ the field--theory limits of the functions $F^\si_\pi$ 
are enough to determine the coefficients of these equations.

The relation \FINDREL\ should be compared with \DECO: While in the first identity
one specific color ordered amplitude is  decomposed  w.r.t.\ to a set of $(N-3)!$ 
independent color ordered amplitudes all referring to the {\it same kinematics}, 
in the second identity one functions referring to one specific kinematics is decomposed to w.r.t.\ 
to a set of  $(N-3)!$ independent kinematics functions all referring to the {\it same color ordering}.
Moreover, as we shall show in subsection 2.3. 
for a fixed color ordering $(1,\ldots,N)$ 
an explicit set of $(N-2)!$ functions $F^\Pi_{(1\ldots N)},\ \Pi
\in S_{N-2}$ can be given, which fulfills \FINDREL\ -- just as a set of $(N-2)!$ YM amplitudes
$\Ac_{YM,\Pi}$ fulfills \DECO\ for a fixed kinematics. 
Since the latter fact is a result of the (imaginary part) field--theory monodromy relations, 
also the relations \FINDREL\ should follow from a system of equations for  the $(N-2)!$ functions.
Relations between functions $F^\Pi_{(1\ldots N)}$ of same color ordering are obtained by 
either partial fraction decomposition of their integrands or applying partial integration 
techniques within their $N-3$ integrals. The partial fraction expansion yields linear 
equations with integer coefficients for the functions $F^\Pi$ 
-- just like the (real part)  field--theory monodromy relations  yield linear identities (e.g. subcyclic 
identities) for the color ordered subamplitudes $\Ac_{YM}$.
On the other hand, the  partial integration techniques applied to the $(N-2)!$ functions $F^\Pi$ 
provides a system of equations of rank $(N-3)!$,  whose solution is given by \FINDREL.
Hence, we have found a complete analogy between the monodromy relations equating subamplitudes
$\Ac_{YM_,\Pi}$ of {\it different color orderings} $\Pi\in S_{N-2}$ at the {\it same kinematics} 
and a system of equations relating functions 
$F^\Pi$ referring to {\it different kinematics}  $\Pi\in S_{N-2}$ at the {\it same color ordering}.

To conclude, behind the expression \SimpleN\ there are two sets of equations: one set, derived from the 
monodromy relations \DUAL\ and equating all subamplitudes of different color orderings and an other 
set, derived from the  partial fraction decomposition and partial integration relations equating all kinematics 
functions $F^\pi$.
Both systems are of rank $(N-3)!$ and allow to express all colored ordered subamplitudes in terms
of a minimal basis or to express all kinematic functions in terms of a minimal basis.

\subsec{Color decomposition of the full  open superstring amplitude}

The  color decomposition of the full $N$--point open superstring amplitude $\Mc_N$ can be expressed by 
$(N-3)!\times (N-3)!$ different functions $F_\Pi^\si$ with $(N-3)!$ YM building blocks 
$\Ac_{YM,\si}$. Firstly, the monodromy relations \DUAL\ allow to decompose each superstring 
subamplitude in a $(N-3)!$ element basis \refs{\StiebergerHQ,\BjerrumBohrRD}\foot{In ref.\ \BOHR, 
systems of equations of this type are neatly rephrased in terms of the so--called momentum kernel 
matrix ${\cal S}_{\alpha'}[\pi |\si]$, which keeps track of relative monodromy phases between two $S_{N-2}$ permutations $\pi$ and $\si$. It has 
non--maximal rank $(N-2)! - (N-3)!$, so the linear relations between color ordered superstring amplitudes can be compactly represented as
$\sum\limits_{\si \in S_{N-2}} {\cal S}_{\alpha'}[\pi |\si] \, {\cal A}(1,2_\si,3_\si,\ldots,(N-1)_\si,N)  =  0 , \ \ \ \pi \in S_{N-2}$.
On the level of the functions, this relation implies:
$\sum\limits_{\si \in S_{N-2}} {\cal S}_{\alpha'}[\pi |\si] \, F^\rho_{(1,2_\si,3_\si,\ldots,(N-1)_\si,N)} = 0\ , \ \ \ \pi,\rho \in S_{N-2}\ .$}
\eqn\DECOS{
\Ac(1_\Pi,\ldots,N_\Pi)=\sum_{\pi\in S_{N-3}} \Kc_\Pi^\pi\ \Ac(1,2_\pi,\ldots,(N-2)_\pi,N-1,N)\ ,}
which generalizes the field--theory equation \DECO\ in the sense that $\Kc_\Pi^\pi(\alpha') \, |_{\alpha'=0} =K^\pi_\Pi$. 
The basis expansion \DECOS\ simplifies the color dressed superstring amplitude to
\eqn\FULLA{
\Mc_N=\sum_{\Pi\in S_{N-1}}
\Tr(T^{a_1}T^{a_{2_\Pi}}\ldots T^{a_{N_\Pi}})\ \sum_{\si\in S_{N-3}}
\Ac_{YM,\si}\ \sum_{\pi\in S_{N-3}} \Kc_\Pi^\pi\ F_\pi^\si\ ,}
with:
\eqn\withF{
F_\pi^\si:=F_{(1,\pi(2),\ldots,\pi(N-2),N-1,N)}^\si(\ap)\ .}
In the sum \FULLA\ the same set of basis elements $\Ac_{YM,\si}$ is used for all color 
orderings $\Pi$.
This enables to reorganize the color decomposition sum and to interchange the two sums over color and kinematics:
\eqn\FULLB{
\Mc_N=\sum_{\si\in S_{N-3}}\Ac_{YM,\si}
\sum_{\Pi\in S_{N-1}}\Tr(T^{a_1}T^{a_{2_\Pi}}\ldots T^{a_{N_\Pi}})\ \sum_{\pi\in S_{N-3}}
\Kc_\Pi^\pi\ \ F_\pi^\si\ .}
Now in \FULLB\ the role of color and kinematics is swapped. While \FULLA\ represents a color 
decomposition in terms of $(N-1)!/2$ color ordered subamplitudes, the sum \FULLB\ is a 
decomposition w.r.t.\ to $(N-3)!$ kinematical factors $\Ac_{YM,\si}$.
The sum \FULLB\ could be the string theory realization of the 
recently found observation, that in the color decomposition of a gauge theory amplitude the role
of color and kinematics may be swapped \BernIA. In these lines the sum over $\Pi$ may represent the pre-version of a dual amplitude 
${\cal A}_N^{\rm dual}$, in which all kinematical factors $\Ac_{YM,\Pi}$ are replaced by color 
traces. However, further studies are necessary to establish a clear dictionary between 
Yang--Mills building blocks $\Ac_{YM,\Pi}$ and the kinematic analogue $\tau_{(12\ldots N)}$ 
of color traces: On the one hand, our $\Ac_{YM,\Pi}$ have the required cyclicity property, on 
the other hand, they still carry the kinematic poles which should ultimately be outsourced from 
the local  $\tau_{(12\ldots N)}$ into the dual amplitudes
${\cal A}_N^{\rm dual}$.

\subsec{Yang--Mills subamplitudes}

Compact expressions for $\Ac_{YM}(1,2_\si,\ldots,(N-2)_\si,N-1,N)$ in $D=10$ are derived in \MafraJQ\
and can be used to describe the YM subamplitudes of \SimpleN.
On the other hand for $D=4$, compact forms 
for $\Ac_{YM}(1,2_\si,\ldots,(N-2)_\si,N-1,N)$ in the spinor helicity basis
can be looked up in the literature: In the maximal helicity violating (MHV) case, the subamplitudes
reduce to the famous Parke--Taylor or Berends--Giele formula \refs{\mhv,\BerendsME}. 
For the general NMHV case, the complete expressions
for $\Ac_{YM}(1,2_\si,\ldots,(N-2)_\si,N-1,N)$ can be found in \DixonIK.

Since in the sum \SimpleN\ the kinematical factors $\Ac_{YM}$ and the functions $F^\si$ encoding the 
string effects are multiplied together, supersymmetric Ward identities established in 
field--theory \refs{\GrisaruVM\ParkePN-\ElvangWD} 
hold also for the full superstring amplitude, \cf also \stieS. 
At any rate, after component expansion the pure spinor result provides the $N$--point amplitude 
involving any member of the SYM vector multiplet (VM) \PSS.

\subsec{Minimal basis of multiple hypergeometric functions $F^\si$}

The system of $(N-3)!$ multiple hypergeometric functions $F^\si$ 
appearing in \SimpleN\ are given as generalized Euler integrals \PARTONE\foot{In contrast to \PARTONE\ here we
use momenta redefined by a factor of $i$. As a consequence the signs of the kinematic invariants
are  flipped, \eg $|z_{il}|^{-s_{il}}\ra|z_{il}|^{s_{il}}$.}
\eqn\revol{
F^{(23\ldots N-2)}(s_{ij}) = (-1)^{N-3}\int\limits_{z_i<z_{i+1}}  
\prod_{j=2}^{N-2} dz_j\  \lf(\prod_{i<l} |z_{il}|^{s_{il}}\ri) \ 
\left\{\ \prod_{k=2}^{N-2}  \sum_{m=1}^{k-1} \fc{ s_{mk} }{z_{mk}} \ \right\}\  ,}
with permutations $\si\in S_{N-3}$ acting on all indices within the curly brace.
Integration by parts admits to simplify the  integrand in \revol. As a result the length of 
the sum over $m$ becomes shorter for $k > [N/2]$:
\eqnn\revoli
$$\eqalignno{
F^{(23\ldots N-2)}(s_{ij}) &= (-1)^{N-3} \int\limits_{z_i<z_{i+1}}  
\prod_{j=2}^{N-2} dz_j \ \lf(\prod_{i<l} |z_{il}|^{s_{il}}\ri) & \revoli \cr\crr 
&\times\left( \ \prod_{k=2}^{[N/2]} \  \sum_{m=1}^{k-1} \ \fc{ s_{mk}}{z_{mk}} \ \ri)\ \lf(\  
\prod_{k=[N/2]+1}^{N-2} \ \sum_{n=k+1}^{N-1} \ \fc{ s_{kn}}{z_{kn}} \ \right)\  .}$$
In the above, $[\ldots ]$ denotes the Gauss bracket $[x] = \max_{n \in \IZ,n \leq x} n$, which picks out
the nearest integer smaller than or equal to its argument.

The result \SimpleN\ is manifestly gauge invariant as a consequence of gauge invariance of the 
YM subamplitudes $\Ac_{YM}$. Hence, gauge invariance does not impose further restrictions on the 
$(N-3)!$ functions $F^\si_{(1,\ldots,N)}$, which would further reduce the basis.
The set \revol\ of $(N-3)!$ functions represents a minimal basis for the set of multiple 
Gaussian hypergeometric functions or Euler integrals appearing at $N$--point and referring to the same
color ordering $(1,\ldots N)$ or integration region $z_1<\ldots <z_N$.
Any function of this ordering can be expressed in terms of this basis.

The lowest terms of the $\ap$--expansion of the functions $F^\si$ assume the form
\eqn\LOW{\eqalign{
F^\si&=1+\ap^2\ p_2^\si\ \zeta(2)+\ap^3\ p_3^\si\ \zeta(3) +\ldots\ \ \ ,\ \ \ \si=(23\ldots N-2)\ ,\cr
F^\si&=\ap^2\ p_2^\si\ \zeta(2)+\ap^3\ p_3^\si\ \zeta(3)+\ldots\ \ \ ,\ \ \ \si\neq(23\ldots N-2)\ ,}}
with some polynomials $p_n^\si$ of degree $n$ in the dimensionful kinematic invariants 
$\hat s_{ij}=(k_i+k_j)^2 = s_{ij}/\alpha'$ and $\hat s_{i\ldots l}=(k_i+\ldots+k_l)^2 = s_{i\ldots l}/ \alpha'$.
Note that starting at $N\geq 7$ subsets of $F^\si$ start at even higher order in $\ap$, 
\ie $p_2^\si,\ldots,p_\nu^\si=0$ for some $\nu\geq2$, \cf section~3 and appendix~\appC\ for 
further details.
Hence, only the first term of \SimpleN\ contributes to the field--theory limit of the full 
$N$--point superstring amplitude. 
The power series expansions \LOW\ in $\ap$ is such,  that to each power $\ap^n$
a transcendental function of degree $n$ shows up. More precisely,
a set of multizeta values (MZVs) of fixed weight $n$ appears at $\ap^n$. The latter are multiplied by 
a polynomial $p_n^\si$ of degree $n$ in the kinematic invariants $\hat s$ with rational coefficients. 
We refer the reader to subsection 3.1 and appendix \appA\ for more details on $\ap$--expansions and MZVs.
{}From \LOW\ we conclude, 
that the whole pole structure of the amplitude \SimpleN\ is encoded in the YM subamplitudes
$\Ac_{YM}$, while the functions $F^\si$ are finite, \ie do not have poles in the kinematic 
invariants. A detailed account on multiple Gaussian hypergeometric functions can be found in \Srivastava.

\subsec{Extended set of multiple hypergeometric functions $F^\Pi$}

A system of $(N-2)!$ functions $F^\Pi$, which fulfills \FINDREL\ can be given as follows
\eqn\revoll{
F^{(23\ldots N-1)}(s_{ij}) = \int\limits_{z_i<z_{i+1}}  
\prod_{j=2}^{N-2} dz_j\  \lf(\prod_{i<l} |z_{il}|^{s_{il}}\ri) \ 
\left\{\ \fc{(-1)^{N-3}}{z_{N-1}-z_1}\ \prod_{k=2}^{N-2}  \sum_{m=1}^{k-1} \fc{ s_{mk} }{z_{mk}} \ \right\},}
with permutations $\Pi\in S_{N-2}$ acting on all indices within the curly brace.
The set of $(N-2)!$ functions \revoll\ can be expressed in terms of the basis \revol\ as a
consequence of the relations \FINDREL. This allows to express $(N-2)!-(N-3)!=(N-3)\times(N-3)!$ functions of \revoll\ in terms of \revol.
This will be demonstrated at some examples in the next subsection.

In contrast to the minimal set of functions $F^\sigma, \ \sigma \in S_{N-3}$, some elements of the extended set $F^\Pi, \ \Pi \in S_{N-2}$ might have poles in individual Mandelstam invariants.

\subsec{Examples}

\subsubsec{$N=4$}

The unique integral appearing in \SimpleN\ for the four--point amplitude is
\eqnn\exiv
$$\eqalignno{
F^{(2)}&=-\int_0^1 dz_2\  \lf(\prod_{i<l} |z_{il}|^{s_{il}}\ri)\ 
\fc{s_{12}}{z_{12}}=\fc{\Gamma(1+s_{12})\ \Gamma(1+s_{23})}{\Gamma(1+s_{12}+s_{23})}  \cr
&=1-\zeta(2)\ s_{12}s_{23}+\zeta(3)\ s_{12} s_{13}s_{23}+\Oc(\ap^4)\ . &\exiv}$$
The extended set of two functions consists of \exiv\ (with $F^{(2)}\equiv F^{(23)}$)  and the additional function \revoll:
\eqnn\Exiv
$$\eqalignno{
F^{(32)}&=-\int_0^1 dz_2\  \lf(\prod_{i<l} |z_{il}|^{s_{il}}\ri)\ 
\fc{1}{z_{21}}\ \fc{s_{13}}{z_{13}}=\fc{s_{13}}{s_{12}}\ 
\fc{\Gamma(1+s_{12})\ \Gamma(1+s_{23})}{\Gamma(1+s_{12}+s_{23})}\cr
&=\fc{s_{13}}{s_{12}}-\zeta(2)\ s_{13}s_{23}+\zeta(3)\ s_{13}^2 s_{23}+\Oc(\ap^4)\ . &\Exiv }$$
With this extended set of two functions we may explicitly verify the relation \FINDREL.
For the new basis $\pi=\{(1,3,2,4)\}$ in \eqq \DECO\ we have 
\eqn\KSP{
K_\pi^\si=\fc{s_{12}}{s_{13}}}
w.r.t.\ the reference basis $\si=\{(1,2,3,4,5)\}$ as a consequence of the
field--theory relation $\Ac_{YM}(1,3,2,4)=\fc{s_{12}}{s_{13}}\Ac_{YM}(1,2,3,4)$.
According to \FINDREL\ the following identity indeed holds:
\eqn\holdi{
F^{(32)}=K^{-1}\ F^{(23)}\ .}

\subsubsec{$N=5$}

The set of two basis functions appearing in \SimpleN\ and following from \revol\ is:
\eqnn\exv
$$\eqalignno{
F^{(23)}&=\int\limits_{0<z_2<z_3<1} dz_2 dz_3\  \lf(\prod_{i<l} |z_{il}|^{s_{il}}\ri)\ 
\fc{s_{12}}{z_{12}}\ \lf(\fc{s_{13}}{z_{13}}+\fc{s_{23}}{z_{23}}\ri)\cr
&=\int\limits_{0<z_2<z_3<1} dz_2 dz_3\  
 \lf(\prod_{i<l} |z_{il}|^{s_{il}}\ri)\ \fc{s_{12}}{z_{12}}\ \fc{s_{34}}{z_{34}}=1+\zeta(2)\ (s_1s_3-s_3s_4-s_1s_5)\cr
&-\zeta(3)\ (s_1^2 s_3+2 s_1 s_2 s_3+s_1 s_3^2-s_3^2 s_4-s_3 s_4^2-s_1^2 s_5-s_1 s_5^2)+
\Oc(\ap^4),\cr
F^{(32)}&=\int\limits_{0<z_2<z_3<1} dz_2 dz_3\    \lf(\prod_{i<l} |z_{il}|^{s_{il}}\ri)\ 
\fc{s_{13}}{z_{13}}\ \lf(\fc{s_{12}}{z_{12}}+\fc{s_{32}}{z_{32}}\ri)\cr
&=\int\limits_{0<z_2<z_3<1} dz_2 dz_3\  
\lf(\prod_{i<l} |z_{il}|^{s_{il}}\ri)\ \fc{s_{13}}{z_{13}}\ \fc{s_{24}}{z_{24}}\cr
&=\zeta(2)\ s_{13}\ s_{24}-\zeta(3)\ s_{13}\ s_{24}\ (s_1+s_2+s_3+s_4+s_5)+\Oc(\ap^4)\ ,&\exv}
$$
where $s_i \equiv \alpha'(k_i+k_{i+1})^2$ subject to cyclic identification $k_{i+N} \equiv k_i$.

The extended set of six functions consists of \exv, with 
\eqn\withVI{\eqalign{
F^{(234)}:=F^{(23)}\ \ \ ,\ \ \ F^{(324)}:=F^{(32)}\ ,}}
and the additional four functions \revoll:
\eqnn\Exv
$$\eqalignno{
F^{(423)}&=\int\limits_{0<z_2<z_3<1} dz_2 dz_3\  
 \lf(\prod_{i<l} |z_{il}|^{s_{il}}\ri)\ \fc{1}{z_{31}}\  
 \fc{s_{14}}{z_{14}}\ \fc{s_{23}}{z_{23}}\ ,\cr
F^{(243)}&=\int\limits_{0<z_2<z_3<1} dz_2 dz_3\  
 \lf(\prod_{i<l} |z_{il}|^{s_{il}}\ri)\ \fc{1}{z_{31}}\  
 \fc{s_{12}}{z_{12}}\ \fc{s_{34}}{z_{43}}\ ,\cr 
F^{(432)}&=\int\limits_{0<z_2<z_3<1} dz_2 dz_3\  
 \lf(\prod_{i<l} |z_{il}|^{s_{il}}\ri)\ \fc{1}{z_{21}}\  
 \fc{s_{14}}{z_{14}}\ \fc{s_{23}}{z_{32}}\ ,\cr 
F^{(342)}&=\int\limits_{0<z_2<z_3<1} dz_2 dz_3\  
 \lf(\prod_{i<l} |z_{il}|^{s_{il}}\ri)\ \fc{1}{z_{21}}\  
 \fc{s_{13}}{z_{13}}\ \fc{s_{24}}{z_{42}}\ . &\Exv }
$$
With this extended set of six functions we may explicitly verify the relation \FINDREL.
For the new basis $\pi=\{(1,4,2,3,5),\ (1,2,4,3,5)\}$ in \eqq \DECO\ we have 
\eqn\KSPi{
K_\pi^\si=\fc{1}{s_{14}\ s_{35}}
\pmatrix{s_{12}\ s_{34}&-s_{13}\ (s_{34}+s_{45})\cr
s_{14}\ (s_{12}-s_{45})&-s_{14}\ s_{13}}}
w.r.t.\ the reference basis $\si=\{(1,2,3,4,5),\ (1,3,2,4,5)\}$.
According to \FINDREL\ the following identity indeed holds (with $K^\ast=(K^{-1})^t$):
\eqn\holdi{
\lf({F^{(423)}\atop F^{(243)}}\ri)=K^\ast\ \lf({F^{(234)}\atop F^{(324)}}\ri)\ .}
On the other hand, for the new basis $\pi~=~\{(1,4,3,2,5),\ (1,3,4,2,5)\}$ we have
\eqn\KSPii{
K_{\pi}^\si=\fc{1}{s_{14}\ s_{35}}
\pmatrix{s_{12}\ (s_{14}+s_{34})&s_{13}\ s_{24}\cr
-s_{12}\ s_{14}&-s_{14}\ (s_{12}+s_{23})}\ ,}
and the following relation can be checked:
\eqn\holdii{
\lf({F^{(432)}\atop F^{(342)}}\ri)=K^\ast\ \lf({F^{(234)}\atop F^{(324)}}\ri)\ .}
Hence, the relations \holdi\ and \holdii\ allow to express the additional set of functions \Exv\
in terms of the minimal basis \exv.

\subsubsec{$N=6$}

The set of six basis functions appearing in \SimpleN\ and following from \revol\ is
\eqnn\exvi
$$\eqalignno{
F^{(234)}&=-\hskip-0.5cm\int\limits_{0<z_2<z_3<z_4<1} \hskip-0.5cm dz_2 dz_3 dz_4\ 
\lf(\prod_{i<l} |z_{il}|^{s_{il}}\ri)\  \fc{s_{12}}{z_{12}}\ \fc{s_{45}}{z_{45}}\ 
\lf(\fc{s_{13}}{z_{13}}+\fc{s_{23}}{z_{23}}\ri)=\cr
&\hskip0.5cm=1-\zeta(2)\ (s_4s_5+s_1s_6-s_4t_1-s_1t_3+t_1t_3)\cr
&\hskip0.5cm+\zeta(3)\ \lf(2 s_1 s_2 s_4+2 s_1 s_3 s_4+s_4^2 s_5+s_4 s_5^2+s_1^2 s_6+s_1 s_6^2-2 s_3 s_4 t_1\ri.\cr
&\hskip0.5cm\lf.-s_4^2 t_1-s_4 t_1^2-2 s_1 s_4 t_2-s_1^2 t_3-2 s_1 s_2 t_3+t_1^2 t_3-s_1 t_3^2+
t_1 t_3^2\ri)+\Oc(\ap^4),\cr
F^{(324)}&=-\hskip-0.5cm\int\limits_{0<z_2<z_3<z_4<1}  \hskip-0.5cm dz_2 dz_3 dz_4\ 
\lf(\prod_{i<l} |z_{il}|^{s_{il}}\ri)\  \fc{s_{13}}{z_{13}}\ \fc{s_{45}}{z_{45}}\ 
\lf(\fc{s_{12}}{z_{12}}+\fc{s_{32}}{z_{32}}\ri)=\cr
&\hskip0.5cm=-\zeta(2)\ s_{13}d_9+\zeta(3)\ s_{13}\ 
\lf(s_1 s_2+s_2^2-2 s_2 s_4-2 s_3 s_4-s_1 s_6-s_6^2\ri.\cr
&\hskip0.5cm\lf.+s_2 t_1-s_6 t_1+2 s_4 t_2+s_1 t_3+2 s_2 t_3+t_1 t_3+t_3^2\ri)+\Oc(\ap^4),\cr
F^{(432)}&=-\hskip-0.5cm\int\limits_{0<z_2<z_3<z_4<1}  \hskip-0.5cm dz_2 dz_3 dz_4\ 
\lf(\prod_{i<l} |z_{il}|^{s_{il}}\ri)\  \fc{s_{14}}{z_{14}}\ \fc{s_{25}}{z_{25}}\ 
\lf(\fc{s_{13}}{z_{13}}+\fc{s_{43}}{z_{43}}\ri)\cr
&\hskip0.5cm=-\zeta(2)\ s_{14}s_{25}+\zeta(3)\ s_{14}s_{25}
\ \lf(-s_2-s_3+s_5+s_6+t_1+t_2+t_3\ri) +\Oc(\ap^4),\cr
F^{(342)}&=-\hskip-0.5cm
\int\limits_{0<z_2<z_3<z_4<1}  \hskip-0.5cm dz_2 dz_3 dz_4\ 
\lf(\prod_{i<l} |z_{il}|^{s_{il}}\ri)\  \fc{s_{13}}{z_{13}}\ \fc{s_{25}}{z_{25}}\ 
\lf(\fc{s_{14}}{z_{14}}+\fc{s_{34}}{z_{34}}\ri)\cr
&\hskip0.5cm=\zeta(2)\ s_{13}s_{25}+\zeta(3)\ s_{13}s_{25}
\lf(-s_1+s_2+2 s_3-s_6-t_1-2 t_2-t_3\ri)+\Oc(\ap^4),\cr
F^{(423)}&=-\hskip-0.5cm\int\limits_{0<z_2<z_3<z_4<1}  \hskip-0.5cm dz_2 dz_3 dz_4\ 
\lf(\prod_{i<l} |z_{il}|^{s_{il}}\ri)\  \fc{s_{14}}{z_{14}}\ \fc{s_{35}}{z_{35}}\ 
\lf(\fc{s_{12}}{z_{12}}+\fc{s_{42}}{z_{42}}\ri)\cr
&\hskip0.5cm=\zeta(2)\  s_{14}s_{35}+\zeta(3)\ s_{14}s_{35}
\ \lf(2 s_2+s_3-s_4-s_5-t_1-2 t_2-t_3\ri)+\Oc(\ap^4),\cr
F^{(243)}&=-\hskip-0.5cm\int\limits_{0<z_2<z_3<z_4<1}  \hskip-0.5cm dz_2 dz_3 dz_4\ 
\lf(\prod_{i<l} |z_{il}|^{s_{il}}\ri)\  \fc{s_{12}}{z_{12}}\ \fc{s_{35}}{z_{35}}\ 
\lf(\fc{s_{14}}{z_{14}}+\fc{s_{24}}{z_{24}}\ri)=\cr
&\hskip0.5cm=-\zeta(2)\ s_{35}d_1+\zeta(3)\ s_{35}\ \lf(-2 s_1 s_2-2 s_1 s_3+s_3^2+s_3 s_4-s_4 s_5-s_5^2\ri.\cr
&\hskip0.5cm\lf.+2 s_3 t_1+s_4 t_1+t_1^2+2 s_1 t_2+s_3 t_3-s_5 t_3+t_1 t_3\ri)+\Oc(\ap^4),&\exvi}
$$
with $d_1=s_3-s_5+t_1$ and $d_9=s_2-s_6+t_3$.
The extended set of $24$ functions consists of \exvi\ with
\eqn\withVI{\eqalign{
F^{(2345)}&:=F^{(234)}\ \ \ ,\ \ \ F^{(3245)}:=F^{(324)}\ \ \ ,\ \ \ F^{(4325)}:=F^{(432)}\ ,\cr
F^{(3425)}&:=F^{(342)}\ \ \ ,\ \ \ F^{(4235)}:=F^{(423)}\ \ \ ,\ \ \ F^{(2435)}:=F^{(243)}\ ,}}
and the additional $18$ functions \revoll, which are listed in the appendix \appB.

For the new basis $\pi=\{(1,2,3,5,4,6),\ (1,3,2,5,4,6),\ (1,5,3,2,4,6),\ (1,3,5,2,4,6),$
$(1,5,2,3,4,6),\ (1,2,5,3,4,6)\}$ in \eqq \DECO\ we have 
\eqn\KSIXa{\eqalign{
&K^\si_\pi=s_{46}^{-1}\cr
&\times\pmatrix{\ss{s_5-t_1}&\ss{0}&\ss{0}&\ss{0}&\ss{s_{14}}&\ss{-d_1}\cr\crr
\ss{0}&\ss{s_5-t_1}&\ss{s_{14}}&\ss{s_3+s_{14}}&\ss{0}&\ss{0}\cr\crr
\ss{\fc{s_1s_4d_0}{s_{15}t_{246}}}&\ss{\fc{s_4s_{13}(s_{25}-s_{46})}{s_{15}t_{246}}}&
\ss{\fc{-s_{13}s_{14}s_{25}}{s_{15}t_{246}}}&\ss{\fc{-s_{13}s_{25}(s_3+s_{14})}{s_{15}t_{246}}}&
\ss{\fc{s_{14}(s_{46}-s_1)d_0}{s_{15}t_{246}}}&
\ss{\fc{s_1(s_3+s_4)d_0}{s_{15}t_{246}}}\cr\crr
\ss{\fc{-s_1s_4}{t_{246}}}&\ss{\fc{-s_4(s_1+s_2)}{t_{246}}}&
\ss{\fc{s_{14}d_4}{t_{246}}}& 
\ss{\fc{(s_{14}+s_3)d_4}{t_{246}}}  &
\ss{\fc{s_{14}(s_1-s_{46})}{t_{246}}} &\ss{\fc{-s_1(s_3+s_4)}{t_{246}}}\cr\crr
\ss{\fc{s_1s_4(s_{35}-s_{46})}{s_{15}t_{125}}}&
\ss{\fc{s_4s_{13}d_3}{s_{15}t_{125}}}&
\ss{\fc{(s_{46}-s_{13})d_3s_{14}}{s_{15}t_{125}}}&
\ss{\fc{(s_4+s_{24})s_{13}d_3}{s_{15}t_{125}}}&
\ss{\fc{-s_1s_{14}s_{35}}{s_{15}t_{125}}}&
\ss{\fc{s_1s_{35}d_1}{s_{15}t_{125}}}\cr\crr
\ss{\fc{s_4(s_1-t_1)}{t_{125}}}&\ss{\fc{-s_4s_{13}}{t_{125}}}&
\ss{\fc{s_{14}(s_{13}-s_{46})}{t_{125}}}&\ss{\fc{-s_{13}(s_4+s_{24})}{t_{125}}}&
\ss{\fc{-s_{14}d_2}{t_{125}}}&\ss{\fc{d_1d_2}{t_{125}}}\cr\crr}}}
w.r.t.\ the reference basis 
$\si=\{(1,2,3,4,5,6),\ (1,3,2,4,5,6),\ (1,4,3,2,5,6),\ (1,3,4,2,5,6),$
$(1,4,2,3,5,6),\ (1,2,4,3,5,6)\}$.
According to \FINDREL\ the following identity indeed holds:
\eqn\holdi{
\lf(\matrix{
F^{(2354)}\cr
F^{(3254)}\cr
F^{(5324)}\cr
F^{(3524)}\cr
F^{(5234)}\cr
F^{(2534)}}\ri)=K^\ast\ 
\lf(\matrix{
F^{(2345)}\cr
F^{(3245)}\cr
F^{(4325)}\cr
F^{(3425)}\cr
F^{(4235)}\cr
F^{(2435)}}\ri)\ .}
In the above matrix \KSIXa\ we have introduced $d_0=s_{15}+s_{35},\ 
d_2=s_1-s_4-s_5,\ d_3=s_3-s_5-t_3,\ d_4=s_4+s_5-s_{13}$ and
$t_{ijk}= \ap(k_ik_j+k_ik_k+k_jk_k)$.
The other two sets of basis $\pi$ and their relations \FINDREL\ to the reference basis 
$\si$ are displayed in appendix \appB.

\subsec{Properties of the full amplitude}

The factorization properties of tree--level amplitudes are well studied in field--theory
\MANGO. These properties represent an important test of our string result.

\subsubsec{Soft limit}

According to subsection 2.2 it is sufficient to focus on the $N$--gluon amplitude. 
We consider the limit $k_{N-2}\ra 0$. In this limit the amplitude \SimpleN\  behaves as\foot{The vectors $\xi$
and $k$ refer to the transverse polarization and momentum of the soft--gluon, respectively. 
Furthermore, $k_j$ denote the external momenta of remaining legs. 
One could also express the kinematic dependent factor as soft or eikonal factor written 
\eg in the $D=4$ spinor helicity
basis \refs{\MANGO,\LANCEeff}.}:
\eqn\SOFT{
\Ac(1,\ldots,N)\lra \lf(\fc{\xi k_{N-2}}{k_{N-2}k}-\fc{\xi k_{N-3}}{k_{N-3}k}\ri)\ 
\Ac(1,\ldots,N-1)\ .}
This can be proven by considering the  limits of the individual summands of \SimpleN:
\eqnn\limit
$$\eqalignno{
(i)&\ \ \  \si\in S_{N-4}\ {\rm with}\ (N-3)_\si=N-3: &\limit \cr
&\Ac_{YM}(1,2_\si,\ldots,(N-3)_\si,N-2,N-1,N)\ F^\si(\ap)\cr 
&\lra\lf(\fc{\xi k_{N-2}}{k_{N-2}k}-\fc{\xi k_{N-3}}{k_{N-3}k}\ri) \ 
\Ac_{YM}(1,2_\si,\ldots,(N-3)_\si,N-2,N-1)\ \wtilde F^\si(\ap)\ ,\cr\crr
(ii)&\ \ \ \si\in S_{N-4}\ {\rm with}\ (N-3)_\si\neq N-3:\cr
&\Ac_{YM}(1,2_\si,\ldots,(N-3)_\si,N-2,N-1,N)\ F^\si(\ap)\cr 
&\lra\lf(\fc{\xi k_{N-2}}{k_{N-2}k}-\fc{\xi k_{(N-3)_\si}}{k_{(N-3)_\si}k}\ri) \Ac_{YM}(1,2_\si,\ldots,(N-3)_\si,N-2,N-1)\ \wtilde F^\si(\ap)\ ,\cr\crr
(iii)&\ \ \ \si\in S_{N-4}\ {\rm with}\ N-3\in\{2_\si,\ldots,i_\si\}\ {\rm and}\ i=2,\ldots,N-4,\ {\ie} 
(N-3)_\si\neq N-3:\cr
&\Ac_{YM}(1,2_\si,\ldots,i_\si,N-2,(i+1)_\si,\ldots,(N-3)_\si,N-1,N)\ F^\si(\ap)\cr 
&\lra\lf(\fc{\xi k_{(i+1)_\si}}{k_{(i+1)_\si}k}-\fc{\xi k_{i_\si}}{k_{i_\si}k}\ri) \Ac_{YM}(1,2_\si,\ldots,(N-3)_\si,N-2,N-1)\ \wtilde F^\si(\ap)\ ,\cr\crr
(iv)&\ \ \ \si\in S_{N-4}:\ \ \Ac_{YM}(1,N-2,2_\si,\ldots,(N-3)_\si,N-1,N)\ F^\si(\ap)\lra0\ ,\cr
(v)&\ \ \ \si\in S_{N-4}\ {\rm with}\ N-3\in\{(i+1)_\si,\ldots,(N-3)_\si\}\ {\rm and}\ i=2,\ldots,N-4:\cr
&\Ac_{YM}(1,2_\si,\ldots,i_\si,N-2,(i+1)_\si,\ldots,(N-3)_\si,N-1,N)\ F^\si(\ap)\lra0\ .}
$$
The above functions $\wtilde F^\si$ refer to the $N-1$--point amplitude.
While the $(N-5)!$ summands of case $(i)$ already have the right form \SOFT\ and give rise to
$(N-5)!$ terms of the $N-1$--point amplitude \SimpleN, the remaining non--vanishing limits $(ii)$ and
$(iii)$ for a given $\si\in S_{N-4}$ with $(N-3)_\si\neq N-3$ conspire to comprise the remaining 
$(N-5)(N-5)!$ terms of \SimpleN\ thanks to the relation:
$$\lf(\fc{\xi k_{N-2}}{k_{N-2}k}-\fc{\xi k_{(N-3)_\si}}{k_{(N-3)_\si}k}\ri)+
\sum_{i=2\atop N-3\in\{2_\si,\ldots,i_\si\}}^{N-4}\  \lf(\fc{\xi k_{(i+1)_\si}}{k_{(i+1)_\si}k}-\fc{\xi k_{i_\si}}{k_{i_\si}k}\ri) =\lf(\fc{\xi k_{N-2}}{k_{N-2}k}-\fc{\xi k_{N-3}}{k_{N-3}k}\ri)\ .$$
The remaining $\h(N-3)!$ terms of the cases $(iv)$ and $(v)$ do not contribute in the soft limit.

\subsubsec{Collinear limit}

Again, according to subsection 2.2 it is sufficient to focus on the $N$--gluon amplitude. 
The collinear limit is defined as two adjacent external momenta $k_i$ and $k_{i+1}$, with $i+1$ mod $N$, 
becoming parallel. Due to cyclic symmetry, these can be chosen as $k_{N-3}$ and $k_{N-2}$,
with $k_{N-3}$ carrying the fraction $x$ of the combined momentum $k_{N-3}+k_{N-2}\ra k_{N-3}$. Formally,
\eqn\formally{
k_{N-3}\ra x\ k_{N-3}\ \ \ ,\ \ \ k_{N-2}\to (1-x)\ k_{N-3}\ ,}
where the  momenta appearing in the limits describe the scattering amplitude of $N-1$ gluons.
In this limit the amplitude \SimpleN\  behaves as\foot{One could also express the kinematic dependent 
factor as splitting amplitude written \eg in the $D=4$ spinor helicity
basis \refs{\MANGO,\LANCEeff}.}
\eqn\SOFT{
\Ac(1,\ldots,N)\lra \fc{1}{k_{N-3}k_{N-2}}\ V^i\fc{\p}{\p \xi^i_{N-3}} \ 
\Ac(1,\ldots,N-1)\ ,}
with the three--gluon vertex $V^i=(\xi_{N-3}\xi_{N-2}) (k^i_{N-2}-k^i_{N-3})+2(\xi_{N-2}k_{N-3})\xi_{N-3}^i-2
(\xi_{N-3}k_{N-2})\xi_{N-2}^i$.
This can be proven by considering the  limits of the individual summands of \SimpleN. First, if the 
two states $N-3$ and $N-2$ are not neighbours, we have:
\eqnn\limiti
$$\eqalignno{
(i)&\ \ \ \si\in S_{N-4}\ {\rm with}\ 2_\si\neq N-3:\ \ \Ac_{YM}(1,N-2,2_\si,\ldots,(N-3)_\si,N-1,N)\lra0 \ ,\cr
(ii)&\ \ \ \si\in S_{N-4}\ {\rm with}\ i_\si,(i+1)_\si\neq N-3\ {\rm and}\ i=2,\ldots,N-4:\cr
&\Ac_{YM}(1,2_\si,\ldots,i_\si,N-2,(i+1)_\si,\ldots,(N-3)_\si,N-1,N)\lra0\ . &\limiti}
$$
On the other hand, the remaining $2(N-4)!$ terms of \SimpleN\ pair up into $(N-4)!$ tuples $(\si,\tilde\si)$
each giving rise to one of the $(N-4)!$ terms of the $N-1$--point amplitude  \SimpleN:
\eqnn\limitii
$$\eqalignno{
&\ \ \ \si,\tilde\si\in S_{N-4}\ {\rm with}\ i_\si=(i+1)_{\tilde\si}=N-3\ {\rm and}\ i=2,\ldots,N-4: &\limitii\cr
&\Ac_{YM}(1,2_\si,\ldots,i_\si,N-2,(i+1)_\si,\ldots,(N-3)_\si,N-1,N)\ F^\si(\ap)\cr
+&\; \Ac_{YM}(1,2_{\tilde\si},\ldots,i_{\tilde\si},N-2,(i+1)_{\tilde\si},\ldots,(N-3)_{\tilde\si},N-1,N)\ 
F^{\tilde\si}(\ap)\cr 
&\ra \fc{1}{k_{N-3}k_{N-2}}\ V^i\fc{\p}{\p \xi^i_{N-3}} \ 
\Ac_{YM}(1,2_\si,\ldots,(N-3)_\si,N-2,N-1)\ F^\si(\ap)\ .}$$
Note that in the above combination  the $x$--dependent parts of the two functions $F^\si$ and 
$F^{\tilde\si}$, which stems from the  limit \formally, add up to zero.

\subsubsec{Cyclic invariance}

While the YM constituent $\Ac_{YM}(1,\ldots,N)$ of \SimpleN\ is invariant 
under cyclic transformations of its labels $i\ra i+1\ \mod\ N$,
all others transform non--trivially. 
More precisely, the set 
$\{ \Ac_{YM}(1,2_\si,\ldots,(N-2)_\si,N-1,N)\ |\ \si\in S_{N-3}\}$ is mapped to
the set $\{ \Ac_{YM}(1,2,3_\si,\ldots,(N-2)_\si,(N-1)_\si,N)\ |\ \si\in S_{N-3}\}$ 
by virtue of the cyclic properties of the $\Ac_{YM}$.
The latter set belongs to the extended $S_{N-2}$ family 
$\{ \Ac_{YM}(1,2_\Pi,\ldots,(N-1)_\Pi,N)\ |\ \Pi\in S_{N-2}\}$, which can be expanded
in terms of the original basis $\Ac_{YM}(1,2_\si,\ldots,(N-2)_\si,N-1,N)$ according to 
\DECO. The cyclic transformation properties of the minimal basis functions $F^\si$
are such that the change of $\Ac_{YM,\si}$ into $\Ac_{YM,\Pi(\si)}=\sum\limits_{\pi\in S_{N-3}}
K^\pi_{\Pi(\si)} \Ac_{YM,\pi}$ is compensated:
\eqn\Hammer{
\lf. F^\si\ \ri|_{k_i\ra k_{i+1}}=F^{\Pi(\si)}=
\sum_{\rho\in S_{N-3}}(K^{-1})^{\Pi(\si)}_\rho\    F^\rho\ .}
The map $\Pi(\si)$ is defined by $(2_{\Pi(\si)},\ldots,(N-1)_{\Pi(\si)})=
(2,2_\si+1,\ldots,(N-2)_\si+1)$.

\newsec{The module of multiple hypergeometric functions}
\def\nt{{\tilde n}}

The functions $F^\si$ describing the full $N$--point amplitude \SimpleN\ 
have been introduced in \eqqs \revol\ and \revoll\ and 
are given by generalized Euler integrals. Generalized Euler integrals
appear in any higher--point open string amplitude computation.
Therefore, we find it useful in this section to investigate the properties of these 
integrals on general grounds.

\subsec{Generalized Euler integrals and multiple hypergeometric functions}

For the color ordering $(1,\ldots,N)$ the integrals of interest can be written
\eqn\GENERI{
B_N\lf[\nt \ri]=\int\limits_{z_i<z_{i+1}} \lf(\prod_{j=2}^{N-2} d z_j\ri)\ 
\prod\limits_{1\leq i<j\leq N-1} \ |z_{ij}|^{s_{ij}}\  z_{ij}^{\nt_{ij}}\ ,}
with some set $\nt$ of integers $\nt_{ij}\in\IZ$.
The latter must fulfill the conditions\foot{Note that the integrands of \revol\ and \revoll\
can always be completed to meet this condition.}
\eqn\conditions{\sum_{i< j}^N \nt_{ij}+\sum_{i>j}^N\nt_{ji}=-2\ \ \ ,\ \ \ j=1,\ldots,N}
as a result of conformal invariance on the string world--sheet.
After fixing three of the vertex positions as
\eqn\FIX{
z_1=0\ \ \ ,\ \ \ z_{N-1}=1\ \ \ ,\ \ \ z_N=\infty\ ,}
and parameterizing  the integration region $z_2<\ldots<z_{N-2}$ as
\eqn\PARA{
z_k=\prod_{l=k-1}^{N-3}x_l\ \ \ ,\ \ \ k=2,\ldots,N-2\ ,}
with $0<x_i<1$ the integrand in \revol\ takes the generic form:
\eqn\GENERIC{
B_N\lf[n\ri]=\lf(\prod_{i=1}^{N-3} \ \int^1_0 d x_i\ri) \ 
\prod_{j=1}^{N-3} \ x_j^{s_{12...j+1}+n_{j} } \ 
\prod_{l=j}^{N-3} \ \lf( \ 1 \ - \ \prod_{k=j}^l x_k \ \right)^{ s_{j+1,l + 2}+n_{jl}}\ ,}
with the set of $\h N(N-3)$ integers $n_j,n_{jl}\in\IZ$ and $s_{i,j}\equiv s_{ij}$:
\eqn\INTEGERS{\eqalign{
n_{jl}&=\tilde n_{j+1,l+2}\ \ \ ,\ \ \ j\leq l\ ,\cr
n_j&=j-1+\sum_{i<j}^{j+1} \tilde n_{il}\ \ \ ,\ \ \ 1\leq j\leq N-3\ .}}
The integrals represent generalized Euler integrals and integrate to multiple
Gaussian hypergeometric functions \STii.

With \conditions\ and \INTEGERS\ from a rational function 
$$R(x_i)=\prod\limits_{j=1}^{N-3} x_j^{n_j}\ \prod\limits_{l=j}^{N-3} (1-\prod\limits_{k=j}^l x_k)^{n_{jl}}$$ 
in the $N-3$ variables $x_i$ multiplying the integrand of \GENERIC\  an other rational function 
$$\wtilde R(z_{ij})=\prod\limits_{1\leq i<j\leq N-1}  z_{ij}^{\nt_{ij}}$$ 
depending on the $N-1$ variables $z_i$ and multiplying the integrand of \GENERI\ can be computed.
In the following we write this correspondence as: 
\eqn\CORRES{
R(x_i)\simeq \wtilde R(z_{ij})\ .}

\subsec{Partial fraction decomposition and finding a basis}

There are many relations 
among integrals \GENERI\ with different sets $\nt$ of integers
as a result\foot{In fact, these tools have allowed to boil down the set of functions appearing in the 
open superstring $N$--point amplitude \PARTONE\ to the set \revol.} of partial fraction decomposition
\eqn\PARTFR{
\fc{1}{z_{ij}z_{jk}}+\fc{1}{z_{ik}z_{kj}}=\fc{1}{z_{ij}z_{ik}}}
and partial integration of their integrands:
\eqnn\PARTIN
$$\eqalignno{
0&=\int \prod_{j=2}^{N-2} d z_j\
\fc{\p}{\p z_k}\ 
\prod\limits_{1\leq i<j\leq N-1} \ |z_{ij}|^{s_{ij}}\  z_{ij}^{\nt_{ij}} &\PARTIN\cr
&=\int \prod_{j=2}^{N-2} d z_j\
\prod\limits_{1\leq i<j\leq N-1} \ |z_{ij}|^{s_{ij}}\  z_{ij}^{\nt_{ij}}\ 
\lf(\sum_{m<k}\fc{s_{mk}+\nt_{mk}}{z_{mk}}+\sum_{m>k}\fc{s_{km}+\nt_{km}}{z_{km}}\ri)\ .}
$$
Note that in this way any integral \GENERI\ with powers $\nt_{ij}<-1$ can always be expressed
by a chain of integrals with $\nt_{ij}\geq -1$. Hence, in the following it is sufficient
to concentrate on those cases $\nt_{ij}\geq -1$.
A quantitative handiness on finding a minimal set of functions can be obtained by 
performing \br
$(i)$ a classification of the integrals \GENERI\ according to their pole 
structure in the kinematic invariants $s_{ij}$ and \br
$(ii)$ a Gr\"obner basis analysis for those integrals \GENERI\ without poles.
 
Any partial fraction decomposition of an Euler integral with poles can be arranged 
according to its pole structure (modulo finite or subleading pieces) 
and the classification $(i)$ yields a basis for them.
This is achieved by performing a partial fraction expansion of the leading singularity
in the kinematic invariants $s_{ij}$.
On the other hand, the Gr\"obner basis analysis $(ii)$ provides an independent set
of rational functions or monomials in the Euler integrals and any integral \GENERI\ 
without poles can be expanded in terms of this set.
Combining $(i)$ and $(ii)$ yields an independent set of integrals \GENERI\ and any partial fraction 
decomposition of Euler integrals \GENERI\ can be expressed in terms of the basis
obtained this way. In  subsection 3.3. and 3.4 we explicitly construct this 
partial fraction basis for the cases 
$N=4,5$ and $N=6$ and verify its dimension $(N-2)!$.

The first classification $(i)$ of the integrals \GENERI\ is done w.r.t.\  
their pole structure 
in the kinematic invariants $s_{ij}$. The maximum number of possible 
simultaneous poles of an $N$--point amplitude is $N-3$. 
Integrals of this type play an important role, since
they capture the field--theory limit of the full amplitude. 
They assume the following power series expansion in $\ap$:
\eqn\EXPA{\eqalign{
B_N[\nt]&=\ap^{3-N}\ p_{3-N}[\nt]+\ap^{5-N}\ \sum_{m=0}^\infty\ \ap^m
\sum_{i_r\in {\bf N}, i_1>1\atop i_1+\ldots+i_d=m+2}'p^{\bf i}_{5-N+m}[\nt]\ 
\zeta(i_1,\ldots,i_d)\cr
&=\ap^{3-N}\ p_{3-N}[\nt]+\ap^{5-N}\ p_{5-N}[\nt]\ \zeta(2)\ +\ap^{6-N}\ p_{6-N}[\nt]\ 
\zeta(3)\ +\ldots\ .}}
The above rational functions or monomials $p^{\bf i}_{5-N+m}[\nt]$ 
are of degree $5-N+m$ in the dimensionful kinematic invariants $\hat s_{ij} =s_{ij} / \alpha'$
and depend on the integer set $\nt$. Furthermore, we have introduced the MZVs
$$\zeta(i_1,\!\ldots,\!i_d)=
\sum\limits_{n_1>\ldots>n_d>0}\ \prod\limits_{r=1}^dn_r^{-i_r},\ \ \ i_r\in{\bf N},\ i_1>1$$
of transcendentality degree $\sum_{r=1}^di_r=m+2$ and depth $d$, \cf e.g. \GRAV\ for more details and references.
The prime at the sum \EXPA\ means, that the latter runs only over  
a basis of independent MZVs of weight $m+2$.
In \EXPA\ at each order $5-N+m$ in $\ap$ a set of MZVs of a fixed transcendentality degree $m+2$ appears. 
We call such a power series expansion transcendental, \cf appendix \appA\ for a
detailed discussion.
In subsection 3.3 we present a  method to extract the first term of \EXPA\ corresponding to 
integrals \GENERI\ with $N-3$ simultaneous poles. In fact, this method allows  to extract any 
lowest order poles  from integrals \GENERI\  with fewer simultaneous poles. However, as we shall 
demonstrate, their type of integrals generically does not assume the transcendental power series expansion \EXPA.
At any rate, the method of subsection 3.3 determines the lowest order poles
of the integral \GENERI.

The second classification $(ii)$ of the integrals \GENERI\ is appropriate, if the latter 
have no poles, \ie their  power series expansion in $\ap$ starts with some zeta constants. 
In subsection 3.4 we introduce a Gr\"obner basis analysis, which allows to find an independent
set of finite integrals \GENERI, which serves as basis. Any other finite integral \GENERI\
is a $\IR$--linear combination of this basis.

Note that the {\it individual\/} integrals entering the functions \revol\ and \revoll\ are of both 
types -- some of them have $N-3$ simultaneous poles and their $\ap$--expansion assumes the  form \EXPA, 
others have no poles and start with some zeta constants. 
In either case our methods $(i)$ or $(ii)$ can be applied to further reduce them.

\subsec{Structure of multiple resonance exchanges}

Generically, an $N$--point scattering process has multiple resonance exchanges.
As a result, the power series expansion in $\ap$ of the integrals \GENERIC\ may 
have multiple poles in the Mandelstam variables. These poles come from 
regions of the integrand for which $x_i\ra0$ or $x_i\ra1$ for some of the variables $x_i$.
To obtain information on the pole structure of the integrals \GENERIC\ it is useful to transform
the integrand to a different form, in which the poles can be easily extracted.

For an $N$--point scattering process there are $\h N(N-3)$ planar 
channels $(i,j)\in \Pc$ associated to the Mandelstam variable 
$S_{i,j}=\ap(k_i+k_{i+1}+\cdots+k_j)^2$, with 
\eqn\Partition{
\Pc=\{\ (1,j)\ |\ 2\leq j\leq N-2\ \}\cup\{\ (p,q)\ |\ 2\leq p<q\leq N-1\ \}}
for the color ordering $(1,2,\ldots,N)$.
The channels $(i,j)$ with states from $i,\ldots,j$ and 
$(j+1,i-1)$ with states from $j+1,\ldots,N,1,\ldots,i-1$ are identical.
The set of $N-3$ kinematic invariants, which can simultaneously 
appear in the denominator of the $\ap$--expansion 
of the $N$--point amplitude, describe the allowed (planar) 
channels of the underlying field--theory diagram involving cubic vertices. 
Not all combinations of channels are allowed. 
E.g.\ adjacent channels as $(i,i+1)$ and $(i+1,i+2)$ cannot appear simultaneously in 
denominators (dual or incompatible channels).
On the other hand, for non--dual channels coincident poles are possible.  
A geometric way to find all compatible channels is to draw a convex $N$--polygon of $N$ sides 
representing momentum conservation. 
The number of ways of cutting this polygon into $N-2$ triangles with $N-3$ 
non--intersecting straight lines gives the number of distinct sets of allowed channels. 
According to Euler's polygon division problem
this number is given by $C_{N-2}=\fc{2^{N-2}\ (2N-5)!!}{(N-1)!}$, with the
Catalan number $C_n=\fc{1}{n+1}\lf(2n\atop n\ri)$.
The $N-3$ diagonals of this polygon represent the momenta of possible intermediate states.
To each of the $\h N(N-3)$ channels $(i,j)$ a variable $u_{i,j}\in (0,1)$ may be ascribed, 
with $u_{i,j}\equiv u_{j+1,i-1}$. For an account and references on the multiparticle dual model
see \Frampton.
\ifig\multiperipheral{Multiperipheral configuration and corresponding dual diagram for $N=8$.}
{\epsfxsize=0.9\hsize\epsfbox{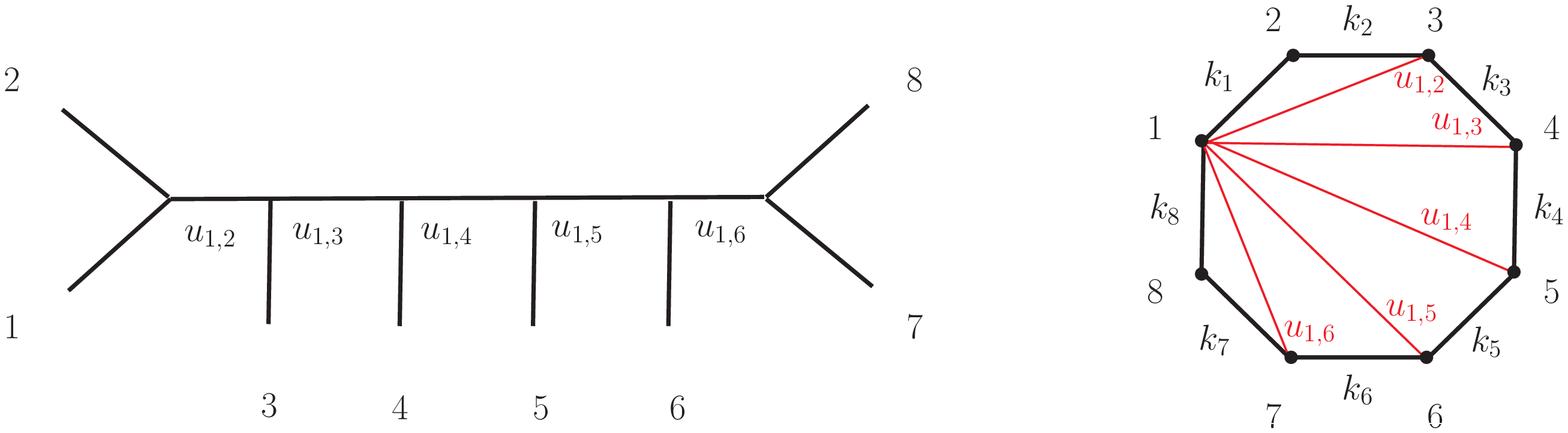}}
\noindent
For a given channel $(i,j)$ with $u_{i,j}=0$ all incompatible channels $(p,q)$ are required to have 
$u_{p,q}=1$. This property is described by the $\h N(N-3)$ duality constraint equations
\eqn\constraint{
u_{i,j}=1-\prod_{1\leq p<i\atop i\leq q<j} u_{p,q}\prod_{i<r\leq j\atop j<s\leq N-1} 
u_{r,s}\ \ \ ,\ \ \ 1\leq i<j\leq N\ ,}
which are sufficient for excluding simultaneous poles in incompatible channels.
We define $u_{i,i}=0,\ u_{1,N-1}=1$ and have $u_{k,N}=u_{1,k-1},\ k\geq3$.
Only $\h(N-2)(N-3)$ of these equations \constraint\ 
are independent, leaving $N-3$ variables $u_{i,j}$ 
out of the set of $\h N(N-3)$ variables free. The set of $N-3$ independent variables
$u_{i,j}$ can be  associated to the inner lines of one of the $C_{N-2}$ sliced $N$--polygon. 
In particular, as a canonical choice we may define 
\eqn\dualchoice{
u_{1,j+1}=x_j\ \ \ ,\ \ \ j=1,\ldots,N-3}
as a set of $N-3$ independent variables corresponding to \multiperipheral.
Hence, each of the  internal lines of the polygon corresponds to an independent variable $x_j$ 
in the integral \GENERIC.
Choosing the inner lines of an other sliced $N$--polygon results in a different integral representation \GENERIC.
As a consequence of \constraint\ and \dualchoice\ we have\foot{  
The inverse solution to the duality constraint \constraint\ may be found as
($p=2,3,\ldots,N-2; q=3,4,\ldots,N-1$ and  $p<q$):
\eqn\solution{
u_{p,q}=\cases{
\fc{\lf(1-\prod\limits_{m=p}^{q-1} u_{1,m}\ri)\ \lf(1-\prod\limits_{n=p-1}^{q} u_{1,n}\ri)}
{\lf(1-\prod\limits_{r=p-1}^{q-1} u_{1,r}\ri)\ \lf(1-\prod\limits_{s=p}^{q} u_{1,s}\ri)}\ ,&
$q\neq N-1$\cr\crr
\fc{\lf(1-\prod\limits_{m=p}^{q-1} u_{1,m}\ri)}
{\lf(1-\prod\limits_{r=p-1}^{q-1} u_{1,r}\ri)}\ ,&
$q= N-1$\ .}}  }
:
\eqnn\consequence
$$\eqalignno{
1-x_j&=\prod_{0<r\leq j\atop j<s\leq N-2} u_{r+1,s+1}\ \ \ ,\ \ \ j=1,\ldots,N-3\ ,&\consequence\cr
1-\prod_{k=i}^jx_k&=\prod_{1\leq p\leq i\atop j+1\leq q\leq N-2} u_{p+1,q+1}  \ \ \ ,\ \ \ 1\leq i\leq j\leq N-3\ .}$$
With \consequence\ and the Jacobian $\prod\limits_{2\leq i<j\leq N-1} u_{i,j}^{j-i-1}$, the 
integral \GENERIC\ translates into an integral over all $\h N(N-3)$ variables $u_\Pc$ related 
to the partitions $\Pc$ given in \Partition
\eqn\GENERICC{\eqalign{
B_N[n]&=\prod_{(i,j)\in \Pc}\ \int_0^1\  du_{i,j}\ u_{i,j}^{S_{i,j}+n_{i,j}}\ 
\prod_{\Pc'\notin(1,j)}\delta\lf(u_{\Pc'}-1+\prod_{\tilde\Pc}u_{\tilde\Pc}\ri)\ ,}}
with the assignments:
\eqnn\RELbesser
$$\eqalignno{
n_{1,j+1}&=n_j\ \ \ ,\ \ \ n_{j+1,j+2}=n_{jj}\ \ \ ,\ \ \ j=1,\ldots,N-3\ ,\cr
n_{i,j}&=j-i-1+\sum_{i-1\leq k\leq l}^{j-2}n_{kl}\ \ \ ,\ \ \ 1<i<j<N\ .&\RELbesser}$$
In \GENERICC\ the integration is constrained by the duality conditions \constraint\ resulting in
a product of $\h(N-2)(N-3)$ independent $\delta$--functions.
In this form \GENERICC\ many properties of  the integrals \GENERIC\ like the pole structure or 
cyclicity become manifest. Later this will be elucidated with  examples.

We can introduce  a fundamental set of $C_{N-2}$ integrals $B_N$ 
\eqn\FundamentalSet{
\bigcup_{(i_l,j_l)\in\Pc}\ 
\lf\{\ \prod_{(i,j)\in \Pc}\ \int_0^1\  du_{i,j}\ u_{i,j}^{S_{i,j}}
\lf(\prod\limits_{l=1}^{N-3} u_{i_l,j_l}\ri)^{-1}\ 
\prod_{\Pc'\notin(1,j)}\delta\lf(u_{\Pc'}-1+\prod_{\tilde\Pc}u_{\tilde\Pc}\ri)\ \ri\}\ ,}
with $(i_l,j_l)$ running over all $C_{N-2}$ allowed channels\foot{As pointed out before, these integrals appear as constituents of some of the functions $F^\sigma$. The poles in their $B_N$ combinations are cancelled by the corresponding $s_{ij}$ factors in the numerator of the $F^\sigma$ such that they are rendered local.}. 
The $\ap$--expansion of each of the elements \FundamentalSet\ assumes the form \EXPA\ 
with $\prod\limits_{l=1}^{N-3} S_{i_l,j_l}^{-1}$ as its lowest order term.
Any other integral \GENERIC\ with $N-3$ simultaneous poles can be expressed as $\IR$--linear 
combination of the basis \FundamentalSet\  modulo less singular terms.
In case of a sum of $N-3$ simultaneous poles this is achieved by partial fraction decomposition 
of the polynomials according to their leading singular term and associating the latter with the basis \FundamentalSet.

A special role is played by the integral:
\eqn\CYC{\eqalign{
B_N[n=-1]&=\prod_{(i,j)\in \Pc}\ \int_0^1\  du_{i,j}\ u_{i,j}^{S_{i,j}-1}\ 
\prod_{\Pc'\notin(1,j)}\delta\lf(u_{\Pc'}-1+\prod_{\tilde\Pc}u_{\tilde\Pc}\ri)\ .}}
By construction it is manifestly invariant under cyclic transformations
$S_{i,j}\ra S_{i+1,j+1}$, with $i\equiv i+ N,\ j\equiv j+ N$. Furthermore, it
furnishes all $C_{N-2}$ sets of allowed channels at the lowest order, \ie
\eqn\LOWSPECIAL{
B_N[n=-1]=\sum_{(i_l,j_l)\in\Pc}
\fc{1}{\prod\limits_{l=1}^{N-3} S_{i_l,j_l}}+\cdots\ ,}
with the sum running over all $C_{N-2}$ allowed channels.
In terms of \GENERIC, \eqq \CYC\ takes the form:
\eqn\GENERIc{\eqalign{
B_N\lf[n_i=-1\atop n_{ii}=-1\ri]&=\lf(\prod_{i=1}^{N-3} \ \int^1_0 d x_i\ri) \ 
\prod_{j=1}^{N-3} \, x_j^{s_{12...j+1}-1} \  (1-x_j)^{s_{j+1,j+2}-1}\cr  
&\times\prod_{l=j+1}^{N-3} \ \left( \ 1 \ - \ \prod_{k=j}^l x_k \ \right)^{ s_{j+1,l + 2}}}}
Obviously, \CYC\ can be expanded in terms of the basis \FundamentalSet.

\subsubsec{$N=4$}

In the case of $N=4$ we have the two planar channels $(1,2)$ and $(2,3)\equiv (1,4)$ related to the 
two variables $u_{1,2}$ and $u_{2,3}$, respectively.
After choosing the independent variable $u_{1,2}=x_1:=x$ and following the steps 
\consequence\ the integral \GENERIC\
\eqn\EXfour{
B_4[n]=\int_0^1 dx\ x^{s_{12}+n_1}\ (1-x)^{s_{23}+n_{11}}}
takes the form \GENERICC
\eqn\EXFour{
B_4[n]=\int_0^1 du_{1,2}\ \int_0^1 du_{2,3}\ u_{1,2}^{s_{12}+n_{1,2}}\ u_{2,3}^{s_{23}+n_{2,3}}\ 
\delta(u_{1,2}+u_{2,3}-1)\ ,}
with \RELbesser, \ie $n_{1,2}=n_1$ and $n_{2,3}=n_{11}$.

The fundamental objects \FundamentalSet\ correspond to the two rational functions
\eqn\fundamentaliv{
\fc{1}{u_{1,2}}\ \ \ ,\ \ \ \fc{1}{u_{2,3}}\ ,}
which furnish the $C_2=2$ poles $s_{12}^{-1}$ and 
$s_{23}^{-1}$ as single poles, respectively. 
The  cyclically invariant integral \CYC\ is given by 
\eqn\CYCiv{
B_4\lf[{n_1=-1\atop n_{11}=-1}\ri]=\int_0^1 dx\ x^{s_{12}-1}\ (1-x)^{s_{23}-1}=B(s_{12},s_{23})=\fc{1}{s_{12}}+
\fc{1}{s_{23}}+\cdots}
and exhibits both poles in its power series expansion.

\subsubsec{$N=5$}

In this case we have the five planar channels $(1,2),\ (2,3),\ (3,4),\ (1,3)\equiv (4,5)$
and $(2,4)\equiv (1,5)$ related to the 
five variables $u_{1,2}, u_{2,3}, u_{3,4}, u_{4,5}\equiv u_{1,3}$ and $u_{5,1}\equiv u_{2,4}$, respectively.
The five--point integral \GENERIC\ becomes
\eqn\EXfive{
B_5[n]=\int_0^1 dx_1\ \int_0^1 dx_2\ x_1^{s_{1}+n_1}\ x_2^{s_{4}+n_2}\ (1-x_1)^{s_{2}+n_{11}}\ 
(1-x_2)^{s_{3}+n_{22}}\ (1-x_1x_2)^{s_{24}+n_{12}}\ ,}
with $s_i=\ap(k_i+k_{i+1})^2,\ i=1,\ldots,5$ subject to the cyclic identification 
$i + 5 \equiv i$.
To transform \EXfive\ into the form \GENERICC\ according to \dualchoice\ we choose the two 
independent variables $u_{1,2}=x_1$ and $u_{1,3}=x_2$. Then, with \consequence\ the integral \EXfive\ takes the form
\eqn\EXFive{\eqalign{
B_5[n]&=\int_0^1 du_{1,2}\int_0^1 du_{2,3}\int_0^1 du_{3,4}\int_0^1 du_{4,5} 
\int_0^1 du_{1,5}\ 
u_{1,2}^{s_1+n_{1,2}}\ u_{2,3}^{s_2+n_{2,3}}\ u_{3,4}^{s_3+n_{3,4}}\ u_{4,5}^{s_4+n_{1,3}}\cr
&\times u_{1,5}^{s_5 +n_{2,4}}\
\delta(u_{2,3}+u_{1,2}u_{3,4}-1)\ \delta(u_{2,4}+u_{1,2}u_{4,5}-1)\ 
\delta(u_{3,4}+u_{2,3}u_{4,5}-1)\ ,}}
with the assignment \RELbesser.

In what follows it is convenient to introduce
\eqn\Ifive{
I_5(x,y)=x^{s_{4}}\ y^{s_{1}}\ (1-x)^{s_{3}}\ (1-y)^{s_{2}}\ (1-xy)^{s_{24}}}
arising from \EXfive\ with the identifications $x_1:=y$ and $x_2:=x$.
Furthermore, we use the following shorter notation for the dual variables $u_{i,j}$:
\eqn\Uxyv{
X_i=u_{i,i+1}\ \ \ ,\ \ \ i=1,\ldots,5\ \ \ ,\ \ \ i+5 \equiv i}
and define:
\eqn\Jfive{
J_5(X)=\lf(\prod_{i=1}^5 X_i^{s_i}\ri) \ \delta(X_2+X_1X_3-1)\  \delta(X_3+X_2X_4-1)\ 
\delta(X_5+X_1X_4-1)\ .}
\noindent
Let us now discuss a few examples. The pole structure of the integral
\eqn\exei{
\int_0^1dx\int_0^1dy\ \fc{I_5(x,y)}{(1-y)\ (1-x y)}}
can be easily deduced after transforming it into the form \EXFive
\eqn\exeii{
\lf(\prod_{i=1}^5\int_0^1 dX_i\ri)\  J_5(X)\ \fc{1}{X_2X_5}=\fc{1}{s_2s_5}+\cdots\ .}
Hence, the only simultaneous pole is at $X_2,X_5\ra0$ with the product of $\delta$--functions 
yielding the constraints for the three variables $X_1,X_3,X_4\ra1$.
In the sequel we list a few  non--trivial examples:
\eqn\RESI{
\matrix{
{\underline{rational\ function}\atop\underline{in\ eq.\ \GENERI}}&
{\underline{rational\ function}\atop\underline{in\ eq.\ \EXfive}}&
{\underline{rational\ function}\atop\underline{in\ eq.\ \EXFive}}&
\ss{\underline{lowest\ order\ poles}}\cr\cr
\fc{z_{15}}{z_{12} z_{13} z_{14} z_{25} z_{35} z_{45}} &\fc{1}{x\ y}&\fc{X_5}{X_1X_4}&\fc{1}{s_1s_4}\ ,\cr\cr
\fc{1}{z_{12} z_{13} z_{24} z_{35} z_{45}} &\fc{1}{x\ y\ (1-xy)}&\fc{1}{X_1X_4}&\fc{1}{s_1s_4}\ ,\cr\cr
\fc{1}{z_{13} z_{14} z_{23} z_{25} z_{45}} &\fc{1}{x\ (1-y)}&\fc{1}{X_2X_4}&\fc{1}{s_2s_4}\ ,\cr\cr
\fc{1}{z_{14} z_{15} z_{23} z_{25} z_{34}}  &\fc{1}{(1-x)\ (1-y)}&\fc{1}{X_2X_3X_5}&\fc{1}{s_2s_5}\ss{+}\fc{1}{s_3s_5}\ ,\cr\cr
\fc{1}{z_{12} z_{15} z_{24} z_{34} z_{35}} &\fc{1}{(1-x)\ y\ (1-xy)}&\fc{1}{X_1X_3X_5}&\fc{1}{s_1s_3}\ss{+}\fc{1}{s_3s_5}\ .\cr\cr}}

The fundamental objects \FundamentalSet\ correspond to the five rational functions
\eqn\fundamentalv{
\fc{1}{X_1X_3}\ ,\ \fc{1}{X_2X_4}\ ,\ \fc{1}{X_3X_5}\ ,\ \fc{1}{X_1X_4}\ , \fc{1}{X_2X_5}\ ,}
which furnish the $C_3=5$ poles
\eqn\Fundamentalv{
\fc{1}{s_1s_3}\ ,\ \fc{1}{s_2s_4}\ ,\ \fc{1}{s_3s_5}\ ,\ \fc{1}{s_1s_4}\ , \fc{1}{s_2s_5}\ ,}
as single poles, respectively.
In the basis \EXfive\ the rational functions become
\eqn\Rationalv{
\fc{1}{(1-x)y}\ ,\ \fc{1}{x(1-y)}\ ,\ \fc{1}{(1-x)(1-xy)}\ ,\ \fc{1}{xy(1-xy)}\ , 
\fc{1}{(1-y)(1-xy)}\ ,}
respectively.
The cyclically invariant integral \CYC\ is given by
\eqn\CYCv{
B_5\lf[n_i=-1\atop n_{ii}=-1\ri]
=\int_0^1dx\int_0^1dy\ \fc{I_5(x,y)}{x\ (1-x)\ y\ (1-y)}=\fc{1}{s_1s_3}+\fc{1}{s_2s_4}+
\fc{1}{s_3s_5}+\fc{1}{s_1s_4}+\fc{1}{s_2s_5}+\cdots\ ,}
and exhibits all five poles \Fundamentalv\ in its power series expansion.

Finally, as we shall see in the next subsection there is one rational function without 
poles and its series expansion starts at $\zeta(2)$:
\eqn\RESii{
\matrix{
{\underline{rational\ function}\atop\underline{in\ eq.\ \GENERI}}&
{\underline{rational\ function}\atop\underline{in\ eq.\ \EXfive}}&
{\underline{monomial}\atop\underline{in\ eq.\ \EXFive}}&
\ss{\underline{lowest\ order}}\cr\cr
\fc{1}{z_{13} z_{14} z_{24} z_{25} z_{35}}&\fc{1}{(1-xy)}&1&\zeta(2)\ .}}
The function \RESii\ together with \RESI\ furnishes the six dimensional partial fraction basis of $N=5$ integrals. It may be added to \fundamentalv\ to give rise  to another fundamental set
\eqn\Afundamentalv{
\fc{X_2}{X_1X_3}\ ,\ \fc{X_3}{X_2X_4}\ ,\ \fc{X_4}{X_3X_5}\ ,\ \fc{X_5}{X_1X_4}\ , \fc{X_1}{X_2X_5}\ ,}
subject to the constraints \Jfive\ and with the same poles \Fundamentalv, respectively.
In the basis \GENERIC\ the latter rational functions correspond to
\eqn\ARationalv{
\fc{1-y}{(1-x)y(1-xy)}\ ,\ \fc{1-x}{x(1-y)(1-xy)}\ ,\ \fc{x}{(1-x)(1-xy)}\ ,\ \fc{1}{xy}\ , 
\fc{y}{(1-y)(1-xy)}\ ,}
respectively. Since we have
$$\eqalign{
\fc{X_3}{X_1X_4}&\simeq \fc{1}{xy}\fc{1-x}{(1-xy)^2}\simeq\fc{z_{25}z_{34}}{z_{12}z_{13}z_{24}^2z_{35}^2z_{45}}\ ,
\ \fc{X_2}{X_1X_4}\simeq \fc{1}{xy}\fc{1-y}{(1-xy)^2}\simeq\fc{z_{14}z_{23}}{z_{12}z_{13}^2z_{24}^2z_{35}z_{45}}\ ,\cr
\fc{X_3X_5}{X_1X_4}&\simeq \fc{1}{xy}\fc{1-x}{(1-xy)}\simeq\fc{z_{15}z_{34}}{z_{12}z_{13}z_{14}z_{24}z_{35}^2z_{45}}\ ,
\ \fc{X_2X_5}{X_1X_4}\simeq \fc{1}{xy}\fc{1-y}{(1-xy)}\simeq\fc{z_{15}z_{23}}{z_{12}z_{13}^2z_{24}z_{25}z_{35}z_{45}},}$$
the two rational functions $\fc{1}{X_1X_4}$ and $\fc{X_5}{X_1X_4}$ are the only 
possibilities to realize the  poles $\fc{1}{s_1s_4}$ without double poles in the denominator
of \GENERI. Due to cyclicity these arguments take over to the other four poles \Fundamentalv\
and their rational functions \fundamentalv\ and \Afundamentalv.
Generally, rational functions other than the latter
give rise to double powers in the denominator of \GENERI, e.g.:
$$\eqalign{
\fc{1}{X_1}&\simeq \fc{1}{y(1-xy)}\simeq\fc{1}{z_{12}z_{14}z_{24}z_{35}^2}\ ,\cr
\fc{X_1}{X_2}&\simeq \fc{y}{1-y}\simeq\fc{z_{12}}{z_{13}z_{14}^2z_{23}z_{25}^2}\ .}$$
Similarly, as we shall see in the next subsection monomials in the variables $X_i$ other than the trivial 
case \RESii\ yield to double powers in the denominator of \GENERI, e.g.:
$$\eqalign{
X_1X_4&\simeq\fc{xy}{1-xy}\simeq\fc{z_{12}z_{45}}{z_{13}z_{14}^2z_{24}z_{25}^2z_{35}}\ ,\cr
X_1&\simeq \fc{y}{1-xy}\simeq\fc{z_{12}}{z_{13}^2z_{14}z_{24}z_{25}^2}\ ,\cr
X_3X_5&\simeq\fc{1-x}{1-xy}\simeq\fc{z_{15}z_{34}}{z_{13}z_{14}^2z_{24}z_{25}z_{35}^2}\ ,\cr
X_2X_3&\simeq\fc{(1-x)(1-y)}{(1-xy)^3}\simeq\fc{z_{23}z_{34}}{z_{13}^2z_{24}^3z_{35}^2}\ .\cr}
$$

\subsubsec{$N=6$}

In this case we have the nine planar channels $(1,2),\ (1,3),\ (1,4)\equiv (5,6),\ (2,3),\ (2,4)$,
$(2,5)\equiv (1,6),\ (3,4),\ (3,5)$
and $(4,5)$ related to the nine variables $u_{1,2}, u_{1,3}, u_{1,4}, u_{2,3}, u_{2,4}$, 
$u_{2,5}, u_{3,4}$, $u_{3,5}$ and $u_{4,5}$, respectively.
The six--point integral \GENERIC\ becomes
\eqn\EXsix{\eqalign{
B_6[n]&=\int_0^1 dx_1 \int_0^1 dx_2 \int_0^1dx_3\ 
x_1^{s_1+n_1}\ x_2^{t_1+n_2}\ x_3^{s_5+n_3}\ 
(1-x_1)^{s_2+n_{11}}\ (1-x_2)^{s_3+n_{22}}\cr 
&\times (1-x_3)^{s_4+n_{33}}\ (1-x_1x_2)^{s_{24}+n_{12}}\ (1-x_2x_3)^{s_{35}+n_{23}}\ (1-x_1x_2x_3)^{s_{25}+n_{13}}\ ,}}
with $s_i=\ap(k_i+k_{i+1})^2,\ i=1,\ldots,6$ subject to the cyclic identification $i+6\equiv i$
and $t_j=\ap(k_j+k_{j+1}+k_{j+2})^2,\ j=1,\ldots,3$.

To bring \EXsix\ into the form \GENERICC\ according to \dualchoice, we choose the three independent 
variables $u_{1,2}=x_1, u_{1,3}=x_2$ and $u_{1,4}=x_3$. Then, with \consequence\ the integral \EXsix\ takes the form
\eqn\EXSix{\eqalign{
B_6[n]&=\int_0^1 du_{1,2}\int_0^1 du_{1,3} \int_0^1 du_{1,4} \int_0^1 du_{2,3} 
\int_0^1 du_{2,4} \int_0^1 du_{2,5} \int_0^1 du_{3,4} \int_0^1 du_{3,5} \int_0^1 du_{4,5}\cr 
&\times u_{1,2}^{s_1+n_{1,2}}\ u_{2,3}^{s_2+n_{2,3}}\ u_{3,4}^{s_3+n_{3,4}}\  
u_{4,5}^{s_4+n_{4,5}}\ u_{1,4}^{s_5 +n_{1,4}}\ u_{2,5}^{s_6 +n_{2,5}}\ u_{1,3}^{t_1+n_{1,3}}\ 
u_{2,4}^{t_2+n_{2,4}}\ u_{3,5}^{t_3+n_{3,5}}\cr 
&\times \delta(u_{2,3}+u_{1,2}u_{3,4}u_{3,5}-1)\ \delta(u_{2,4}+u_{1,2}u_{1,3}u_{3,5}u_{4,5}-1)\ \delta(u_{2,5}+u_{1,2}u_{1,3}u_{1,4}-1)\cr
&\times \delta(u_{3,4}+u_{1,3}u_{2,3}u_{4,5}-1)\ \delta(u_{3,5}+u_{1,3}u_{1,4}u_{2,3}u_{2,4}-1)\ \delta(u_{4,5}+u_{1,4}u_{2,4}u_{3,4}-1)\ ,}}
with the assignment \RELbesser.

Similarly as in the five--point case, it is convenient to introduce
\eqn\Isix{
I_6(x,y,z)=x^{s_5}\ y^{t_1}\ z^{s_1}\ (1-x)^{s_4}\ (1-y)^{s_3}\ (1-z)^{s_2}\ (1-xy)^{s_{35}}\ (1-yz)^{s_{24}}\ (1-xyz)^{s_{25}}}
which arises from \EXsix\ with the identifications $x_1:=z, x_2:=y$ and $x_3:=x$.
Furthermore, we define
\eqn\Uxyvi{
X_i=u_{i,i+1}\ \ \ ,\ \ \ i=1,\ldots,6\ \ \ ,\ \ \ i+6 \equiv i\ \ \ ,\ \ \ Y_j=u_{j,j+2}\ \ \ ,
\ \ \ j=1,\ldots,3}
and
\eqn\Jsix{\eqalign{
J_6(X,Y)=\lf(\prod_{i=1}^6 X_i^{s_i}\ri)\ \lf(\prod_{j=1}^3 Y_j^{t_j}\ri) &\ 
\delta(X_2+X_1X_3Y_3-1)\ \delta(Y_2+X_1X_4Y_1Y_3-1)\cr 
\times &\delta(X_6+X_1X_5Y_1-1)\ \delta(X_3+X_2X_4Y_1-1)\cr
\times &\delta(Y_3+X_2X_5Y_1Y_2-1)\ \delta(X_4+X_3X_5Y_2-1)\ .}}
\noindent
Let us now discuss a few examples. The pole structure of the integral
\eqn\exei{
\int_0^1dx\int_0^1dy\int_0^1dz\ \fc{I_6(x,y,z)}{(1-x)\ (1-xy)\ (1-xyz)}}
can be easily deduced after transforming it into the form \EXSix
\eqn\exeii{
\lf(\prod_{i=1}^6\int_0^1 dX_i\ri) \ \lf(\prod_{j=1}^3\int_0^1 dY_j\ri) J_6(X,Y)
\ \fc{Y_2}{X_4X_6Y_3}=\fc{1}{s_4s_6t_3}+\cdots\ .}
Hence, the only simultaneous pole is at $X_4,X_6,Y_3\ra0$ with the product of $\delta$--functions 
yielding the constraints for the six variables $X_1,X_2,X_3,X_5,Y_1,Y_2\ra1$.
Note, that by construction a set of three poles in \EXSix\ does not necessarily yield a compatible 
set of channels, \eg the integral
\eqn\exeiii{\eqalign{
\int_0^1dx\int_0^1dy\int_0^1dz&\fc{I_6(x,y,z)}{(1-x)\ (1-y)}=\lf(\prod_{i=1}^6\int_0^1 dX_i\ri) \ \lf(\prod_{j=1}^3\int_0^1 dY_j\ri) 
\ \fc{J_6(X,Y)}{X_3X_4Y_3} \cr
&=\fc{1}{s_3t_3}+\fc{1}{s_4t_3}+\fc{1}{s_3}+\fc{1}{s_4}-\fc{s_1}{s_3t_3}-\fc{s_1}{s_4t_3}
-\fc{s_6}{s_3t_3}-\fc{s_6}{s_4t_3}+\cdots}}
does not give rise to a triple pole as $(3,4),(4,5)$ and $(3,5)$ are not compatible channels.
Similarly, for 
\eqnn\exeiv
$$\eqalignno{
\int_0^1dx\int_0^1dy\int_0^1dz&\fc{I_6(x,y,z)}{z(1-z)(1-xy)(1-xyz)}=\lf(\prod_{i=1}^6\int_0^1 dX_i\ri) \ \lf(\prod_{j=1}^3\int_0^1 dY_j\ri) 
\ \fc{J_6(X,Y)}{X_1X_2X_6}\cr
&=\fc{1}{s_2s_6}+\fc{1}{s_2}+\fc{1}{s_6}-\fc{s_4}{s_2s_6}-\fc{t_2}{s_2s_6}+\fc{\zeta(2)}{s_1}+\cdots &\exeiv}$$
the channels $(1,2),(2,3)$ and $(6,1)$ are not compatible.
In the following table we list a few  non--trivial examples:
\eqn\RESI{
\matrix{
{\underline{rational\ function}\atop\underline{in\ eq.\ \GENERI}}&
{\underline{rational\ function}\atop\underline{in\ eq.\ \EXsix}}&
{\underline{rational\ function}\atop\underline{in\ eq.\ \EXSix}}&
\ss{\underline{lowest\ order\ poles}}\cr\cr
\fc{z_{16}^2}{z_{12} z_{13} z_{14} z_{15} z_{26} z_{36} z_{46} z_{56}}& \fc{1}{x\ y\ z}&\fc{X_6^2Y_2Y_3}{X_1X_5Y_1}&\fc{1}{s_1s_5t_1}\ ,\cr\cr
\fc{z_{16}}{z_{12} z_{13} z_{15} z_{26} z_{36} z_{45} z_{46}}  &\fc{1}{(1-x)\ y\ z}&\fc{X_6Y_3}{X_1X_4Y_1}&\ss{-}\fc{1}{s_1s_4t_1}\ ,\cr\cr
\fc{1}{z_{13} z_{15} z_{23} z_{26} z_{45} z_{46}}   &\fc{1}{(1-x)\ y\ (1-z)}&\fc{1}{X_2X_4Y_1}&\fc{1}{s_2s_4t_1}\ ,\cr\cr
\fc{1}{z_{12} z_{14} z_{25} z_{34} z_{36} z_{56}}   &\fc{1}{x\ (1-y)\ z\ (1-xyz)}&\fc{1}{X_1X_3X_5}&\fc{1}{s_1s_3s_5}\ ,\cr\cr
\fc{z_{13}z_{45}}{z_{12} z_{14}^2 z_{25} z_{34} z_{35} z_{36} z_{56}}   &\fc{y\ (1-x)}{x\ (1-y)\ z\ (1-xy)\ (1-xyz)}&\fc{X_4Y_1}{X_1X_3X_5}&\fc{1}{s_1s_3s_5}\ ,\cr\cr
\fc{1}{z_{14} z_{15} z_{23} z_{26} z_{34} z_{56}}   &\fc{1}{x (1-y) (1-z)}&\fc{1}{X_2X_3X_5Y_2}&\fc{1}{s_2s_5t_2}\ss{+}\fc{1}{s_3s_5t_2}\ ,\cr\cr
\fc{1}{z_{12} z_{15} z_{26} z_{34} z_{36} z_{45}}   &\fc{1}{z\ (1-x)\ (1-y)}&\fc{1}{X_1X_3X_4Y_3}&\fc{1}{s_1s_3t_3}\ss{+}\fc{1}{s_1s_4t_3}\ ,\cr\cr
\fc{1}{z_{15} z_{16} z_{24} z_{26} z_{34} z_{35}}   &\fc{y}{(1-y)\ (1-xy)\ (1-yz)}&\fc{Y_1}{X_3X_6Y_2Y_3}&\fc{1}{s_3s_6t_2}\ss{+}\fc{1}{s_3s_6t_3}\ ,\cr\cr
\fc{1}{z_{15} z_{16} z_{23} z_{26} z_{34} z_{45}}   &\fc{1}{(1-x)\ (1-y)\ (1-z)}&\fc{1}{X_2X_3X_4X_6Y_2Y_3}&\ss{-}\fc{1}{s_2s_4s_6}\ss{-}\fc{1}{s_2s_6t_2}\ss{-}\fc{1}{s_3s_6t_2}\cr
&&&\ss{-}\fc{1}{s_3s_6t_3}\ss{-}\fc{1}{s_4s_6t_3}\ .\cr\crr}}

The fundamental objects \FundamentalSet\  correspond to  the $14$ rational functions
\eqnn\fundamentalvi
$$\eqalignno{
\fc{1}{X_1 X_3 X_5},\ &\fc{1}{X_2 X_4 X_6},\ \fc{1}{X_1 X_4 Y_1},\ \fc{1}{X_2 X_5 Y_2},\ 
\fc{1}{X_3 X_6 Y_3},\ \fc{1}{X_2 X_5 Y_1},\ \fc{1}{X_3 X_6 Y_2},\ \fc{1}{X_1 X_4 Y_3}\ ,\cr
&\fc{1}{X_2 X_4 Y_1},\ \fc{1}{X_3 X_5 Y_2},\ \fc{1}{X_4 X_6 Y_3}\ ,\ \fc{1}{X_1 X_5 Y_1}\ ,\ 
\fc{1}{X_2 X_6 Y_2}\ ,\ \fc{1}{X_1 X_3 Y_3},\ &\fundamentalvi}$$
which furnish the $C_4=14$ poles
\eqnn\Fundamentalvi
$$\eqalignno{
\fc{1}{s_1 s_3 s_5},\ \fc{1}{s_2 s_4 s_6},\ &\fc{1}{s_1 s_4 t_1},\ \fc{1}{s_2 s_5 t_2},\ 
\fc{1}{s_3 s_6t_3},\ \fc{1}{s_2 s_5 t_1},\ \fc{1}{s_3 s_6 t_2},\ \fc{1}{s_1 s_4 t_3}\ ,\cr
&\fc{1}{s_2 s_4 t_1},\ \fc{1}{s_3 s_5 t_2},\ \fc{1}{s_4 s_6 t_3},\ \fc{1}{s_1 s_5 t_1},\ 
\fc{1}{s_2 s_6 t_2},\ \fc{1}{s_1 s_3 t_3},\ &\Fundamentalvi}$$
as single poles in the denominator of \GENERI, respectively.
\comment{In the basis \EXsix\ the rational functions become
\eqn\Rationalvi{\eqalign{
&\fc{1}{x(1-y)z(1-xyz)},\ \fc{1}{(1-x) (1-z) (1-x y z)},\ \fc{1}{(1-x) y z (1-y z)},\cr 
&\fc{1-xyz}{x (1-x y) (1-z) (1-y z)},\ \fc{1}{(1-y) (1-x y) (1-x y z)},\ \fc{1}{x y (1-x y) (1-z)},\cr
&\fc{1}{(1-y) (1-y z) (1-x y z)},\ \fc{1-xyz}{z(1-x) (1-x y)  (1-y z)},\ \fc{1}{(1-x) y (1-z)}\ ,\ \fc{1}{x (1-y) (1-y z)},\cr
&\fc{1}{(1-x) (1-x y) (1-y z)},\ \fc{1}{x y (1-x y) z (1-y z)},\ \fc{1}{(1-x y) (1-z) (1-y z)},\ \fc{1}{(1-y) (1-x y) z},}}
respectively.}
The cyclically invariant integral \CYC\ is given by
\eqnn\CYCvi
$$\eqalignno{
B_6\lf[n_i=-1\atop n_{ii}=-1\ri]&=\int_0^1dx\int_0^1dy\int_0^1dz\ \fc{I_6(x,y,z)}{x\ (1-x)\ y\ (1-y)\ z\ (1-z)}=
\fc{1}{s_1 s_3 s_5} +\fc{1}{s_2 s_4 s_6}\cr
&+\fc{1}{s_1 s_4 t_1}+\fc{1}{s_2 s_5 t_2}+\fc{1}{s_3 s_6t_3}
+\fc{1}{s_2 s_5 t_1}+\fc{1}{s_3 s_6 t_2}+\fc{1}{s_1 s_4 t_3} \cr
&+\fc{1}{s_2 s_4 t_1}+\fc{1}{s_3 s_5 t_2}+\fc{1}{s_4 s_6 t_3}+\fc{1}{s_1 s_5 t_1}+\fc{1}{s_2 s_6 t_2}+\fc{1}{s_1 s_3 t_3}+\cdots\ ,&\CYCvi}$$
and exhibits all fourteen poles \Fundamentalvi\ in its power series expansion.
After triple poles, for a (transcendental) $N=6$ integral the next leading order to start 
with are single poles. They always come  with a $\zeta(2)$.
In analogy to \fundamentalvi\ for the latter we may 
introduce a fundamental set of rational functions\foot{Note, that the rational functions 
$\fc{1}{Y_i}$ giving rise to the single poles $\sim t_i^{-1}$ have double poles in \GENERI, 
\ie $\nt_{ij}=-2$ for some $z_{ij}$.} furnishing the six single poles 
$\fc{\zeta(2)}{s_i},\ i=1,\ldots,6$:
\eqn\fundamentalvind{
\matrix{
{\underline{rational\ function}\atop\underline{in\ eq.\ \GENERI}}&
{\underline{rational\ function}\atop\underline{in\ eq.\ \EXsix}}&
{\underline{rational\ function}\atop\underline{in\ eq.\ \EXSix}}&
\ss{\underline{lowest\ order\ poles}}\cr\cr
\fc{1}{z_{12} z_{15} z_{24} z_{35} z_{36} z_{46}}&\fc{1}{(1-xy)\ z\ (1-yz)}&\fc{1}{X_1}&\fc{\zeta(2)}{s_1}\ ,\cr\cr
\fc{1}{z_{14} z_{15} z_{23} z_{26} z_{35} z_{46}}&\fc{1}{(1-z)\ (1-xy)}&\fc{1}{X_2}&\fc{\zeta(2)}{s_2}\ ,\cr\cr
\fc{1}{z_{13} z_{15} z_{25} z_{26} z_{34} z_{46}}&\fc{1}{(1-y)\ (1-xyz)}&\fc{1}{X_3}&\fc{\zeta(2)}{s_3}\ ,\cr\cr
\fc{1}{z_{13} z_{15} z_{24} z_{26} z_{36} z_{45}}&\fc{1}{(1-x)\ (1-yz)}&\fc{1}{X_4}&\fc{\zeta(2)}{s_4}\ ,\cr\cr
\fc{1}{z_{13} z_{14} z_{24} z_{26} z_{35} z_{56}}&\fc{1}{x\ (1-xy)\ (1-yz)}&\fc{1}{X_5}&\fc{\zeta(2)}{s_5}\ ,\cr\cr
\fc{1}{z_{13} z_{16} z_{24} z_{25} z_{35} z_{46}}&\fc{1}{(1-xy)\ (1-yz)\ (1-xyz)}&\fc{1}{X_6}&\fc{\zeta(2)}{s_6}\ .}}
All (transcendental) integrals with single poles can be decomposed w.r.t.\ the 
basis \fundamentalvind\ modulo finite pieces to be discussed in a moment. Subject to 
\constraint\ we have e.g.:
\eqnn\DECOONE
$$\eqalignno{
\fc{1}{z\ (1-xy)}\simeq \fc{X_6Y_2}{X_1}&=\fc{1}{X_1}-Y_1\ ,\cr\crr
\fc{y}{(1-y)\ (1-xyz)}\simeq\fc{Y_1}{X_3}&=\fc{1}{X_3}-X_6Y_2Y_3\ ,\cr\crr
\fc{x}{(1-x)\ (1-xyz)}\simeq\fc{X_5Y_2}{X_4}&=\fc{1}{X_4}-Y_3\ ,&\DECOONE\cr\crr
\fc{1}{x\ (1-yz)}\simeq\fc{X_6Y_3}{X_5}&=\fc{1}{X_5}-Y_1\ .}$$
After single poles, for an $N=6$ integral the next leading order to start with are constants. 
They always come  with a $\zeta(2)$ or $\zeta(3)$, e.g.:
\eqn\RESiii{
\matrix{
{\underline{rational\ function}\atop\underline{in\ eq.\ \GENERI}}&
{\underline{rational\ function}\atop\underline{in\ eq.\ \EXsix}}&
{\underline{monomial}\atop\underline{in\ eq.\ \EXSix}}&
\ss{\underline{lowest\ order}}\cr\cr
\fc{1}{z_{14} z_{15} z_{24} z_{26} z_{35} z_{36}}&\fc{y}{(1-xy)\ (1-yz)}&Y_1&2\ \zeta(3)\ ,\cr\cr
\fc{1}{z_{13} z_{14} z_{25} z_{26} z_{35} z_{46}}&\fc{1}{(1-xy)\ (1-xyz)}&Y_2&2\ \zeta(3)\ ,\cr\cr
\fc{1}{z_{13} z_{15} z_{24} z_{25} z_{36} z_{46}}&\fc{1}{(1-yz)\ (1-xyz)}&Y_3&2\ \zeta(3)\ ,\cr\cr
\fc{z_{16}}{z_{13} z_{14} z_{15} z_{25} z_{26} z_{36} z_{46}}&\fc{1}{1-xyz}&X_6Y_2Y_3&\zeta(3)\ ,\cr\cr
\fc{z_{56}}{z_{14} z_{15} z_{25} z_{26} z_{35} z_{36} z_{46}}&\fc{xy}{(1-xy)\ (1-xyz)}&X_5Y_1Y_2&\zeta(3)\ .}}
Again, we may add the functions \RESiii\ to \fundamentalvi\ to obtain  other fundamental 
sets subject to the constraints \Jsix\ and with the same poles \Fundamentalvi, cf.\ the next 
subsection for more details.

\subsubsec{$N=7$}

In this case we have the $14$ planar channels $(1,2),(1,3),(1,4)\equiv(5,7),(1,5)\equiv (6,7)$,
$(2,3),(2,4),(2,5),(2,6) \equiv (1,7),(3,4),(3,5),(3,6),(4,5),(4,6)$
and $(5,6)$ related to the $14$ variables 
$u_{1,2},u_{1,3},u_{1,4},u_{1,5},u_{2,3},u_{2,4}$
$u_{2,5},u_{2,6},u_{3,4},u_{3,5},u_{3,6},u_{4,5},u_{4,6}$ and $u_{5,6}$, respectively.
The seven--point integral \GENERIC\ becomes
\eqn\EXseven{\eqalign{
B_7[n]&=\int_0^1 dx_1 \int_0^1 dx_2 \int_0^1dx_3\int_0^1dx_4\ 
x_1^{s_1+n_1}\ x_2^{t_1+n_2}\ x_3^{t_5+n_3}\ x_4^{s_6+n_4}
(1-x_1)^{s_2+n_{11}}\cr 
&\times (1-x_2)^{s_3+n_{22}}\ (1-x_3)^{s_4+n_{33}}\ (1-x_4)^{s_5+n_{44}}\ 
(1-x_1x_2)^{s_{24}+n_{12}} \   (1-x_2x_3)^{s_{35}+n_{23}}\cr 
&\times (1-x_1x_2x_3)^{s_{25}+n_{13}}
\ (1-x_3x_4)^{s_{46}+n_{34}}\ (1-x_2x_3x_4)^{s_{36}+n_{24}} (1-x_1x_2x_3x_4)^{s_{26}+n_{14}}\ ,}}
with $s_i=\ap(k_i+k_{i+1})^2,\ t_j=\ap(k_j+k_{j+1}+k_{j+2})^2,\ i,j=1,\ldots,7$ subject to the 
cyclic identifications $i+7\equiv i$ and $j+7\equiv j$, respectively.

To bring \EXseven\ into the form \GENERICC\ according to \dualchoice\ we choose the three 
independent variables $u_{1,2}=x_1, u_{1,3}=x_2, u_{1,4}=x_3$ and $u_{1,5}=x_4$. Then, with 
\consequence\ the integral \EXseven\ assumes the form \GENERICC
\eqnn\EXSeven
$$\eqalignno{
B_7[n]&=\int_0^1 du_{i,j}\ 
 u_{1,2}^{s_1+n_{1,2}}\ u_{2,3}^{s_2+n_{2,3}}\ u_{3,4}^{s_3+n_{3,4}}\  
u_{4,5}^{s_4+n_{4,5}}\ u_{5,6}^{s_5+n_{5,6}}\ u_{1,5}^{s_6 +n_{1,5}}\ u_{2,6}^{s_7 +n_{2,6}}\ 
u_{1,3}^{t_1+n_{1,3}}\cr 
&\times u_{2,4}^{t_2+n_{2,4}}\ u_{3,5}^{t_3+n_{3,5}}\ u_{4,6}^{t_4+n_{4,6}}\ 
u_{1,4}^{t_5+n_{1,4}}\ u_{2,5}^{t_6+n_{2,5}}\ u_{3,6}^{t_7+n_{3,6}}\cr 
&\times \delta(u_{2,3}+u_{1,2}u_{3,4}u_{3,5}u_{3,6}-1)\ \delta(u_{2,4}+u_{1,2}u_{1,3}u_{3,5}u_{3,6}u_{4,5}u_{4,6}-1)\cr
&\times \delta(u_{2,5}+u_{1,2}u_{1,3}u_{1,4}u_{3,6}u_{4,6}u_{5,6}-1)\ 
\ \delta(u_{2,6}+u_{1,2}u_{1,3}u_{1,4}u_{1,5}-1)\cr 
&\times \delta(u_{3,4}+u_{1,3}u_{2,3}u_{4,5}u_{4,6}-1)\ 
\ \delta(u_{3,5}+u_{1,3}u_{1,4}u_{2,3}u_{2,4}u_{4,6}u_{5,6}-1)\cr
&\times\delta(u_{3,6}+u_{1,3}u_{1,4}u_{1,5}u_{2,3}u_{2,4}u_{2,5}-1)  
\ \delta(u_{4,5}+u_{1,4}u_{2,4}u_{3,4}u_{5,6}-1)\cr
&\times \delta(u_{4,6}+u_{1,4}u_{1,5}u_{2,4}u_{2,5}u_{3,4}u_{3,5}-1)  
\ \delta(u_{5,6}+u_{1,5}u_{2,5}u_{3,5}u_{4,5}-1)\ ,&\EXSeven}$$
with the assignment \RELbesser.

Using the identifications $x_1:=w,\ x_2:=z,\ x_3:=y$ and $x_4:=x$ in \EXseven, it
is convenient to introduce
\eqnn\Iseven
$$\eqalignno{
I_7(x,y,z,w)&=x^{s_6}\ y^{t_5}\ z^{t_1}\ w^{s_1}\ (1-x)^{s_5}\ (1-y)^{s_4}\ (1-z)^{s_3}\ 
(1-w)^{s_2}&\Iseven\cr 
&\times  (1-xy)^{s_{46}}\ (1-wz)^{s_{24}}\ (1-yz)^{s_{35}}\  (1-xyz)^{s_{36}}\ (1-yzw)^{s_{25}}\ 
(1-xyzw)^{s_{26}}}$$
and use the following notation for the dual variables $u_{i,j}$
\eqn\Uxyvii{
X_i=u_{i,i+1}\ \ \ ,\ \ \ Y_j=u_{j,j+2}\ \ \ ,\ \ \ i,j=1,\ldots,7\ \ \ ,\ \ \ 
i+7 \equiv i\ ,\ i,j=1,\ldots,7\ .}
Furthermore, we define:
\eqn\Jseven{\eqalign{
J_7(X,Y)&=\lf(\prod_{i=1}^7 X_i^{s_i}\ri) \ \lf(\prod_{j=1}^7 Y_j^{t_j}\ri)\   
\delta(X_2+X_1X_3Y_3Y_7-1)\ \delta(Y_2+X_1X_4Y_1Y_3Y_4Y_7-1)\cr
&\times \ \delta(Y_6+X_1X_5Y_1Y_4Y_5Y_7-1)\ \delta(X_7+X_1X_6Y_1Y_5-1)\ 
\delta(X_3+X_2X_4Y_1Y_4-1)\cr 
&\times 
\delta(X_4+X_3X_5Y_2Y_5-1)\ \delta(Y_4+X_3X_6Y_2Y_3Y_5Y_6-1)  \ 
\delta(X_5+X_4X_6Y_3Y_6-1)
\cr  
&\times \delta(Y_3+X_2X_5Y_1Y_2Y_4Y_5-1) \ 
\delta(Y_7+X_2X_6Y_1Y_2Y_5Y_6-1) \ .}}

Let us now discuss a few examples. The pole structure of the integral
\eqn\exei{
\int_0^1dx\int_0^1dy\int_0^1dz\int_0^1dw\ \fc{I_7(x,y,z,w)}{x\ (1-y)\ (1-wz)\ (1-yz)}}
can be easily deduced after transforming it into the form \EXSix
\eqn\exeii{
\lf(\prod_{i=1}^7\int_0^1 dX_i\ri) \ \lf(\prod_{j=1}^7\int_0^1 dY_j\ri) 
\ \fc{J_7(X,Y)}{X_4X_6Y_3Y_6}=\fc{1}{s_4s_6t_3t_6}+\cdots\ .}
Hence, the only simultaneous pole is at $X_4,X_6,Y_3,Y_6\ra0$ with the product of $\delta$--functions 
yielding the constraints for the ten variables $X_1,X_2,X_3,X_5,X_7,Y_1,Y_2,Y_4,Y_5,Y_7\ra1$.
Note, that by construction a set of four poles in \EXSeven\ does not necessarily yield a compatible 
set of channels, \eg the integral
\eqn\exEiii{\eqalign{
\int_0^1dx\int_0^1dy\int_0^1dz&
\int_0^1dw\ \fc{I_7(x,y,z,w)}{w\ (1-x)\ (1-z)\ (1-wyz)\ (1-wxyz)}\cr
&=\lf(\prod_{i=1}^7\int_0^1 dX_i\ri) \ \lf(\prod_{j=1}^7\int_0^1 dY_j\ri) 
\ \fc{J_7(X,Y)}{X_1X_3X_5X_7}=\fc{1}{s_1s_3s_5}+\cdots}}
does not give rise to a quadruple pole as $(1,2),(3,4),(5,6)$ and $(7,1)$ are not compatible channels.
Subsequently, in the sequel we list a few  non--trivial examples:
\eqn\RESI{
\matrix{
{\underline{rational\ function}\atop\underline{in\ eq.\ \GENERI}}&
{\underline{rational\ function}\atop\underline{in\ eq.\ \EXseven}}&
{\underline{rational\ function}\atop\underline{in\ eq.\ \EXSeven}}&
\ss{\underline{lowest\ order\ poles}}\cr\cr
\fc{z_{17}^3}{z_{12}z_{13}z_{14}z_{15}z_{16}z_{27}z_{37}z_{47}z_{57}z_{67}}&\fc{1}{x\ y\ z\ w}&\fc{X_7^3Y_2Y_3Y_4Y_6^2Y_7^2}{X_1X_6Y_1Y_5}&\fc{1}{s_1s_6t_1t_5}\ ,\cr\cr
\fc{z_{17}^2}{z_{12} z_{14} z_{15} z_{16} z_{27} z_{34} z_{37} z_{57} z_{67}}&\fc{1}{x\ y\ (1-z)\ w}&\fc{X_7^2Y_4Y_6Y_7}{X_1X_3X_6Y_5}&\fc{1}{s_1s_3s_6t_5}\ ,\cr\cr
\fc{z_{17}^2}{z_{12} z_{13} z_{14} z_{16} z_{27} z_{37} z_{47} z_{56} z_{57}}&\fc{1}{(1-x)\ y\ z\ w}&\fc{X_7^2Y_2Y_3Y_6^2Y_7}{X_1X_5Y_1Y_5}&\fc{1}{s_1s_5t_1t_5}\ ,\cr\cr
\fc{z_{17}z_{67}}{z_{12} z_{13} z_{16} z_{27} z_{37} z_{46} z_{47} z_{56} z_{57}}&\fc{x}{w\ z\ (1-x)\ (1-xy)}&\fc{X_6X_7Y_2Y_3Y_6^2}{X_1X_5Y_1Y_4}&\fc{1}{s_1s_5t_1t_4}\ ,\cr\cr
\fc{z_{17}}{z_{12} z_{14} z_{16} z_{27} z_{34} z_{37} z_{56} z_{57}}&\fc{1}{y\ w\ (1-x)\ (1-z)}&\fc{X_7Y_6}{X_1X_3X_5Y_5}&\fc{1}{s_1s_3s_5t_5}\ ,\cr\cr
\fc{z_{17}}{z_{12} z_{13} z_{16} z_{27} z_{37} z_{45} z_{47} z_{56}}&\fc{1}{z\ w\ (1-x)\ (1-y)}&\fc{X_7Y_3Y_6}{X_1X_4X_5Y_1Y_4}&\fc{1}{s_1s_4t_1t_4}\ss{+}\fc{1}{s_1s_5t_1t_4}\ ,\cr\cr
\fc{z_{17}}{z_{14} z_{15} z_{16} z_{23} z_{27} z_{34} z_{57} z_{67}}&\fc{1}{x\ y\ (1-z)\ (1-w)}&\fc{X_7Y_4Y_7}{X_2X_3X_6Y_2Y_5}&\fc{1}{s_2s_6t_2t_5}\ss{+}\fc{1}{s_3s_6t_2t_5}\ ,\cr\cr
\fc{z_{67}}{z_{12} z_{16} z_{27} z_{36} z_{37} z_{45} z_{47} z_{56}}&\fc{x\ y}{w(1-x)(1-y)(1-xyz)}&\fc{X_6Y_2Y_5Y_6}{X_1X_4X_5Y_4Y_7}&\fc{1}{s_1s_4t_4t_7}\ss{+}\fc{1}{s_1s_5t_4t_7}\ ,\cr\cr
\fc{1}{z_{12} z_{16} z_{27} z_{34} z_{37} z_{45} z_{56}}&\fc{1}{w(1-x)(1-y)(1-z)}&\fc{1}{X_1X_3X_4X_5Y_3Y_4Y_7}&\fc{1}{s_1s_3s_5t_7}\ss{+}\fc{1}{s_1s_3t_3t_7}\ss{+}\fc{1}{s_1s_4t_3t_7}\cr
&&&\ss{+}\fc{1}{s_1s_4t_4t_7}\ss{+}\fc{1}{s_1s_5t_4t_7}\ .\cr\crr}}
After quadruple poles, for an $N=7$ integral the next leading order to start with are double poles. 
They always come  with a $\zeta(2)$, e.g.: 
\eqn\RESiia{
\matrix{
{\underline{rational\ function}\atop\underline{in\ eq.\ \GENERI}}&
{\underline{rational\ function}\atop\underline{in\ eq.\ \EXseven}}&
{\underline{rational\ function}\atop\underline{in\ eq.\ \EXSeven}}&
\ss{\underline{lowest\ order\ poles}}\cr\cr
\fc{1}{z_{12} z_{15} z_{24} z_{35} z_{37} z_{46} z_{67}}&\fc{1}{w x (1-x y) (1-w z) (1-y z)}&\fc{1}{X_1X_6}&\fc{\zeta(2)}{s_1s_6}\ ,\cr\cr
\fc{z_{14}}{z_{12} z_{13} z_{16} z_{24} z_{35} z_{46} z_{47} z_{57}}&\fc{1}{w (1-x y) z (1-w z) (1-y z)}&\fc{1}{X_1Y_1}&\fc{\zeta(2)}{s_1t_1}\ ,\cr\cr
\fc{1}{z_{15} z_{16} z_{26} z_{27} z_{34} z_{37} z_{45}}&
\fc{yz}{(1-y) (1-z) (1-w x y z)}&\fc{Y_1Y_5}{X_3X_4Y_3}&
\fc{\zeta(2)}{s_3t_3}+\fc{\zeta(2)}{s_4t_3}\ .\cr\cr}}
After double poles, for an $N=7$ integral the next leading order to start with are single poles. 
They are always accompanied  by   $\zeta(2)$ or $\zeta(3)$ factors:
\eqn\RESiib{
\matrix{
{\underline{rational\ function}\atop\underline{in\ eq.\ \GENERI}}&
{\underline{rational\ function}\atop\underline{in\ eq.\ \EXseven}}&
{\underline{rational\ function}\atop\underline{in\ eq.\ \EXSeven}}&
\ss{\underline{lowest\ order\ poles}}\cr\cr
\fc{1}{z_{12} z_{16} z_{24} z_{35} z_{37} z_{46} z_{57}}&\fc{1}{w (1-x y) (1-w z) (1-y z)}&\fc{1}{X_1}&\fc{2\zeta(2)}{s_1}\ ,\cr\cr
\fc{z_{15}}{z_{12} z_{14} z_{16} z_{25} z_{35} z_{37} z_{46} z_{57}}&\fc{1}{w(1-xy)(1-yz)(1-wyz)}&\fc{Y_2}{X_1}&\fc{2\zeta(2)}{s_1}\ ,\cr\cr
\fc{1}{z_{12} z_{16} z_{24} z_{35} z_{36} z_{47} z_{57}}&\fc{1}{w (1-w z) (1-y z) (1-x y z)}&\fc{Y_4}{X_1}&\fc{2\zeta(3)}{s_1}\ ,\cr\cr
\fc{1}{z_{12} z_{15} z_{16} z_{24} z_{35} z_{37} z_{46} z_{47}}&\fc{y}{w (1-x y) (1-w z) (1-y z)}&\fc{Y_5}{X_1}&\fc{2\zeta(3)}{s_1}\ ,\cr\cr
\fc{z_{15}z_{23}}{z_{12} z_{13} z_{16} z_{24} z_{25} z_{35} z_{37} z_{46} z_{57}}&\fc{1-w}{w (1-x y) (1-w z) (1-y z) (1-w y z)}&\fc{X_2Y_2}{X_1}&\fc{2\zeta(2)}{s_1}\ .\cr\crr}}
After single poles, for an $N=7$ integral the next leading order to start with are the zeta constants $\zeta(2), \zeta(3)$ or $\zeta(4)$.
First, we display examples without poles and whose series expansion starts
at $\zeta(2)$ or $\zeta(3)$:
\eqn\RESiv{
\matrix{
{\underline{rational\ function}\atop\underline{in\ eq.\ \GENERI}}&
{\underline{rational\ function}\atop\underline{in\ eq.\ \EXseven}}&
{\underline{monomial}\atop\underline{in\ eq.\ \EXSeven}}&
\ss{\underline{lowest\ order}}\cr\cr
\fc{z_{47}}{z_{14} z_{16} z_{24} z_{27} z_{35} z_{37} z_{46} z_{57}}&\fc{z}{(1-xy)\ (1-yz)\ (1-wz)}&Y_1&2\ \zeta(2)+2\ \zeta(3)\ ,\cr\cr
\fc{z_{14}z_{37}}{z_{13} z_{15} z_{16} z_{24} z_{27} z_{35} z_{36} z_{47}^2}&\fc{y}{(1-yz)\ (1-wz)\ (1-xyz)}&Y_4Y_5&\fc{3}{2}\ \zeta(2)+\fc{3}{2}\ \zeta(3)\ ,\cr\cr
\fc{z_{15}z_{37}}{z_{13} z_{14} z_{16} z_{25} z_{27} z_{35} z_{36} z_{47} z_{57}}&\fc{1}{(1-yz)\ (1-xyz)\ (1-wyz)}&Y_2Y_4&\fc{5}{2}\ \zeta(4)+4\ \zeta(3)-2\ \zeta(3)\ ,\cr\cr
\fc{1}{z_{13} z_{14} z_{25} z_{27} z_{36} z_{46} z_{57}}&\fc{1}{(1-xy)\ (1-wyz)\ (1-xyz)}&Y_2Y_3Y_6&3\ \zeta(3)\ ,}}
Finally, we give examples without poles and whose series expansion starts at $\zeta(4)$:
\eqn\RESiii{
\matrix{
{\underline{rational\ function}\atop\underline{in\ eq.\ \GENERI}}&
{\underline{rational\ function}\atop\underline{in\ eq.\ \EXseven}}&
{\underline{monomial}\atop\underline{in\ eq.\ \EXSeven}}&
\ss{\underline{lowest\ order}}\cr\cr
\fc{1}{z_{13} z_{16} z_{24} z_{27} z_{35} z_{46} z_{57}}&\fc{1}{(1-xy)\ (1-yz)\ (1-wz)}&1&\fc{27}{4}\ \zeta(4)\ ,\cr\cr
\fc{1}{z_{14} z_{16} z_{24} z_{27} z_{35} z_{36} z_{57}}&\fc{z}{(1-yz)\ (1-wz)\ (1-xyz)}&Y_1Y_4&\fc{17}{4}\ \zeta(4)\ ,\cr\cr
\fc{z_{37}}{z_{13} z_{14} z_{26} z_{27} z_{35} z_{36} z_{47} z_{57}}&\fc{1}{(1-yz)\ (1-xyz)\ (1-wxyz)}&Y_2Y_4Y_6&\ 3\ \zeta(4)\ ,\cr\cr
\fc{1}{z_{13} z_{14} z_{25} z_{26} z_{37} z_{46} z_{57}}&\fc{1}{(1-xy)\ (1-wyz)\ (1-wxyz)}&Y_2Y_3Y_6Y_7&\ \fc{5}{2}\ \zeta(4)\ ,\cr\cr
\fc{z_{16}}{z_{13} z_{14} z_{15} z_{26} z_{27} z_{36} z_{46} z_{57}}&\fc{1}{(1-xy)\ (1-xyz)\ (1-wxyz)}&Y_6^2Y_2Y_3&\ 3\ \zeta(4)\ .}}
Again, we may add the functions \RESiii\ to the $42$ fundamental quadruple poles to obtain  
other fundamental sets subject to the constraints \Jseven, cf.\ the next subsection for more details.

\subsubsec{$N=8$}

In this case we have the $20$ planar channels $(1,2),(1,3)\equiv(4,8),(1,4)\equiv(5,8),(1,5)$
$\equiv(6,8),(1,6)\equiv (7,8),(2,3),(2,4),(2,5),(2,6),(2,7) \equiv (1,8),(3,4),(3,5),(3,6),(3,7),(4,5),$
$(4,6),(4,7),(5,6),(5,7)$ and $(6,7)$ related to the $20$ variables 
$u_{1,2},u_{2,3},u_{3,4},u_{4,5},u_{5,6},u_{6,7}$,
$u_{1,6}=u_{7,8},u_{2,7},u_{1,3},u_{2,4},u_{3,5},u_{4,6},u_{5,7},u_{6,8},u_{2,6},u_{3,7},u_{1,4},u_{2,5},u_{3,6},u_{4,8}$.
The eight--point integral \GENERIC\ becomes
\eqnn\EXeight
$$\eqalignno{
B_8[n]&=\int_0^1 dx_1 \int_0^1 dx_2 \int_0^1dx_3\int_0^1dx_4\ \int_0^1dx_5\ 
x_1^{s_1+n_1}\ x_2^{t_1+n_2}\ x_3^{u_1+n_3}\ x_4^{t_6+n_4}\ x_5^{s_7+n_5}\cr 
&\times (1-x_1)^{s_2+n_{11}}\ (1-x_2)^{s_3+n_{22}}\ (1-x_3)^{s_4+n_{33}}\ (1-x_4)^{s_5+n_{44}}\ (1-x_5)^{s_6+n_{55}}\cr 
&\times (1-x_1x_2)^{s_{24}+n_{12}}\ 
 (1-x_2x_3)^{s_{35}+n_{23}}\ (1-x_3x_4)^{s_{46}+n_{34}}\ (1-x_4x_5)^{s_{57}+n_{45}}\cr 
&\times (1-x_1x_2x_3)^{s_{25}+n_{13}}\ (1-x_2x_3x_4)^{s_{36}+n_{24}}\ 
(1-x_3x_4x_5)^{s_{47}+n_{35}} &\EXeight\cr
&\times (1-x_1x_2x_3x_4)^{s_{26}+n_{14}}\ 
(1-x_2x_3x_4x_5)^{s_{37}+n_{25}}\ (1-x_1x_2x_3x_4x_5)^{s_{27}+n_{15}}\ ,}$$
with $s_i=\ap(k_i+k_{i+1})^2,\ t_j=\ap(k_j+k_{j+1}+k_{j+2})^2,\ i,j=1,\ldots,8$ subject to 
the cyclic identifications $i+8\equiv i,\ j+8\equiv j$, respectively and 
$u_l=\ap(k_l+k_{l+1}+k_{l+2}+k_{l+3})^2$, for $l=1$ to $4$.

To bring \EXeight\ into the form \GENERICC\ according to \dualchoice\ we choose the three 
independent variables $u_{1,2}=x_1, u_{1,3}=x_2, u_{1,4}=x_3$ and $u_{1,5}=x_4$. Then, with 
\consequence\ the integral \EXeight\ assumes the form \GENERICC.
\eqnn\EXEight
$$\eqalignno{
B_8[n]&=\int_0^1 du_{i,j}\ 
 u_{1,2}^{s_1+n_{1,2}}\ u_{2,3}^{s_2+n_{2,3}}\ u_{3,4}^{s_3+n_{3,4}}\  
u_{4,5}^{s_4+n_{4,5}}\ u_{5,6}^{s_5+n_{5,6}}\ u_{6,7}^{s_6+n_{6,7}}\ u_{1,6}^{s_7 +n_{1,6}}\ u_{2,7}^{s_8 +n_{2,7}}\cr 
&\times u_{1,3}^{t_1+n_{1,3}}\ u_{2,4}^{t_2+n_{2,4}}\ u_{3,5}^{t_3+n_{3,5}}\ 
u_{4,6}^{t_4+n_{4,6}}\ u_{5,7}^{t_5+n_{5,7}}\ 
u_{1,5}^{t_6+n_{1,5}}\ u_{2,6}^{t_7+n_{2,6}}\ u_{3,7}^{t_8+n_{3,7}}\cr 
&\times u_{1,4}^{u_1+n_{1,4}}\ u_{2,5}^{u_2+n_{2,5}}\ u_{3,6}^{u_3+n_{3,6}}\ 
u_{4,7}^{u_4+n_{4,7}}\cr
&\times \delta(u_{2,3}+u_{1,2} u_{3,4} u_{3,5} u_{3,6} u_{3,7}-1)
\ \delta(u_{2,4}+u_{1,2} u_{1,3} u_{3,5} u_{3,6} u_{3,7} u_{4,5} u_{4,6} u_{4,7}-1)\cr
&\times\delta(u_{2,5}+u_{1,2} u_{1,3} u_{1,4} u_{3,6} u_{3,7} u_{4,6} u_{4,7} u_{5,6} u_{5,7}-1)
\ \delta(u_{2,7}+u_{1,2} u_{1,3} u_{1,4} u_{1,5} u_{1,6}-1)\cr
&\times\delta(u_{2,6}+u_{1,2} u_{1,3} u_{1,4} u_{1,5} u_{3,7} u_{4,7} u_{5,7} u_{6,7}-1)
\ \delta(u_{3,4}+u_{1,3} u_{2,3} u_{4,5} u_{4,6} u_{4,7}-1)\cr
&\times\delta(u_{3,5}+u_{1,3} u_{1,4} u_{2,3} u_{2,4} u_{4,6} u_{4,7} u_{5,6} u_{5,7}-1)
\ \delta(u_{6,7}+u_{1,6} u_{2,6} u_{3,6} u_{4,6} u_{5,6} -1)\cr
&\times\delta(u_{3,7}+u_{1,3} u_{1,4} u_{1,5} u_{1,6} u_{2,3} u_{2,4} u_{2,5} u_{2,6}-1)
\ \delta(u_{4,5}+u_{1,4} u_{2,4} u_{3,4} u_{5,6} u_{5,7}-1)\cr
&\times\delta(u_{4,6}+u_{1,4} u_{1,5} u_{2,4} u_{2,5} u_{3,4} u_{3,5} u_{5,7} u_{6,7}-1)
\ \delta(u_{5,6}+u_{1,5} u_{2,5} u_{3,5} u_{4,5} u_{6,7}-1)\cr
&\times\delta(u_{4,7}+u_{1,4} u_{1,5} u_{1,6} u_{2,4} u_{2,5} u_{2,6} u_{3,4} u_{3,5} u_{3,6}-1)\cr
&\times\delta(u_{5,7}+u_{1,5} u_{1,6} u_{2,5} u_{2,6} u_{3,5} u_{3,6} u_{4,5} u_{4,6}-1)\cr
&\times\delta(u_{3,6}+u_{1,3} u_{1,4} u_{1,5} u_{2,3} u_{2,4} u_{2,5} u_{4,7} u_{5,7} u_{6,7}-1)\ , &\EXEight}$$
with the assignment \RELbesser.

In what follows it is convenient to introduce
\eqn\Ieight{\eqalign{
I_8(x,y,z,w,v)&=x^{s_7}\ y^{t_6}\ z^{u_1}\ w^{t_1}\ v^{s_1}\ (1-x)^{s_6}\ (1-y)^{s_5}\ (1-z)^{s_4}\  (1-w)^{s_3}\  (1-v)^{s_2}\cr 
&\times (1-xy)^{s_{57}}\ (1-yz)^{s_{46}}\ (1-wz)^{s_{35}}\ (1-vw)^{s_{24}}\ (1-xyz)^{s_{47}}\ 
(1-wyz)^{s_{36}}\cr 
&\times (1-vwz)^{s_{25}}\ (1-wxyz)^{s_{37}}\ (1-vwyz)^{s_{26}}\ (1-vwxyz)^{s_{27}}}}
arising from \EXeight\ with the identifications $x_1:=v,\ x_2:=w,\ x_3:=z,\ x_4=y$ and $x_5:=x$.
Similarly as in the previous subsections, the following shorter notation for the dual variables $u_{i,j}$ is used
\eqn\Uxyviii{\eqalign{
X_i&=u_{i,i+1}\ \ \ ,\ \ \ Y_j=u_{j,j+2}\ \ \ ,\ \ \ i,j=1,\ldots,8\ \ \ ,\ \ \ 
i+8 \equiv i\ ,\ j+8 \equiv j\ ,\cr
Z_k&=u_{k,k+3}\ \ \ ,\ \ \ k=1,\ldots,4\ ,}}
and we also define
\eqnn\Jeight
$$\eqalignno{
J_8(X,Y,Z)&=\lf(\prod_{i=1}^8 X_i^{s_i}\ri)\ \lf(\prod_{j=1}^8 Y_j^{t_j}\ri)\ \lf(\prod_{k=1}^4 Z_k^{u_k}\ri)\  \delta(X_2+X_1 X_3 Y_3 Y_8 Z_3-1)\cr
&\times\delta(Y_2+X_1 X_4 Y_1 Y_3 Y_4 Y_8 Z_3 Z_4-1)\ 
\delta(Z_2+X_1 X_5 Y_1 Y_4 Y_5 Y_8 Z_1 Z_3 Z_4-1)\cr
&\times\delta(Y_7+X_1 X_6 Y_1 Y_5 Y_6 Y_8 Z_1 Z_4-1)\ \delta(X_8+X_1 X_7 Y_1 Y_6 Z_1-1)\cr
&\times \delta(X_3+X_2 X_4 Y_1 Y_4 Z_4-1)\ \delta(Y_3+X_2 X_5 Y_1 Y_2 Y_4 Y_5 Z_1 Z_4-1)\cr
&\times\delta(Z_3+X_2 X_6 Y_1 Y_2 Y_5 Y_6 Z_1 Z_2 Z_4-1)\ 
\delta(Y_8+X_2 X_7 Y_1 Y_2 Y_6 Y_7 Z_1 Z_2-1)\cr
&\times \delta(X_4+X_3 X_5 Y_2 Y_5 Z_1-1)\ \delta(Y_4+X_3 X_6 Y_2 Y_3 Y_5 Y_6 Z_1 Z_2-1)\cr
&\times \delta(Z_4+X_3 X_7 Y_2 Y_3 Y_6 Y_7 Z_1 Z_2 Z_3-1)\ \delta(X_5+X_4 X_6 Y_3 Y_6 Z_2-1)\cr
&\times \delta(Y_5+X_4 X_7 Y_3 Y_4 Y_6 Y_7 Z_2 Z_3-1)\ 
\delta(X_6+X_5 X_7 Y_4 Y_7 Z_3 -1)\ .&\Jeight}$$

\noindent
Let us now discuss a few examples. The pole structure of the integral
\eqn\exei{
\int_0^1dx\int_0^1dy\int_0^1dz\int_0^1dw\int_0^1dv\ \fc{I_8(x,y,z,w,v)}{w\ (1-v) \ (1-z)\ (1-x y) \ (1-y z)}}
can be easily deduced after transforming it into the form \EXEight
\eqn\exeii{
\lf(\prod_{i=1}^8\int_0^1 dX_i\ri) \ \lf(\prod_{j=1}^8\int_0^1 dY_j\ri) 
\lf(\prod_{k=1}^4\int_0^1 dZ_k\ri) 
\ \fc{J_8(X,Y,Z)}{X_2 X_4 Y_1 Y_4 Z_4}=\fc{1}{s_2 s_4 t_1 t_4 u_4}+\cdots\ .}
Hence, the only simultaneous pole is at $X_2,X_4,Y_1,Y_4,Z_4\ra0$ with the product of 
$\delta$--functions yielding the constraints for the $15$ variables $X_1,X_3,X_5,X_6,X_7,X_8,Y_2,Y_3,Y_5,Y_6$,
$Y_7,Y_8,Z_1,Z_2,Z_3\ra1$.
Subsequently, in the sequel we list a few  non--trivial examples:
$$\matrix{
{\underline{rational\ function}\atop\underline{in\ eq.\ \GENERI}}&
{\underline{rational\ function}\atop\underline{in\ eq.\ \EXeight}}&\ss{\ldots}\cr\cr
\fc{z_{18}^4}{z_{12} z_{13} z_{14} z_{15} z_{16} z_{17} z_{28} z_{38} z_{48} z_{58} z_{68} z_{78}}&\fc{1}{x\ y\ z\ w\ v}&\ss{\ldots}\cr\cr
\fc{z_{18}^3}{z_{12} z_{13} z_{14}  z_{16} z_{17} z_{28} z_{38} z_{45} z_{58} z_{68} z_{78}}&\fc{1}{x\ y\ (1-z)\ w\ v}&\ss{\ldots}\cr\cr
\fc{1}{z_{17} z_{18} z_{23} z_{24} z_{35} z_{46} z_{57} z_{68}}&\fc{1}{(1-v)\ (1-x y)\ (1-w z)\ (1-y z)\ (1-v w)}&\ss{\ldots}\cr\cr
\fc{1}{z_{12} z_{17} z_{28} z_{34} z_{36} z_{47} z_{56} z_{58}}&\fc{y}{v\ (1-y) \ (1-w)\ (1-xyz)\ (1-wyz)}&\ss{\ldots}\cr\cr
\fc{1}{z_{17} z_{18} z_{24} z_{26} z_{35} z_{37} z_{45} z_{68}}&\fc{w\ z}{(1-z) \ (1-v w)\ (1-w z)\ (1-v w y z)\ (1-w x y z)}&\ss{\ldots}\cr\cr
\fc{z_{18}^2}{z_{12} z_{15} z_{16} z_{17} z_{28} z_{34} z_{38} z_{45} z_{68} z_{78}}&\fc{1}{x\ y\ v\ (1-z)\ (1-w)}&\ss{\ldots}\cr\cr
\fc{1}{z_{12} z_{17} z_{24} z_{34} z_{38} z_{56} z_{57} z_{68}}&\fc{1}{v\ z\ (1-y)\ (1-w)\ (1-xy)\ (1-vw)}&\ss{\ldots}\cr\cr
\fc{1}{z_{13} z_{17} z_{23} z_{25} z_{45} z_{46} z_{68} z_{78}}&\fc{1}{x\ y\ w\ (1-z)\ (1-v)\ (1-yz)\ (1-vwz)}&\ss{\ldots}\cr\cr}$$

\eqn\RESI{
\matrix{
\ss{\ldots}&{\underline{rational\ function}\atop\underline{in\ eq.\ \EXEight}}&
\ss{\underline{lowest\ order\ poles}}\cr\cr
\ss{\ldots}&\fc{X_8^4Y_2Y_3Y_4Y_5Y_7^3Y_8^3Z_2^2Z_3^2Z_4^2}{X_1X_7Y_1Y_6Z_1}&
\fc{1}{s_1s_7t_1t_6u_1}\ ,\cr\cr
\ss{\ldots}&\fc{X_8^3Y_2Y_5Y_7^2Y_8^2Z_2Z_3^2Z_4}{X_1X_4X_7Y_1Y_6}&\fc{1}{s_1s_4s_7t_1t_6}\ ,\cr\cr
\ss{\ldots}&\fc{1}{X_2 X_8 Y_2 Y_7 Z_2}&\fc{1}{s_2s_8t_2t_7 u_2}\ ,\cr\cr
\ss{\ldots}&\fc{Y_6 Z_2}{X_1 X_3X_5Y_8Z_3}&\fc{1}{s_1 s_3 s_5 t_8  u_3}\ ,\cr\cr
\ss{\ldots}&\fc{Y_1 Y_5 Z_1 Z_4}{X_4 X_8Y_3Y_7 Z_2}&\fc{1}{s_4 s_8 t_3 t_7 u_2}\ ,\cr\cr
\ss{\ldots}&\fc{X_8^2Y_5Y_7Y_8Z_4}{X_1X_3X_4X_7Y_3Y_6}&\fc{1}{s_1s_3s_7t_3t_6}\ss{+}
\fc{1}{s_1s_4s_7t_3t_6}\ ,\cr\cr
\ss{\ldots}&\fc{1}{X_1X_3X_5Y_2Y_5Z_1}&\fc{1}{s_1 s_3 s_5 t_5 u_1}\ss{+}\fc{1}{s_3 s_5 t_2 t_5 u_1} \ ,\cr\cr
\ss{\ldots}&\fc{Y_5Y_8}{X_3X_4X_7Y_1Y_4Y_6Y_7Z_2}&
\fc{1}{s_2 s_4 s_7 t_1 t_4}\ss{+}\fc{1}{s_2 s_4 s_7 t_1 t_6}  \ss{+}\fc{1}{s_2 s_4 s_7 t_4 t_7}\ss{+}\fc{1}{s_2 s_4 s_7 t_6 u_2}\ss{+}\fc{1}{s_2 s_4 s_7 t_7 u_2}\cr\cr}}


\subsec{Polynomial relations and Gr\"obner basis reduction}

For $n_{i,j}\geq 0$
the representation \GENERICC\ in the dual variables $u_{i,j}$ gives rise to a polynomial 
ring $\IR[u_\Pc]$ describing polynomials in $u_{i,j},\ (i,j)\in\Pc$ with coefficients in $\IR$.
This ring is suited to perform a Gr\"obner basis analysis to find a minimal basis for the polynomials in the integrand.
The set of integrals \GENERICC\ with $n_{i,j}\geq 0$ describe all integrals without 
poles in their $\ap$--expansion.
Due to the constraints \constraint, which give rise to the $\delta$--functions in \GENERICC,
many polynomials in the variables $u_{i,j}$ referring to different choices of the 
integers $n_{i,j}$ yield to the same integral $B_N$. The constraints \constraint\ 
define a monomial ideal $I$ in the polynomial ring $\IR[u_\Pc]$. Hence, 
we consider the quotient space $\IR[u_\Pc]/I$ and the Gr\"obner basis method is well
appropriate to choose a basis in the ideal $I$ and generate independent
sets of polynomials in the quotient ring $\IR[u_\Pc]/I$.
We are interested in simple representatives of equivalence classes 
for congruence modulo $I$. The properties of an ideal are reflected in the form of 
the elements of the Gr\"obner basis \refs{\Cox,\Sturmfels}.

Given a monomial ordering\foot{As monomial ordering we may choose lexicographic order 
or graded lexicographic order. Then, a monomial ordering of two polynomials 
$f=\sum_\al a_\al x^\al$ and $g=\sum_\bet b_\al x^\bet$ can be defined as follows:
$(i)$ {\it lexicographic order}:  $\al>_{lex}\bet$, if in the vector difference $\al-\bet\in\IZ^n$
the leftmost nonzero entry is positive ($x^\al>_{lex}x^\bet$ if $\al>_{lex}\bet$).\br
$(ii)$ {\it graded lexicographic order}:  $\al>_{grlex}\bet$, if 
$|\al|=\sum\limits_{i=1}^n\al_i>|\bet|$ and $\al>_{lex}\bet$  
($x^\al>_{grlex}x^\bet$, if $\al>_{grlex}\bet$).} in the ring 
a Gr\"obner basis $G=\{g_1,\ldots,g_d\}$ comprises a finite subset of the ideal $I$ such that the
leading term\foot{The leading term $LT(f)$ of a polynomial $f$ is defined as follows \Cox:
For $f=\sum_\al a_\al x^\al$ a nonzero polynomial in $\IR[x_1,\ldots,x_n]$ and $>$ a
specific monomial order \br
$(i)$ the {\it multidegree} of $f$ is 
$multideg(f):={\rm Max}\{\al\in \IZ^n_{\geq 0}\ |\ a_\al\neq0\ \}$,\br 
$(ii)$ the {\it leading coefficient} of $f$ is: $LC(f):=a_{multideg(f)}\in \IR$,\br 
$(iii)$ the {\it leading monomial}
of $f$ is $LM(f)=x^{multideg(f)}$, with coefficient $1$, and \br
$(iv)$ the {\it leading term} of $f$ is $$LT(f)=LC(f)\ LM(f)\ .$$
As an example we consider $f=xyz+2xy^2z^2+3z^3-7x^5y+3x^2z^2$ with $>$ the lexicographic
order. Then we have: $multideg(f)=(5,1,0),\ LC(f)=-7, LM(f)=x^5y$ and $LT(f)=-7x^5y$.}
 of any element of the ideal $I$ is divisible by a
leading term $LT(g_i)$ of an element of the subset. 
Alternatively, a finite subset $G$ of an ideal $I$ in a polynomial ring represents a Gr\"obner basis, 
if $\vev{LT(g_1),\ldots,LT(g_d)}=\vev{LT(I)}$ \refs{\Cox,\Sturmfels}.
Buchberger's algorithm generates the unique {\it reduced\/} Gr\"obner basis $G$, 
in which no monomial in a polynomial $p\in G$ of this basis is
divisible by a leading term of the other polynomials in
the basis and $LC(p)=1$.

The main idea is, that after dividing a polynomial 
$p\in \IR[x_1,\ldots,x_n]$ by a 
Gr\"obner basis $G=\{g_1,\ldots,g_d\}$ 
for the ideal $I\subset\IR[x_1,\ldots,x_n]$ the remainder $\ov p^G$
is uniquely fixed by the polynomial~$p$,
\cf Chapter 5, \S 3 of \Cox. More precisely according to the Proposition 1 therein we have:
For a given monomial ordering on $\IR[x_1,\ldots,x_n]$ and an ideal 
$I\subset\IR[x_1,\ldots,x_n]$,\br
$(i)$ Every $f\in \IR[x_1,\ldots,x_n]$ is congruent modulo 
$I$ to a unique polynomial $r$, which is a $\IR$--linear combination of 
the monomials in the complement of $\vev{LT(I)}$.\br
$(ii)$ The elements $\{x^\al\ |\ x^\al\notin \vev{LT(I)}\}$ are linearly independent
modulo $I$, \ie if $\sum_\al c_\al x^\al=0\ \mod\ I$, where the $x^\al$ are all in the complement 
of $\vev{LT(I)}$, then $c_\al=0$ for all $\al$.
As a consequence, for any given $f\in \IR[x_1,\ldots,x_n]$ the remainder $\ov f^G$ is a $\IR$--linear 
combination of the monomials contained in the complement of $LT(I)$, \ie
$\ov f^G\in Span\lf(x^\al\ |\ x^\al\notin\vev{LT(I)}\ri)$:
\eqn\CONSEQ{
f=x^a\equiv x_1^{a_1}\ldots x_n^{a_n}=\sum_{i=1}^d c_i\ g_i+\sum_{x^\al\notin\vev{LT(I)}}
r_{\al}\ x^\al\ .}

In the following with the Gr\"obner basis method we want to construct a basis for those polynomials,
which are independent on the constraints \constraint. This basis is determined by  the complement 
of $\vev{LT(I)}$ w.r.t.\ a Gr\"obner basis $G$.
Note, that the representation of this basis (and also of $\vev{LT(I)}$ and the remainders)  may 
depend on the chosen monomial ordering. At any rate, there is always the {\it same\/} number of 
monomials in the complement of $\vev{LT(I)}$. In addition, on the degree of the basis monomials  
we impose a condition to ensure, that in  the denominator of the integrands of \GENERI\ the $z_{ij}$ 
only appear with powers of at most one, \ie $\nt_{ij}\geq-1$. This restriction is useful to take into account
the relations stemming from partial integrations \PARTFR.  We illustrate the method with the following examples.

\subsubsec{$N=4$}

We work with the two coordinates $X_1=u_{1,2}$ and $X_2=u_{2,3}$ and consider the polynomial ring $\IR[X_1,X_2]$.
{}From \EXFour\ we can read off the constraints \constraint\ giving rise to the monomial ideal:
\eqn\idealiv{
I=\vev{\ X_1+X_2-1\ }\subset\IR[X_1,X_2]\ .}
w.r.t.\ lexicographic order we find for the  Gr\"obner basis of \idealiv:
\eqn\Griv{
G=\{g_1\}=\{\ X_1+X_2-1\ \}\ .}
Hence w.r.t.\ lexicographic order the leading term of this monomial gives rise to:
\eqn\LTv{
LT(I)=X_1\ .}
Therefore, the set of possible remainders modulo $I$ is the set of all $\IR$--linear
combinations of the following monomials:
\eqn\basisiv{
\{1,\ X_2,\ X_2^2,\ X_2^3,\ldots\ \}\ .}
For some examples let us determine their remainders on dividing them by the Gr\"obner basis 
\Griv:
\eqnn\exiv
$$\eqalignno{
X_1&=g_1+1-X_2\simeq 1-X_2\ ,&\exiv\cr
X_2&=0\ g_1+X_2\simeq X_2\ ,\cr
X_1X_2&=X_2\ g_1+X_2-X_2^2\simeq X_2-X_2^2\ ,\cr
X_1^2&=(1+X_1-X_2)\ g_1+1-2\ X_2+X_2^2\simeq 1-2\ X_2+X_2^2\ ,\cr
X_1^2X_2&=X_2\ (1+X_1-X_2)\ g_1+X_2-2\ X_2^2+X_2^3\simeq X_2-2\ X_2^2+X_2^3\ .}$$
Indeed, the remainders (displayed after the $\simeq$ sign) are generated by the basis \basisiv.

In \EXFour\ the monomials $X_2^{n_{11}},\ n_{11}=0,1,\ldots$ of \basisiv\
give rise to the following integrals 
\EXfour:
\eqn\Giveriseiv{
B_4[n]=\int_0^1 dx\ x^{s_{12}}\ (1-x)^{s_{23}+n_{11}}\ .} 
The integrals \EXfour\ without poles in their field--theory expansions are given by 
the integers $n_1,n_{11}=0,1\ldots$.
According to our construction all these integrals \EXfour\ can be generated 
 by $\IR$--linear combinations of the basis \Giveriseiv.
However according to \CORRES\ we have
\eqn\corrvi{
(1-x)^{n_{11}}\simeq \fc{z_{14}^{n_{11}}\ z_{23}^{n_{11}}}{z_{13}^{2+n_{11}}z_{24}^{2+n_{11}}}\ ,}
\ie all finite integrals \Giveriseiv\ in \GENERI\ imply some powers $\nt_{ij}$ 
with $\nt_{ij}<-1$. As a consequence the set of integrals \Giveriseiv\
cannot serve as a basis and \fundamentaliv\ are the only elements of the partial fraction
basis. Note, that this basis is two--dimensional, \ie $(N-2)!=2$ for $N=4$.

\subsubsec{$N=5$}

We work with the five coordinates \Uxyv\ and consider the polynomial ring
$\IR[X_1,\ldots,X_5]$.
{}From \Jfive\ we can read off the constraints \constraint\ giving rise to the monomial ideal:
\eqn\idealv{
I=\vev{\ X_2+X_1X_3-1,\ X_3+X_2X_4-1,\ X_5+X_1X_4-1\ }\subset\IR[X_1,\ldots,X_5]\ .}
w.r.t.\ lexicographic order we find for the (reduced) Gr\"obner basis of \idealv\ the three elements:
\eqn\Grv{
G=\{g_1,g_2,g_3\}=\{\ X_1+X_2X_5-1,\ X_3+X_2X_4-1,\ X_4+X_3X_5-1\ \}\ .}
Hence w.r.t.\ lexicographic order the leading terms of these three monomials give rise to:
\eqn\LTv{
LT(I)=\{\ X_1,\ X_2X_4,\ X_3X_5\ \}\ .}
Therefore, the set of possible remainders modulo $I$ is  the set of all $\IR$--linear
combinations of the following monomials:
\eqn\basisv{
\bigcup_{m,n=0}^\infty\{\ X_2^mX_3^n,\ X_2^mX_5^n,\ X_3^mX_4^n,\ X_4^mX_5^n\ \}\ .}
For some examples let us determine their remainders on dividing them by the Gr\"obner 
basis \Grv:
\eqnn\exv
$$\eqalignno{
X_1&=g_3+1-X_2X_5\simeq 1-X_2X_5\ ,&\exv\cr
X_1X_4&=g_1-X_5\ g_2+X_4\ g_3+1-X_5\simeq 1-X_5\ ,\cr
X_3X_5&=g_1+1-X_4\simeq 1-X_4\ ,\cr
X_3X_5^2&=X_5\ g_1+X_5-X_4X_5\simeq X_5-X_4X_5\ ,\cr
X_1X_2&=X_2\ g_3+X_2-X_2^2X_5\simeq X_2-X_2^2X_5\ ,\cr
X_2X_3X_5&=X_2\ g_1-g_2-1+X_2+X_3\simeq -1+X_2+X_3\ .}$$
Indeed, the remainders (displayed after the $\simeq$--sign) are generated by the basis \basisv.

We have the following dictionary
\eqn\DICv{
\matrix{
{\underline{monomial}\atop\underline{in\ eq.\ \EXFive}}&{\underline{rational\ function}\atop\underline{in\ eq.\ \EXfive}}&
{\underline{rational\ function}\atop\underline{in\ eq.\ \GENERI}}\cr\cr
1&\fc{1}{1-xy}&\fc{1}{z_{13}z_{14}z_{24}z_{25}z_{35}}\ ,\cr\cr
X_2&\fc{1-y}{(1-xy)^2}&\fc{z_{23}}{z_{13}^2z_{24}^2z_{25}z_{35}}\ ,\cr\cr
X_3&\fc{1-x}{(1-xy)^2}&\fc{z_{34}}{z_{13}z_{14}z_{24}^2z_{35}^2}\ ,\cr\cr
X_4&\fc{x}{(1-xy)}&\fc{z_{45}}{z_{14}^2z_{24}z_{25}z_{35}^2}\ ,\cr\cr
X_5&1&\fc{z_{15}}{z_{13}z_{14}^2z_{25}^2z_{35}}\ ,\cr\cr
X_2X_3&\fc{(1-x)(1-y)}{(1-xy)^3}&\fc{z_{23}z_{34}}{z_{13}^2z_{24}^3z_{35}^2}\ ,\cr\cr
X_2X_5&\fc{1-y}{1-xy}&\fc{z_{15}z_{23}}{z_{13}^2z_{14}z_{24}z_{25}^2z_{35}}\ ,\cr\cr
X_3X_4&\fc{x(1-x)}{(1-xy)^2}&\fc{z_{34}z_{45}}{z_{14}^2z_{24}^2z_{35}^3}\ ,\cr\cr
X_4X_5&x&\fc{z_{15}z_{45}}{z_{14}^3z_{25}^2z_{35}^2}\ ,}}
between monomials in the integral \EXFive, the polynomial in \EXfive, and the representation
\GENERI.
According to the list \DICv\ from the generators \basisv\ of the complement 
$\overline{\vev{LT(I)}}$ only the element $1$ does not give rise to higher powers
of $z_{ij}$ in the denominator of the integrand \GENERI, \ie $\nt_{ij}\geq -1$. Therefore, we 
dismiss all other basis elements and the integral
\eqn\Giverisev{
\int_0^1dx\int_0^1dy\ \fc{I_5(x,y)}{1-xy}=\zeta(2)+\cdots}
is left as the only basis element without poles. The integral \Giverisev\ yields a transcendental
power series in $\ap$, \cf appendix \appA.
Together with the fundamental set \fundamentalv\ we obtain a six--dimensional partial fraction basis, \ie $(N-2)!=6$ for $N=5$.

\subsubsec{$N=6$}

Using the coordinates \Uxyvi\ we consider the polynomial ring 
$\IR[X_1,\ldots,X_6,Y_1,\ldots,Y_3]$.
{}From \Jsix\ we can read off the constraints \constraint\ giving rise to the monomial ideal:
\eqn\idealvi{\eqalign{
I=&\vev{\ X_2+X_1X_3Y_3-1\ ,\ X_3+X_2X_4Y_1-1\ ,\ X_4+X_3X_5Y_2-1\ ,\cr 
&X_6+X_1X_5Y_1-1\ ,\ Y_2+X_1X_4Y_1Y_3-1\ ,\ Y_3+X_2X_5Y_1Y_2-1}\ .}}
W.r.t.\ lexicographic order we find for the (reduced) Gr\"obner basis of \idealvi\ the $13$ elements:
\eqnn\Grvi
$$\eqalignno{
G=&\{\ 1-Y_1+X_6 Y_1-X_6 Y_2-X_6 Y_3+X_6^2 Y_2 Y_3,\ 
-1+X_5 Y_1+X_6 Y_3, &\Grvi\cr 
&1-X_5-X_6+X_5 X_6 Y_2,\ -1+X_4 Y_3+X_5 Y_2,\ -1+X_4 Y_1+X_3 Y_2,\cr
&X_4-X_6-X_4 Y_1+X_4 X_6 Y_1+X_6 Y_2-X_4 X_6 Y_2,\ -1+X_2 Y_1+X_3 Y_3,\cr 
&X_3-X_4+X_6-X_3 Y_1+X_4 Y_1-X_6 Y_1+X_3 X_6 Y_1-X_3 X_6 Y_3-X_6 Y_2+X_4 X_6 Y_2,\cr
&-X_3+X_3 X_5-X_6+X_3 X_6+X_4 X_6,\ 1-X_2-X_3 Y_3-X_6 Y_3+X_2 X_6 Y_3+X_3 X_6 Y_3,\cr
&-1+X_2+X_5-X_2 Y_3+X_3 Y_3-X_5 Y_3+X_2 X_5 Y_3+X_6 Y_3-X_3 X_6 Y_3-X_2 X_5 Y_2,\cr
&-X_2+X_3+X_2 X_4-X_5+X_2 X_5+X_6-X_3 X_6-X_4 X_6,-1+X_1+X_2 X_6 Y_2\ \}\ .}$$
Hence w.r.t.\ lexicographic order the leading terms of these $13$ monomials give rise to:
\eqnn\LTvi
$$\eqalignno{
LT(I)=&\{\ X_6^2 Y_2 Y_3,\ X_5 Y_1,\ X_5 X_6 Y_2,\ X_4 Y_3,\ X_4 X_6 Y_1,\ X_3 Y_2,\ X_3 X_6 Y_1,\ X_3 X_5,\ X_2 Y_1,\cr
& X_2 X_6 Y_3,\ X_2 X_5 Y_3,\ X_2 X_4,\ X_1\ \}\ .&\LTvi}$$
We would like to mention that the Gr\"obner basis consists of $18$ elements in the case
of degree lexicographic order.

{}From the set \LTvi\ the monomials generating the complement $\overline{\vev{LT(I)}}$ 
can be determined. Most of these monomials yield to higher powers
of $z_{ij}$ in the denominator of the integrand \GENERI, \ie $\nt_{ij}=-2$ for some $z_{ij}$.
In fact, only
the following five monomials give rise to single powers in their denominators, 
\ie $\nt_{ij}\geq -1$:
\eqn\DICvi{
\matrix{
{\underline{monomial}\atop\underline{in\ eq.\ \EXSix}}&{\underline{rational\ function}\atop\underline{in\ eq.\ \EXsix}}&
{\underline{rational\ function}\atop\underline{in\ eq.\ \GENERI}}\cr\cr
1&\fc{1}{(1-xy)\ (1-yz)}&\fc{1}{z_{13}z_{15}z_{24}z_{26}z_{35}z_{46}}\ ,\cr\cr
Y_1&\fc{y}{(1-xy)\ (1-yz)}&\fc{1}{z_{14}z_{15}z_{24}z_{26}z_{35}z_{36}}\ ,\cr\cr
Y_2&\fc{1}{(1-xy)\ (1-xyz)}&\fc{1}{z_{13}z_{14}z_{25}z_{26}z_{35}z_{46}}\ ,\cr\cr
Y_3&\fc{1}{(1-yz)\ (1-xyz)}&\fc{1}{z_{13}z_{15}z_{24}z_{25}z_{36}z_{46}}\ ,\cr\cr
X_6Y_2Y_3&\fc{1}{(1-xyz)}&\fc{z_{16}}{z_{13}z_{14}z_{15}z_{25}z_{26}z_{36}z_{46}}\ .\cr\crr}}
Therefore, we dismiss all other basis elements of $\overline{\vev{LT(I)}}$. 
All (finite) integrals \GENERI\ with only single powers of $z_{ij}$
in their denominators, 
\ie $\nt_{ij}\geq -1$, are spanned by the following five integrals\foot{Note, that although 
for degree lexicographic order the Gr\"obner basis 
consists of more elements than \Grvi\ the resulting list \DICvi\ of monomials is the same
for any monomial ordering rule.}:
\eqnn\Giverisevi
$$\eqalignno{
G_0&=\int_0^1dx\int_0^1dy\int_0^1dz\ \fc{I_6(x,y,z)}{(1-xy)\ (1-yz)}=2\ \zeta(2)+\cdots\ ,\cr
G_{1}&=\int_0^1dx\int_0^1dy\int_0^1dz\ \fc{y\ I_6(x,y,z)}{(1-xy)\ (1-yz)}=2\ \zeta(3)+\cdots\ ,\cr
G_{2}&=\int_0^1dx\int_0^1dy\int_0^1dz\ \fc{I_6(x,y,z)}{(1-xy)\ (1-xyz)}=2\ \zeta(3)+\cdots\ ,\cr
G_{3}&=\int_0^1dx\int_0^1dy\int_0^1dz\ \fc{I_6(x,y,z)}{(1-yz)\ (1-xyz)}=2\ \zeta(3)+\cdots\ ,\cr
G_{4}&=\int_0^1dx\int_0^1dy\int_0^1dz\ \fc{I_6(x,y,z)}{1-xyz}=\zeta(3)+\cdots\ .&\Giverisevi}$$
E.g.\ we have
\eqnn\EXAMPL
$$\eqalignno{
\int_0^1dx\int_0^1dy\int_0^1dz\ \fc{yz\ I_6(x,y,z)}{(1-yz)\ (1-xyz)}&=G_3-G_4\ ,\cr\cr
\int_0^1dx\int_0^1dy\int_0^1dz\ \fc{y\ (1-z)\ I_6(x,y,z)}{(1-xy)\ (1-yz)\ (1-xyz)}&=G_1-G_3+G_4\ ,\cr\cr
\int_0^1dx\int_0^1dy\int_0^1dz\ \fc{(1-y)\ I_6(x,y,z)}{(1-xy)\ (1-yz)\ (1-xyz)}&=-G_1+G_2+G_3-G_4\ ,\cr\cr
\int_0^1dx\int_0^1dy\int_0^1dz\ \fc{(1-x)\ y\ I_6(x,y,z)}{(1-xy)\ (1-yz)\ (1-xyz)}&=G_1-G_2+G_4\ ,\cr\cr
\int_0^1dx\int_0^1dy\int_0^1dz\ \fc{xy\ I_6(x,y,z)}{(1-xy)\ (1-xyz)}&=G_2-G_4 &\EXAMPL}$$
as result from the identities between their corresponding monomials 
on dividing them by the Gr\"obner basis \Grvi:
\eqnn\REMAINDERvi
$$\eqalignno{
X_1Y_1Y_3&=Y_3-X_6Y_2Y_3\ ,\cr
X_2Y_1Y_2&=Y_1-Y_3+X_6Y_2Y_3\ ,\cr
X_3Y_2Y_3&=-Y_1+Y_2+Y_3-X_6Y_2Y_3\ ,\cr
X_4Y_1Y_3&=Y_1-Y_2+X_6Y_2Y_3\ ,\cr
X_5Y_1Y_2&=Y_2-X_6Y_2Y_3\ . &\REMAINDERvi}$$

To conclude: Any finite integral \GENERI\ with $\nt_{ij}\geq -1$ can be expressed as 
$\IR$--linear combination of the basis \Giverisevi\ as a result of partial fraction decomposition of their integrands.

Except the first integral $G_0$, the other four integrals \Giverisevi\ yield a transcendental 
power series in $\ap$, \cf appendix \appA.
Any partial fraction decomposition, which involves $G_0$ must refer to a non--transcendental
integral \GENERI\ and only partial fraction expansions  involving the basis $G_1,\ldots,G_4$
comprise into a transcendental integral.
In the previous subsection we have found a set of six transcendental integrals 
\fundamentalvind\ with single poles. 
Together with the fundamental set \fundamentalvi\ we obtain a partial fraction basis (of transcendental 
integrals \GENERI)
with $4+6+14=24$ elements, \ie $(N-2)!=24$ for $N=6$.

\subsubsec{$N=7$}

Using the coordinates \Uxyvii\ we consider the polynomial ring 
$\IR[X_1,\ldots,X_7,Y_1,\ldots,Y_7]$.
{}From \Jseven\ we can read off the constraints \constraint\ giving rise to the monomial ideal:
\eqn\idealvii{\eqalign{
I=&\vev{\ X_2+X_1 X_3 Y_3 Y_7-1,\ 
X_3+X_2 X_4 Y_1 Y_4-1,\ 
X_4+X_3 X_5 Y_2 Y_5-1,\ 
X_5+X_4 X_6 Y_3 Y_6-1,\cr
&Y_4+X_3 X_6 Y_3 Y_2 Y_5 Y_6-1,\ 
Y_6+X_1 X_5 Y_1 Y_4 Y_5 Y_7-1,\ 
Y_7+X_2 X_6 Y_1 Y_2 Y_5 Y_6-1,\cr
&
X_7+X_1 X_6 Y_1 Y_5-1,\ 
Y_2+X_1 X_4 Y_1 Y_3 Y_4 Y_7-1,\ 
Y_3+X_2 X_5 Y_1 Y_2 Y_4 Y_5-1
}\ .}}
W.r.t.\ lexicographic order we find $84$ elements in the (reduced) Gr\"obner basis of 
\idealvii. On the other hand w.r.t.\ degree lexicographic order we have $184$ basis elements. 
In the following, we determine the monomials generating the complement 
$\overline{\vev{LT(I)}}$ w.r.t.\ to degree lexicographic order as this ordering directly yields a cyclic invariant basis.
Most of the monomials in the complement $\overline{\vev{LT(I)}}$ yield to higher powers
of $z_{ij}$ in the denominator of the integrand \GENERI, \ie $\nt_{ij}=-2$ for some $z_{ij}$. After 
disregarding those, only the following six monomials and their cyclic transformations give rise to 
single powers in their denominators, \ie $\nt_{ij}\geq-1$:
\eqn\DICvii{
\matrix{
{\underline{monomial}\atop\underline{in\ eq.\ \EXSeven}}&{\underline{rational\ function}\atop\underline{in\ eq.\ \EXseven}}&
{\underline{rational\ function}\atop\underline{in\ eq.\ \GENERI}}\cr\cr
1&\fc{1}{(1-xy)\ (1-yz)\ (1-wz)}&      \fc{1}{z_{13}z_{16}z_{24}z_{27}z_{35}z_{46}z_{57}}\ ,\cr\cr
Y_1Y_4&\fc{z}{(1-yz)\ (1-wz)\ (1-xyz)}&\fc{1}{z_{14}z_{16}z_{24}z_{27}z_{35}z_{36}z_{57}}\ ,\cr\cr
Y_1Y_3Y_6&\fc{z}{(1-xy)\ (1-wz)\ (1-xyz)}&\fc{z_{47}}{z_{14}z_{15}z_{24}z_{27}z_{36}z_{37}z_{46}z_{57}}\ ,\cr\cr
Y_1Y_2Y_5&\fc{yz}{(1-xy)\ (1-yz)\ (1-wyz)}&\fc{1}{z_{14}z_{16}z_{25}z_{27}z_{35}z_{37}z_{46}}\ ,\cr\cr
Y_2Y_4&\fc{1}{(1-yz)\ (1-wyz)\ (1-xyz)}&\fc{z_{15}z_{37}}{z_{13}z_{14}z_{16}z_{25}z_{27}z_{35}
z_{36}z_{47}z_{57}}\ ,\cr\cr
Y_1&\fc{z}{(1-xy)\ (1-wz)\ (1-yz)}&\fc{z_{47}}{z_{14}z_{16}z_{24}z_{27}z_{35}z_{37}z_{46}z_{57}}
\ .\cr\cr}}
Therefore, in total we have a basis of $36$ elements and
all (finite) integrals \GENERI\ with only single powers 
in their denominators $z_{ij}$, \ie  $\nt_{ij}\geq-1$, are spanned by the following six integrals
\eqnn\Giverisevii
$$\eqalignno{
G_0&=\int_0^1dx\int_0^1dy\int_0^1dz\int_0^1dw\ \fc{I_7(x,y,z,w)}{(1-xy)\ (1-yz)\ (1-wz)}=\fc{27}{4} \zeta(4)+\cdots\ ,&\Giverisevii\cr
G_{1a}&=\int_0^1dx\int_0^1dy\int_0^1dz\int_0^1dw\ \fc{z\ I_7(x,y,z,w)}{(1-yz)\ (1-wz)\ (1-xyz)}=\fc{17}{4}\ \zeta(4)+\cdots\ ,\cr
G_{2b}&=\int_0^1dx\int_0^1dy\int_0^1dz\int_0^1dw\ \fc{z\ I_7(x,y,z,w)}{(1-xy)\ (1-wz)\ (1-xyz)}=3\ \zeta(4)+\cdots\ ,\cr
G_{3a}&=\int_0^1dx\int_0^1dy\int_0^1dz\int_0^1dw\ \fc{yz\ I_7(x,y,z,w)}{(1-xy)\ (1-yz)\ (1-wyz)}=3\ \zeta(3)+\cdots\ ,\cr
G_{4b}&=\int_0^1dx\int_0^1dy\int_0^1dz\int_0^1dw\ \fc{I_7(x,y,z,w)}{(1-yz)\ (1-wyz)\ (1-xyz)}=\fc{5}{2}\ \zeta(4)+4\ \zeta(3)-2\ \zeta(2)+\cdots\ ,\cr
G_{5a}&=\int_0^1dx\int_0^1dy\int_0^1dz\int_0^1dw\ \fc{z\ I_7(x,y,z,w)}{(1-xy)\ (1-wz)\ (1-yz)}=2\ \zeta(3)+2\ \zeta(2)+\cdots\ ,}$$
and their cyclic transformations. 
E.g.\ we have
\eqnn\EXAMPL
$$\eqalignno{
\int_0^1dx\int_0^1dy\int_0^1dz&\int_0^1dw\ \fc{I_7(x,y,z,w)}{(1-xy)\ (1-wz)}=G_0-G_{1b}\ ,&\EXAMPL\cr
\int_0^1dx\int_0^1dy\int_0^1dz&\int_0^1dw\ \fc{yz^2\ I_7(x,y,z,w)}{(1-yz)\ (1-wz)\ (1-xyz)}=-G_0+G_{1a}+G_{1b}+G_{1d}-G_{2b}\ ,\cr
\int_0^1dx\int_0^1dy\int_0^1dz&\int_0^1dw\ \fc{yz\ I_7(x,y,z,w)}{(1-yz)\ (1-wyz)\ (1-xyz)}=G_{5b}-G_{3c}-G_{3f}+G_{4b}\ ,\cr
\int_0^1dx\int_0^1dy\int_0^1dz&\int_0^1dw\ \fc{z\ I_7(x,y,z,w)}{(1-wz)\ (1-xyz)}=G_0-G_{1b}-G_{1d}+G_{2b}\ ,\cr
\int_0^1dx\int_0^1dy\int_0^1dz&\int_0^1dw\ \fc{y\ I_7(x,y,z,w)}{(1-xy)\ (1-wyz)}=G_0-G_{1b}-G_{1f}+G_{2f}\ ,\cr
\int_0^1dx\int_0^1dy\int_0^1dz&\int_0^1dw\ \fc{yz\ I_7(x,y,z,w)}{(1-yz)\ (1-wxyz)}=-G_{1b}+G_{1g}-G_{3d}-G_{3e}-2\ G_{3g}\cr
&+G_{4d}+G_{4g}+G_{5f}\ ,\cr
\int_0^1dx\int_0^1dy\int_0^1dz&\int_0^1dw\ \fc{yz\ I_7(x,y,z,w)}{(1-wyz)\ (1-xyz)}=2\ G_0-2\ G_{1b}-G_{1d}-G_{1f}+G_{2a}\cr
&+G_{2b}+G_{2f}+G_{3a}+G_{3b}+G_{3c}+G_{3f}-G_{4a}-G_{4c}-G_{5c}-G_{5d}\ ,}$$
as results from the identities between their corresponding monomials 
on dividing them by the Gr\"obner basis of \idealvii:
\eqnn\REMAINDERvi
$$\eqalignno{
X_7Y_3Y_6Y_7&=1-Y_1Y_5\ ,\cr
Y_1^2Y_4Y_5&=-1+Y_1 Y_4+Y_1 Y_5+Y_3 Y_6-Y_1 Y_3 Y_6\ ,\cr
Y_1Y_2Y_4Y_5&=Y_3+Y_2 Y_4-Y_2 Y_3 Y_6-Y_3Y_4 Y_7\ ,&\REMAINDERvi\cr
X_7 Y_1 Y_3 Y_4 Y_6 Y_7&=1-Y_1 Y_5-Y_3 Y_6+Y_1 Y_3 Y_6\ ,\cr
X_7 Y_2 Y_3 Y_5 Y_6 Y_7&=1-Y_1 Y_5-Y_3 Y_7+Y_3 Y_5 Y_7\ ,\cr
X_7 Y_1 Y_2 Y_4 Y_5 Y_6 Y_7&=-Y_1 Y_5+Y_6+Y_1 Y_6-Y_2 Y_5 Y_6+Y_7-Y_1 Y_4 Y_7+Y_5Y_7-
2Y_3 Y_6 Y_7\ ,\cr
X_7 Y_1 Y_2 Y_3 Y_4 Y_5 Y_6 Y_7&=
2-Y_1 Y_3-Y_2-Y_4-2 Y_1 Y_5-Y_3 Y_5+Y_1 Y_3 Y_5+Y_1 Y_2 Y_5+Y_1 Y_4 Y_5\cr
&-Y_3 Y_6+Y_1 Y_3 Y_6+Y_2 Y_3 Y_6-Y_3 Y_7+Y_3 Y_4 Y_7+Y_3 Y_5 Y_7\ .}$$
Only $G_0,G_1,G_2$ out of the six integrals in \Giverisevii\ yield a transcendental power series in $\ap$, \cf appendix \appA.

To conclude: Any finite integral \GENERI\ with $\nt_{ij}\geq -1$ can be expressed as 
$\IR$--linear combination of the basis \Giverisevii\ as a result of partial fraction 
decomposition of their integrands.

As a concrete example let us discuss the function $F^{(3452)}$
from the set \revol\ of basis functions for $N=7$.  It is comprised by a sum of 
four integrals:
\eqnn\concrEX
$$\eqalignno{
F^{(3524)}&=s_{13}s_{46}\int\limits_{z_i<z_{i+1}}  
\prod_{j=2}^{N-2} dz_j \lf(\prod_{i<l} |z_{il}|^{s_{il}}\ri) &\concrEX\cr
&\times \ \lf(
\fc{s_{15}s_{24}}{z_{13} z_{15} z_{24} z_{46}}+
\fc{s_{15}s_{26}}{z_{13} z_{15} z_{26} z_{46}}+
\fc{s_{24}s_{35}}{z_{13} z_{24} z_{35} z_{46}}+
\fc{s_{26}s_{35}}{z_{13} z_{26} z_{35} z_{46}}\ri)\ .}$$
Their corresponding rational functions in \EXseven\ and monomials in \EXSeven\ 
are given in the following table:
\eqn\DICix{
\matrix{
{\underline{rational\ function}\atop\underline{in\ eq.\ \GENERI}}&{\underline{rational\ function}\atop\underline{in\ eq.\ \EXseven}}&{\underline{monomial}\atop\underline{in\ eq.\ \EXSeven}}\cr\cr
\fc{z_{17}}{z_{13} z_{15} z_{16} z_{24}z_{27} z_{37}z_{46}z_{57}}&\fc{1}{(1-x y)\ (1-w z)}&
X_7 Y_3 Y_6 Y_7\ ,\cr\cr
\fc{z_{17}z_{67}}{z_{13} z_{15} z_{16}z_{26} z_{27}z_{37}z_{46}z_{47}z_{57}}&\fc{xy}{(1-x y)\ (1-w x y z)}&
X_6 X_7 Y_2 Y_3 Y_5 Y_6^2 Y_7\ ,\cr\cr
\fc{1}{z_{13}z_{16} z_{24} z_{27}z_{35} z_{46}z_{57}}&\fc{1}{(1-x y)\ (1-w z)\ (1-y z)}&
1\ ,\cr\cr
\fc{z_{67}}{z_{13} z_{16}z_{26} z_{27} z_{35}z_{46}z_{47}z_{57}}&
\fc{xy}{(1-x y)\ (1-y z)\ (1-w x y z)}&X_6 Y_2 Y_5 Y_6\ .\cr\cr}}
Their polynomial reduction w.r.t.\ the Gr\"obner basis of  \idealvii\ gives
\eqn\REDGR{\eqalign{
X_7 Y_3 Y_6 Y_7&=1-Y_1 Y_5\ ,\cr
X_6 X_7 Y_2 Y_3 Y_5 Y_6^2 Y_7&=1-Y_1 Y_5+Y_6-Y_2 Y_5 Y_6-Y_4 Y_7+Y_5 Y_7-Y_3 Y_6 Y_7\ ,\cr
X_6 Y_2 Y_5 Y_6&=1-Y_4 Y_7\ ,}}
respectively. The remaining monomials belong to the set \DICvii\ and cyclic transformations 
thereof. Hence, with the lowest expansion coefficients  from \Giverisevii\  we compute:
$$\eqalign{
\int_0^1dx\int_0^1dy\int_0^1dz&\int_0^1dw\ 
\fc{I_7(x,y,z,w)}{(1-xy)\ (1-wz)}=\fc{10}{4}\ \zeta(4)+\cdots\ ,\cr
\int_0^1dx\int_0^1dy\int_0^1dz&\int_0^1dw\ 
\fc{xy\ I_7(x,y,z,w)}{(1-x y)\ (1-w x y z)}=\fc{3}{4}\ \zeta(4)+\cdots\ ,\cr
\int_0^1dx\int_0^1dy\int_0^1dz&\int_0^1dw\ 
\fc{I_7(x,y,z,w)}{(1-x y)\ (1-w z)\ (1-y z)}=\fc{27}{4}\ \zeta(4)+\cdots\ ,\cr
\int_0^1dx\int_0^1dy\int_0^1dz&\int_0^1dw\ 
\fc{xy\ I_7(x,y,z,w)}{(1-x y)\ (1-y z)\ (1-w x y z)}=\fc{10}{4}\ \zeta(4)+\cdots\ .}$$
Eventually for \concrEX\ we obtain:
\eqn\EVENTUALLY{
F^{(3524)}=\fc{1}{4}\ \zeta(4)\ s_{13}\ s_{46}\ 
\lf(10\ s_{15}s_{24}+3\ s_{15}s_{26}+27\ s_{24}s_{35}+10\ s_{26}s_{35}\ri)+\Oc(\ap^5)\ .}
A similar analysis can be done for the other three functions 
$F^{(5324)}, F^{(3542)}$ and $F^{(5342)}$ starting at $\zeta(4)$, c.f. appendix \appC.

\break

\newsec{Concluding remarks}

In the first part of this work \PARTONE\ we derived a strikingly short and compact expression for the 
$N$--point superstring amplitude involving any external massless open string state from the SYM vector multiplet.
The final expression is given in \Simple\ and gives rise to a beautiful harmony of the string amplitudes. 
We have elucidated their implications both from and to field--theory in section~2.
Our result demonstrates how to efficiently compute tree--level superstring amplitudes with an arbitrary 
number of external states. The pure spinor cohomology techniques sketched in \refs{\FTAmps,\MSST} proved to be 
crucial to derive \Simple.
The methods presented in our work should be applicable to tackle any
tree--level disk amplitude computation in any dimensions. 

The availability of the compact expression \Simple\ 
for the superstring $N$--point amplitude
allows a detailed study of possible recursion relations allowing to construct the 
$N$--amplitude from amplitudes with fewer external states and some guiding principle.
Due to the factorized form of \Simple, which separates the YM--part from the string part, 
the basic question is how to combine the field--theory recursions established in the YM
sector~\BCFW\ (see also \MafraJQ) 
to recursions working in the module of hypergeometric functions $B_N$.
For the latter the following recurrence relations may be useful \HopkinsonER
\eqn\UsefulREC{
B_N=\sum B_{n_1}\ B_{n_2}\cdot\ldots\cdot B_{n_k}\ \ \ ,\ \ \ 
\sum_{l=1}^kn_l=N+3\ (k-1)\ ,}
with some partition $\{n_1,\ldots,n_k\}$ into $k$ smaller amplitudes $B_{n_l}$. 
Eq.\ \UsefulREC\ allows to write $B_N$ in terms of products of $(N-3)$ functions $B_4$, \cf
the next Figure.
\ifig\multiperipheral{Partition into products of four--point amplitudes $B_4$.}
{\epsfxsize=0.6\hsize\epsfbox{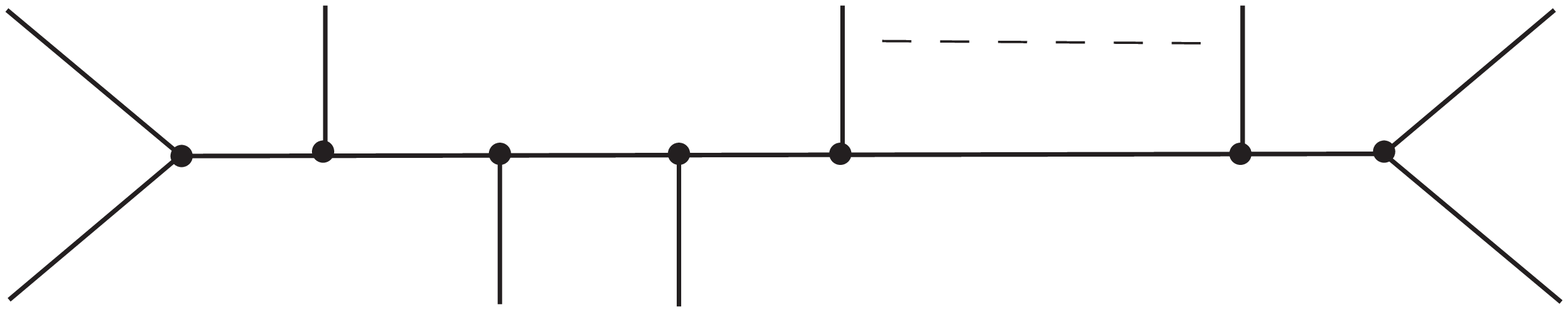}}

The amplitudes  \Simple\ give rise to higher order corrections 
in $\ap$ to the Yang-Mills action, therefore
the YM amplitudes $\Ac_{YM}$ which appear in \Simple\ serve as building blocks to construct 
the higher order terms in the effective action with the expansion coefficients encoded in the
functions $F^\si$.
Moreover, the field--theory amplitudes $\Ac_{YM}$  may be arranged such that only YM three--vertices
contribute \BCJ. Hence, only the latter enter the full superstring amplitude \Simple. 
As a consequence it should be possible to describe the higher order $\ap$--corrections 
in the effective action 
entirely in terms of the fundamental YM three--vertices dressed by the contributions from $F^\si$.

Together with the KLT relations \KawaiXQ, the open string $N$--point amplitudes \Simple\ can be 
used to obtain compact expressions for the $N$--point closed string amplitudes \progress.
The latter give rise to $N$--graviton scattering amplitudes. Their $\ap$--expansions
have been analyzed up to $N\leq 6$ through the order $\ap^8$ in ref.\ \GRAV. These findings proved to be
crucial in constraining possible counterterms in ${\cal N}=8$ supergravity in $D=4$ up
to seven loops \BeisertJX. 
Counterterms invariant under ${\cal N}=8$ supergravity have an unique kinematic structure and
the tree--level closed string amplitudes provide candidates for them, which are compatible 
with SUSY Ward identities and locality.
The absence or restriction on higher order gravitational terms at the order $\ap^l$
together with  their symmetries 
constrain the appearance of possible counter terms  available at $l$--loop.
With the present results it may now be possible to bolster up the results of \GRAV\ and 
to extend the research performed in \refs{\BeisertJX,\ACHT}.

\goodbreak
\vskip 5mm
\centerline{\noindent{\bf Acknowledgments} }

We thank Tristan  Dennen, Gudrun Heinrich, Yu--tin Huang, Sven--Olaf Moch, and Tom Taylor for useful discussions.
Furthermore, we thank the Kavli Institute for Theoretical Physics in Santa Barbara 
for hospitality and financial support during completion of this work.
C.M. also thanks the Werner--Heisenberg--Institut 
in M\"unchen for hospitality  during preparation of this work. O.S. is indebted to  UCLA and in particular to Martin Ammon for hospitality  during preparation of this work.
St.St. is grateful to UCLA, Caltech, and SLAC for  hospitality and financial support during completion of this work.
This research was supported in part by the National Science Foundation under Grant No. NSF PHY05--51164. 
C.M. thanks the partial financial support from
the MPG and acknowledges support by the Deutsch-Israelische
Projektkooperation (DIP H52).
Some diagrams have been created by the program JaxoDraw~\Jaxo.

\break

\appendix{A}{Degree of transcendentality in the $\ap$--expansion}

\subsec{Euler integrals and their power series expansions in $\ap$}

The $\ap$--dependence enters through the kinematic invariants $s_{ij},s_{i\ldots l}$ 
into the integrals \GENERI\ or \GENERIC. 
Hence, in their (integer) power series expansions in $\ap$, which may start at least at the order
$\ap^{3-N}$, each power $\ap^n$ is accompanied
by some rational function or polynomial of degree $n$ in the kinematic invariants 
$\hat s_{ij},\hat s_{i\ldots l}$ .
The latter have rational coefficients multiplied by multizeta values (MZVs) of certain weights.
The maximal weight thereof appearing at a given order $\ap^n$ is related to the power $n$.

One important question is, whether the set of MZVs showing up at a given order $n$ in $\ap$ 
is of a fixed weight. In this case we call the power series expansion {\it transcendental}
(we may also call the integral transcendental).
The power series \EXPA\ is of this kind.
E.g. for $N=6$ we may have the following integral and its power series expansion in $\ap$:
\eqnn\transi
$$\eqalignno{
\int_0^1dx&\int_0^1dy\int_0^1dz\ \fc{I_6(x,y,z)}{xyz}=\fc{1}{s_1s_5t_1}-\zeta(2)\ \lf(\fc{s_3}{s_1s_5}+\fc{s_4}{s_1t_1}+\fc{s_2}{s_5t_1}\ri)&\transi\cr
&+\zeta(3)\ \lf(\fc{s_3+s_4-t_3}{s_1}+\fc{s_2+s_3-t_2}{s_5}+\fc{s_3^2+s_3t_1}{s_1s_5}
+\fc{s_4^2+s_4s_5}{s_1t_1}+\fc{s_2^2+s_1s_2}{s_5t_1}\ri)+\Oc(\ap).}$$
In \transi\ to each power $\ap^n$ in $\ap$ a Riemann zeta constant of fixed weight $n+3$  
 (with $n\geq -1$) appears.
Hence, \transi\ represents a transcendental power series expansion.
On the other hand, the following two integrals
\eqn\transii{\eqalign{
\int_0^1dx\int_0^1dy&\int_0^1dz\ \fc{I_6(x,y,z)}{(1-xyz)^2}=\zeta(2)+\zeta(2)\ 
\lf(s_3+s_6-t_2-t_3\ri)\cr
&-\zeta(3)\ \lf(s_1+s_2+2 s_3+s_4+s_5+2 s_6+t_1-t_2-t_3\ri)+\Oc(\al^2)\ ,}}
\eqn\transiii{\eqalign{
\int_0^1dx\int_0^1dy\int_0^1dz\ &\fc{I_6(x,y,z)}{(1-xy)(1-yz)}=2\ \zeta(2)+
\lf[2\ \zeta(2)-4\ \zeta(3)\ri]\ (t_1+t_2+t_3)\cr
&-\lf[2\ \zeta(2)-\zeta(3)\ri]\ (s_1+s_2+s_3+s_4+s_5+s_6)+\Oc(\al^2)}}
yield  examples of non--transcendent power series.

It would be useful to have a criterion at hand, which allows to infer the transcendentality properties 
of an integral by inspecting its integrand before power series expanding the whole integral. In this 
subsection we present a criterion, which allows to deduce from the structure of the integrand, whether 
we should expect a transcendental power series expansion in $\ap$.
Although this is a mathematical question, it will turn out that
superstring theory provides a satisfying answer to this. 

Transforming the integrals from the representation \GENERIC\ into
the form \GENERI\ subject to \conditions\ will prove to be useful in the following.
Integrals \GENERI, whose integrands are rational functions involving double or higher powers of $z_{ij}$ 
in their denominators, \ie $\nt_{ij}<-1$ for some $z_{ij}$, always give rise to non--transcendental power series.
This can be seen by performing a partial integration within the integrals, \eg for
a double power we have:
\eqn\PARTIAL{ 
\int z_{ij}^{s_{ij}-2}\ r(z_{kl})=\fc{1}{s_{ij}-1} \int r(z_{kl})\ \p_{z_i} z_{ij}^{s_{ij}-1}=-\fc{1}{s_{ij}-1} \int z_{ij}^{s_{ij}-1}\ \p_{z_i} r(z_{kl})\ .}
Regardless of the transcendentality structure of the integral $\int 
z_{ij}^{s_{ij}-1}\ \p_{z_i} r(z_{kl})$
the factor $\fc{1}{s_{ij}-1}=1+s_{ij}+s_{ij}^2+\cdots$ always destroys any 
transcendentality.
This explains, why the  integral \transii\ with the corresponding rational functions
(cf. \eqq \CORRES)
$$\fc{1}{(1-xyz)^2}\simeq \fc{1}{z_{13}z_{14}z_{25}^2z_{36}z_{46}}$$
yields a non--transcendental power series expansion.
On the other hand, the non--transcendentality of the integral \GENERIC\
with the rational function
$[(1-x)(1-y)(1-z)(1-xyz)]^{-1}$ 
can only be seen after transforming it into  the representation \GENERI, in which a rational function 
with a double power in the denominator appears, i.e.:
$$\fc{1}{(1-x)(1-y)(1-z)(1-xyz)}\simeq\fc{1}{z_{16}^2z_{23}z_{25}z_{34}z_{45}}\ .$$
Let us now discuss the integrals \transi\ and \transiii\ and elaborate their differences.
w.r.t.\ to the two representations \GENERIC\ and \GENERI\ we have the following correspondences
\eqn\transiv{\eqalign{
\fc{1}{xyz}&\simeq\fc{z_{16}^2}{z_{12}z_{13}z_{14}z_{15}z_{26}z_{36}z_{46}z_{56}}\ra
\fc{1}{z_{12}z_{13}z_{14}z_{15}}\ ,\cr 
\fc{1}{(1-xy)(1-yz)}&\simeq\fc{1}{z_{13}z_{15}z_{24}z_{26}z_{35}z_{46}}\ra
\fc{1}{z_{13}z_{15}z_{24}z_{35}}\ ,}}
respectively. The last correspondence (denoted by the arrow) follows from the choice \FIX, with 
$z_6=z_\infty=\infty$ and taking into account the $z_\infty^2$ factor of the $c$--ghost factor $\vev{c(z_1)c(z_5)c(z_6)}=z_{15}z_\infty^2$.
We may regard the rational functions \transiv\ as originating from a CFT computation of a six--gluon amplitude.
This fact will be exploited in the next subsection to infer 
the transcendentality properties of an integral \GENERIC\ from the $z_{ij}$--representation of its integrand \GENERI.

\subsec{A transcendentality criterion from gluon amplitude computations}

Gluon disk amplitudes in superstring theory provide transcendental power series when expanding 
them w.r.t.\ to $\ap$. This fact follows from dimensional grounds and the underlying 
effective field theory action describing the reducible and irreducible contributions
of the power series expansions. As a consequence the individual constituents of a gluon 
amplitude describing some kinematical factor must be described by  transcendental integrals \GENERI.
Recall that, in the NSR formalism with the choice \FIX\ 
the color ordered $N$--gluon amplitude $\Ac(1,\ldots,N)$ is computed from
\eqn\NSRCOMP{\eqalign{
\Tr(T^{a_1}\ldots T^{a_N})\ &\Ac(1,\ldots,N)=\Tr(T^{a_1}\ldots T^{a_N})\ 
\vev{c(z_1)c(z_{N-1})c(z_N)}\ \cr
&\times\lf(\prod_{l=2}^{N-2}\int_{z_{l-1}}^1 dz_l\ri)\ 
\vev{V_g^{(-1)}(z_i)\ V_g^{(-1)}(z_j)\ \prod_{l\neq i,j}^N V^{(0)}_g(z_l)}\ ,}}
with the $i$--th and $j$--th gluon vertex operator put into the $(-1)$--ghost picture.
The remaining $N-2$ vertex operators are in the zero--ghost picture in order to guarantee a total ghost charge of $-2$. 
The gluon vertex operator  are given by
\eqn\VERTEX{\eqalign{
V_g^{(-1)}&=g_A\ T^a\ e^{-\phi}\ \xi_\mu\  \psi^\mu\ \ e^{ik_\rho X^\rho},\cr
V_g^{(0)}&=T^a\ \fc{g_A}{(2\ap)^{1/2}}\ 
 \xi_\mu\lf[\ i\p X^\mu+2\ap\ (k_\lambda\psi^\lambda)\ \psi^\mu\ 
\ri]\ e^{ik_\rho X^\rho}\ ,}}
in the $(-1)$-- and zero--ghost picture, respectively.
Above we have  the scalar field $\phi$ bosonizing the superghost system, the coupling constant $g_A$ and 
the Chan--Paton factor $T^a$. In the following
we always stick to the canonical color ordering $(1,\ldots,N)$. 
The assignment of the superghost charges is yet left unspecified. 
The interplay between the bosonic fields 
$\p X^\mu$ and the fermionic  parts $(k\psi)\psi^\mu$ of  the $N-2$ zero--ghost vertices $V_g^{(0)}$ 
will play a crucial role for the following considerations\foot{In the sequel, neither the normalization factors $g_A$ of the gluon vertex operators nor the number of space--time 
dimensions play any role.}.

In a six--gluon amplitude \NSRCOMP\  the integral \transi\ describes the space--time contraction 
$(\xi_1\xi_6)(\xi_2k_1)(\xi_3k_1)(\xi_4k_1)(\xi_5k_1)$, while the integral \transiii\ characterizes  
the contraction $(\xi_2\xi_6)(\xi_1k_3)(\xi_3k_5)(\xi_4k_2)(\xi_5k_1)$.
The crucial difference between the two encountered contractions is, that in \NSRCOMP\ 
the first contraction can only be realized by contracting\foot{
According to Wicks rule the correlator in  \NSRCOMP\ decomposes into products of 
two--point correlators, given by:
$\vev{\p X^\mu(z_1) X^\nu(z_2)}=-\fc{2\ap\eta^{\mu\nu}}{z_{12}},\ 
\vev{\psi^\mu(z_1)\psi^\nu(z_2)}=\fc{\eta^{\mu\nu}}{z_{12}}$.}
$$\xi_1^{\mu_1}\xi_2^{\mu_2}\xi_3^{\mu_3}\xi_4^{\mu_4}\xi_5^{\mu_5}\xi_6^{\mu_6}k_1^{\lambda}k_1^{\sigma}k_1^{\rho}k_1^{\tau}\ \vev{\psi_1^{\mu_1}\psi_6^{\mu_6}}
\vev{\p X_2^{\mu_2} X_1^\lambda}\vev{\p X_3^{\mu_3} X_1^\sigma}\vev{\p X_4^{\mu_4} X_1^\rho}\vev{\p X_5^{\mu_5} X_1^\tau}\ ,$$ 
with the first and sixth gluon vertex operator in the $(-1)$--ghost picture. 
Therefore, the integral \transi\ gives rise to a non--vanishing piece in the full amplitude. Since the 
full amplitude is only comprised by transcendental functions 
multiplying kinematical factors the contribution \transi\ must be a transcendental function.
On the other hand, in \NSRCOMP\ the second contraction can be obtained from: 
$$\xi_1^{\mu_1}\xi_2^{\mu_2}\xi_3^{\mu_3}\xi_4^{\mu_4}\xi_5^{\mu_5}\xi_6^{\mu_6}k_1^{\lambda_1}k_2^{\lambda_2}k_3^{\lambda_3}k_5^{\lambda_5}\ \vev{\psi_2^{\mu_2}\psi_6^{\mu_6}}\vev{\p X_1^{\mu_1} X_3^{\lambda_3}}\vev{\p X_3^{\mu_3} X_5^{\lambda_5}} \vev{\p X_4^{\mu_4} X_2^{\lambda_2}} \vev{\p X_5^{\mu_5} X_1^{\lambda_1}}\ ,$$ 
with the second and sixth gluon vertex operator in the $(-1)$--ghost picture.
Furthermore, we may also obtain the second contraction from the contraction involving fermionic
correlators:
$$\xi_1^{\mu_1}\xi_2^{\mu_2}\xi_3^{\mu_3}\xi_4^{\mu_4}\xi_5^{\mu_5}\xi_6^{\mu_6}k_1^{\lambda_1}k_2^{\lambda_2}k_3^{\lambda_3}k_5^{\lambda_5}\ \vev{\psi_2^{\mu_2}\psi_6^{\mu_6}}  
\vev{\psi_1^{\mu_1} \psi_3^{\lambda_3}}\vev{ \psi_3^{\mu_3} \psi_5^{\lambda_5}} 
\vev{\p X_4^{\mu_4} X_2^{\lambda_2}}  \vev{\psi_5^{\mu_5}\psi_1^{\lambda_1}}\ .$$ 
In fact, after taking into account the anti--commutation symmetry 
of fermions the two contractions sum up to zero in the full amplitude \NSRCOMP:
$$\vev{\p X_1 X_3}\vev{\p X_3 X_5} \vev{\p X_5 X_1}-
\vev{\psi_1 \psi_3}\vev{ \psi_3 \psi_5}  \vev{\psi_5\psi_1}=0\ .$$
Otherwise\foot{Alternatively, we could also consider the kinematics 
$(\xi_2\xi_6)(\xi_1k_5)(\xi_3k_1)(\xi_4k_2)(\xi_5k_3)$. Similar arguments as before would 
yield: $\vev{\p X_1 X_5}\vev{\p X_5 X_3} \vev{\p X_3 X_1}-
\vev{\psi_1 \psi_5}\vev{ \psi_5 \psi_3}  \vev{\psi_3\psi_1}=0$.},
the latter would give rise to non--transcendent contributions to the full amplitude
\NSRCOMP.

To summarize: in order to investigate the transcendentality properties of an Euler integral 
\GENERIC\ we transform it into the form \GENERI\ subject to \conditions. 
If the rational function 
$\wtilde R$ of this integrand  involves  powers higher than one in the denominator the corresponding 
integral yields a non--transcendental power series.
Otherwise, the rational function (more precisely its limit $z_N\ra\infty$ with 
taking into account the $c$--ghost factor with the choice \FIX) is mapped to a gluon contraction of 
the form\foot{With no more than one $(\xi\xi)$ 
scalar product. Otherwise in \GENERIC\ there may be double poles, of which not all  disappear by the
choice \FIX.} $(\xi_r\xi_N) (\xi_ik_j)\ldots(\xi_lk_m)$  arising from an $N$--gluon superstring 
computation \NSRCOMP\ 
with the $r$--th and $N$--th gluon vertex operator in the $(-1)$--ghost picture.
If the contraction under consideration can only be realized by the correlator
$\vev{\psi_r\psi_N}\vev{\p X_i X_j}\ldots\vev{\p X_l X_m}$ the corresponding integral is transcendental. 
If on the other hand, the contraction under consideration can also be realized 
by correlators involving more fermionic contractions, the 
underlying integral is non--transcendental and the two contributions add up to zero.
Hence, in the $N$--gluon  amplitude computation \NSRCOMP\
non--transcendental contributions referring to 
a given kinematics $(\xi_r\xi_N) (\xi_ik_j)\ldots(\xi_lk_m)$  are always accompanied
by  contributions involving a circle of fermionic contractions such, that all contributions add up to zero.
Stated differently, integrals describing a kinematics\foot{Note, that this statement assumes the $r$--th 
and $N$--th gluon vertex operator in the $(-1)$--ghost picture to get rid of the double pole from the 
correlator $\vev{e^{-\phi(z_r)}e^{-\phi(z_N)}}\vev{\psi_r\psi_N}$.} $(\xi_r\xi_N) (\xi_ik_j)\ldots(\xi_lk_m)$, 
which can be realized by several
field contractions, describe non--transcendental functions.

In fact, this criterion rules out the double poles \PARTIAL\ to join into a transcendental integral. 
The latter can be realized
by both bosonic and fermionic contractions. E.g. the power $1/z_{ij}^2$ describes the kinematical 
factor $(\xi_ik_j)(\xi_jk_i)$, which 
may stem from either $\xi_i^{\mu_i}\xi_j^{\mu_j}k_i^{\lambda_i}k_j^{\lambda_j}\vev{\p X_i^{\mu_i} X_j^{\lambda_j}}\vev{\p X_j^{\mu_j} X_i^{\lambda_i}}$ 
or from $\xi_i^{\mu_i}\xi_j^{\mu_j}k_i^{\lambda_i}k_j^{\lambda_j}\vev{\psi_i^{\mu_i} \psi_j^{\lambda_j}}\vev{\psi_j^{\mu_j} \psi_i^{\lambda_i}}$. 
Both contributions add up to zero:
$$\vev{\p X_i X_j}\vev{\p X_j X_i}-\vev{\psi_i \psi_j}\vev{\psi_j \psi_i}=0\ .$$
Note, that kinematics involving the product $(\xi_i\xi_j)$ are realized by both $\xi_i^{\mu_i}\xi_j^{\mu_j}\vev{\p X_i^{\mu_i}\p X_j^{\mu_j}}$ 
and $\xi_i^{\mu_i}\xi_j^{\mu_j}\vev{\psi_i^{\mu_i}\psi_j^{\mu_j}}\ k_i^{\lambda_i}k_j^{\lambda_j}\vev{\psi_i^{\lambda_i}\psi_j^{\lambda_j}}$ 
giving rise to $(1-2\ap k_ik_j) (\xi_i\xi_j) z_{ij}^{-2}$ in the end.
According to \PARTIAL\ the non--transcendentality of the
double pole integral  is then compensated by the $1-s_{ij}$ factor 
in the numerator. Therefore, kinematics involving more than two pairs of $(\xi_i\xi_j)$ 
scalar products always involve double powers in the denominator. This is why kinematics with more than two pairs of $(\xi_i\xi_j)$ 
scalar products cannot provide information on the transcendentality property of the underlying integral.
On the other hand, when mapping an integral to
the kinematics $(\xi_r\xi_N) (\xi_ik_j)\ldots(\xi_lk_m)$ in \NSRCOMP\ we put the $r$--th and $N$--th gluon vertex operator 
in the $(-1)$--ghost picture such that the double pole from the contraction $(\xi_r\xi_N)$ drops.

Let us mention, that the two integrals \exeiii\ and \exeiv\ have non--transcendent power series.
Indeed our criterion confirms this:
In the representation \GENERI\ the integral \exeiii\ gives rise to the rational function 
$\fc{1}{z_{13} z_{15} z_{26}^2 z_{34} z_{45}}$ 
involving a double pole. As a consequence of the latter the $\ap$--expansion in \exeiii\ is not transcendental. On the 
other hand, the integral \exeiv\ leads to the rational function 
$\fc{z_{13} z_{26}}{z_{12} z_{14} z_{16} z_{23} z_{25} z_{35} z_{36} z_{46}}\ra\fc{z_{13}}{z_{12} z_{14} z_{23} z_{25} z_{35} }=\fc{1}{z_{12} z_{14} z_{25} z_{35} }+\fc{1}{z_{14} z_{23} z_{25} z_{35} }.$ 
According to the previous statements the last two fractions correspond to the six--gluon
kinematics $(\xi_1\xi_6)(\xi_2k_1)(\xi_3k_5)(\xi_4k_1)(\xi_5k_2)$ and 
$(\xi_1\xi_6)(\xi_4k_1)\underline{(\xi_2k_3)(\xi_3k_5)(\xi_5k_2)}$, respectively. The underlined part of the 
last kinematics may  also be  realized by  contracting fermions along a circle. Hence the power series in \exeiv\ is non--transcendental.

\subsec{Transcendentality criterion at work}

Let us now apply our criterion for some $N=7$ integral examples.
The following integrals can be associated to only one kinematical factor. Therefore, they
represent integrals with transcendental power series expansions.
\eqn\EXSEVEN{
\matrix{
{\underline{rational\ function}\atop\underline{in\ eq.\ \GENERIC}}&
{\underline{rational\ function}\atop\underline{in\ eq.\ \GENERI}}&
\underline{kinematics}&
{\underline{transcendental}\atop\underline{power\ series}}\cr\cr
\fc{1}{(1-x y) (1-w z) (1-y z)}&\fc{1}{z_{13}z_{16}z_{24}z_{35}z_{46}}&(\xi_1\xi_7)(\xi_2k_4)(\xi_3k_1)(\xi_4k_6)(\xi_5k_3)(\xi_6k_1)&{\rm yes}\ ,\cr\crr
\fc{z}{(1-w z) (1-y z) (1-x y z)}&\fc{1}{z_{14}z_{16}z_{24}z_{35}z_{36}}&(\xi_1\xi_7)(\xi_2k_4)(\xi_3k_6)(\xi_4k_1)(\xi_5k_3)(\xi_6k_1)&{\rm yes}\ ,\cr\crr
\fc{y}{(1-x y) (1-y z) (1-w y z)}&\fc{1}{z_{13}z_{16}z_{25}z_{35}z_{46}}&(\xi_1\xi_7)(\xi_2k_5)(\xi_3k_1)(\xi_4k_6)(\xi_5k_3)(\xi_6k_1)&{\rm yes}\ ,\cr\crr
\fc{1}{(1-y z) (1-x y z) (1-w x y z)}&\fc{1}{z_{13}z_{14}z_{26}z_{35}z_{36}}&(\xi_1\xi_7)(\xi_2k_6)(\xi_3k_1)(\xi_4k_1)(\xi_5k_3)(\xi_6k_3)&{\rm yes}\ ,\cr\crr
\fc{z}{(1-wz) (1-w y z)(1-xyz)}&\fc{1}{z_{14}z_{16}z_{24}z_{25}z_{36}}&(\xi_1\xi_7)(\xi_2k_4)(\xi_3k_6)(\xi_4k_1)(\xi_5k_2)(\xi_6k_1)&{\rm yes}\ ,\cr\crr
\fc{yz}{(1-y z) (1-w x y z)}&\fc{1}{z_{14}z_{15}z_{16}z_{26}z_{35}}&(\xi_1\xi_7)(\xi_2k_6)(\xi_3k_5)(\xi_4k_1)(\xi_5k_1)(\xi_6k_1)&{\rm yes}\ ,\cr\crr
\fc{yz}{(1-w y z) (1-x y z)}&\fc{1}{z_{14}z_{15}z_{16}z_{25}z_{36}}&(\xi_1\xi_7)(\xi_2k_5)(\xi_3k_6)(\xi_4k_1)(\xi_5k_1)(\xi_6k_1)&{\rm yes}\ ,\cr\crr
\fc{yz}{(1-y) (1-z) (1-w x y z)}&\fc{1}{z_{15} z_{16} z_{26} z_{34} z_{45}}&(\xi_3\xi_7)(\xi_1k_5)(\xi_2k_6)(\xi_4k_3)(\xi_5k_4)(\xi_6k_1)&{\rm yes}\ ,\cr\crr
\fc{1}{w(1-xy) (1-wz) (1-y z)}&\fc{1}{z_{12} z_{16} z_{24} z_{35} z_{46}}&(\xi_3\xi_7)(\xi_1k_6)(\xi_2k_1)(\xi_4k_2)(\xi_5k_3)(\xi_6k_4)&{\rm yes}\ ,\cr\crr
\fc{1}{w(1-wz) (1-yz) (1-xy z)}&\fc{1}{z_{12} z_{16} z_{24} z_{35} z_{36}}&(\xi_1\xi_7)(\xi_2k_1)(\xi_3k_6)(\xi_4k_2)(\xi_5k_3)(\xi_6k_1)&{\rm yes}\ .\cr\crr}}
Sometimes, before analyzing the integrands a partial fraction decomposition may be useful. E.g.
according to \CORRES\ we have:
$$\fc{1}{(1-xy)(1-xyz)(1-wz)(1-wxyz)}\simeq\fc{z_{16}}{z_{13}z_{14}z_{15}z_{26}z_{27}z_{36}z_{46}z_{57}}\ra\fc{z_{16}}{z_{13}z_{14}z_{15}z_{26}z_{36}z_{46}}\ .$$
The partial fraction expansion yields:
$$\fc{z_{16}}{z_{13}z_{14}z_{15}z_{26}z_{36}z_{46}}=\fc{1}{z_{13}z_{14}z_{15}z_{26}z_{46}}+
\fc{1}{z_{14}z_{15}z_{26}z_{36}z_{46}}\ .$$
The two rational functions on the r.h.s. correspond to the two kinematical factors 
$(\xi_1\xi_7)(\xi_2k_6)(\xi_3k_1)(\xi_4k_1)(\xi_5k_1)(\xi_6k_4)$ and $(\xi_1\xi_7)(\xi_2k_6)(\xi_3k_6)(\xi_4k_1)(\xi_5k_1)(\xi_6k_4)$,
respectively. Both of them do not allow for additional fermionic contractions.
Hence, the integral under consideration yields a transcendental series.

Furthermore, let us discuss some integrals with non--transcendental power series expansions.
The rational functions of the following integrals describe kinematics, which can be realized in two ways. 
The second possibility involves contractions of several pairs of fermions. The latter are contracted along 
a circle and give rise to the underlined subset of the kinematics.
\eqn\EXSEVENi{
\matrix{
{\underline{rational\ function}\atop\underline{in\ eq.\ \GENERIC}}&
{\underline{rational\ function}\atop\underline{in\ eq.\ \GENERI}}&
\underline{kinematics}&
{\underline{transcendental}\atop\underline{power\ series}}\cr\cr
\fc{z}{(1-x y) (1-w z) (1-y z)}&\fc{1}{z_{14}z_{16}z_{24}z_{35}z_{46}}&(\xi_5\xi_7){\underline{(\xi_1k_6)(\xi_4k_1)(\xi_6k_4)}}\ (\xi_2k_4)(\xi_3k_5)&{\rm no}\ ,\cr\crr
\fc{1}{(1-x y) (1-w  z) (1-wx y z)}&\fc{1}{z_{13}z_{15}z_{24}z_{26}z_{46}}&(\xi_1\xi_7){\underline{(\xi_2k_6)(\xi_6k_4)(\xi_4k_2)}}\ (\xi_3k_1)(\xi_5k_1)&{\rm no}\ ,\cr\crr
\fc{xyz}{(1-xy)(1-wyz) (1-x y z)}&\fc{1}{z_{14}z_{16}z_{25}z_{36}z_{46}}&(\xi_2\xi_7){\underline{(\xi_1k_6)(\xi_6k_4)(\xi_4k_1)}}\ (\xi_3k_6) (\xi_5k_2)&{\rm no}\ .\cr\crr}}
Sometimes, before analyzing the integrands a partial fraction decomposition may be useful. E.g.
according to \CORRES\ we have:
$$\fc{z(1-xyz)}{(1-xy)(1-wz)(1-wyz)(1-xyz)}\simeq\fc{z_{26}z_{47}}{z_{14}z_{16}z_{24}z_{25}z_{27}z_{36}z_{37}z_{46}z_{57}}\ra\fc{z_{26}}{z_{14}z_{16}z_{24}z_{25}z_{36}z_{46}}\ .$$
The partial fraction expansion yields:
$$\fc{z_{26}}{z_{14}z_{16}z_{24}z_{25}z_{36}z_{46}}=\fc{1}{z_{14}z_{16}z_{24}z_{25}z_{36}}+\fc{1}{z_{14}z_{16}z_{25}z_{36}z_{46}}\ .$$
The second term on the r.h.s. corresponds to one of the  rational functions discussed in \EXSEVENi.
Hence, the integral under consideration does not give rise to a transcendental series.
An other example is:
$$\fc{1}{(1-yz)(1-wyz)(1-xyz)}\simeq\fc{z_{15}z_{37}}{z_{13}z_{14}z_{16}z_{25}z_{27}z_{35}z_{36}z_{47}z_{57}}\ra\fc{z_{15}}{z_{13}z_{14}z_{16}z_{25}z_{35}z_{36}}\ .$$
The partial fraction expansion yields:
$$\fc{z_{15}}{z_{13}z_{14}z_{16}z_{25}z_{35}z_{36}}=\fc{1}{z_{13}z_{14}z_{16}z_{25}z_{36}}+
\fc{1}{z_{14}z_{16}z_{25}z_{35}z_{36}}\ .$$
The two rational functions on the r.h.s. correspond to the two kinematical factors 
$(\xi_2\xi_7)\underline{(\xi_1k_6)(\xi_3k_1)(\xi_6k_3)}(\xi_4k_1)(\xi_5k_2)$ and $(\xi_2\xi_7)(\xi_1k_6)(\xi_3k_5)(\xi_4k_1)(\xi_5k_2)(\xi_6k_3)$, respectively.
The first kinematics can also be realized by a fermionic contraction along a circle, which is underlined.
Hence, the integral under consideration does not give rise to a transcendental series.
Finally, the third integral with the integrand
$$\fc{y}{(1-wz)(1-yz)(1-xyz)}\simeq\fc{z_{14}z_{37}}{z_{13}z_{15}z_{16}z_{24}z_{27}z_{35}z_{36}z_{47}^2}$$
yields a non--transcendental power series due to the double pole.

The results \EXSEVENi\ can be anticipated by explicitly computing the integrals:
\eqn\PROOF{\eqalign{
&\int_0^1dx\int_0^1dy\int_0^1dz\int_0^1dw\ \fc{z\ I_7(x,y,z,w)}{(1-x y) (1-w z) (1-y z)}=2\ \zeta(2)+2\ \zeta(3)+\cdots\ ,\cr\crr
&\int_0^1dx\int_0^1dy\int_0^1dz\int_0^1dw\ \fc{I_7(x,y,z,w)}{(1-x y) (1-w  z) (1-wx y z)}=
3\ \zeta(3)+\lf(\fc{19}{4}\ \zeta(4)-3\ \zeta(3)\ri)\ s_7\cr
&\hskip4cm+\fc{4}{5}\ \zeta(2)^2\ (s_1+s_6+t_1+t_5)+\cdots\ ,\cr\crr
&\int_0^1dx\int_0^1dy\int_0^1dz\int_0^1dw\ \fc{xyz\ I_7(x,y,z,w)}{(1-x y) (1-w y z) (1-x y z)}=
-2\ \zeta(2)+4\ \zeta(3)+\cdots\ ,\cr\crr
&\int_0^1dx\int_0^1dy\int_0^1dz\int_0^1dw\ \fc{I_7(x,y,z,w)}{(1-y z) (1-w y z) (1-x y z)}=\fc{5}{2}\ 
\zeta(4)+4\ \zeta(3)-2\ \zeta(2)+\cdots\ ,\cr\crr
&\int_0^1dx\int_0^1dy\int_0^1dz\int_0^1dw\ \fc{y\ I_7(x,y,z,w)}{(1-wz) (1-y z) (1-xy z)}=\fc{3}{2}\ \zeta(2)+\fc{3}{2}\ \zeta(3)+\cdots\ ,\cr\crr}}

\appendix{B}{Extended set of multiple hypergeometric functions for $N=6$}

Here we list all additional $18$ functions \revoll\ for the six--point case and give
their relations \FINDREL\ to the basis \revol.

\eqn\Exvi{\eqalign{
F^{(2354)}&=-\int\limits_{0<z_2<z_3<z_4<1} dz_2 dz_3 dz_4\ 
\lf(\prod_{i<l} |z_{il}|^{s_{il}}\ri)\  \fc{1}{z_{41}}\ \fc{s_{12}}{z_{12}}\ \fc{s_{45}}{z_{54}}\ 
\lf(\fc{s_{13}}{z_{13}}+\fc{s_{23}}{z_{23}}\ri)\ ,\cr
F^{(3254)}&=-\int\limits_{0<z_2<z_3<z_4<1} dz_2 dz_3 dz_4\ 
\lf(\prod_{i<l} |z_{il}|^{s_{il}}\ri)\  \fc{1}{z_{41}}\ \fc{s_{13}}{z_{13}}\ \fc{s_{45}}{z_{54}}\ 
\lf(\fc{s_{12}}{z_{12}}+\fc{s_{23}}{z_{32}}\ri)\ ,\cr
F^{(5324)}&=-\int\limits_{0<z_2<z_3<z_4<1} dz_2 dz_3 dz_4\ 
\lf(\prod_{i<l} |z_{il}|^{s_{il}}\ri)\  \fc{1}{z_{41}}\ \fc{s_{15}}{z_{15}}\ \fc{s_{24}}{z_{24}}\ 
\lf(\fc{s_{13}}{z_{13}}+\fc{s_{35}}{z_{53}}\ri)\ ,\cr
F^{(3524)}&=-\int\limits_{0<z_2<z_3<z_4<1} dz_2 dz_3 dz_4\ 
\lf(\prod_{i<l} |z_{il}|^{s_{il}}\ri)\  \fc{1}{z_{41}}\ \fc{s_{13}}{z_{13}}\ \fc{s_{24}}{z_{24}}\ 
\lf(\fc{s_{15}}{z_{15}}+\fc{s_{35}}{z_{35}}\ri)\ ,\cr
F^{(5234)}&=-\int\limits_{0<z_2<z_3<z_4<1} dz_2 dz_3 dz_4\ 
\lf(\prod_{i<l} |z_{il}|^{s_{il}}\ri)\  \fc{1}{z_{41}}\ \fc{s_{15}}{z_{15}}\ \fc{s_{34}}{z_{34}}\ 
\lf(\fc{s_{12}}{z_{12}}+\fc{s_{25}}{z_{52}}\ri)\ ,\cr
F^{(2534)}&=-\int\limits_{0<z_2<z_3<z_4<1} dz_2 dz_3 dz_4\ 
\lf(\prod_{i<l} |z_{il}|^{s_{il}}\ri)\  \fc{1}{z_{41}}\ \fc{s_{12}}{z_{12}}\ \fc{s_{34}}{z_{34}}\ 
\lf(\fc{s_{15}}{z_{15}}+\fc{s_{25}}{z_{25}}\ri)\ ,}}

\eqn\Exvia{\eqalign{
F^{(2453)}&=-\int\limits_{0<z_2<z_3<z_4<1} dz_2 dz_3 dz_4\ 
\lf(\prod_{i<l} |z_{il}|^{s_{il}}\ri)\  \fc{1}{z_{31}}\ \fc{s_{12}}{z_{12}}\ \fc{s_{35}}{z_{53}}\ 
\lf(\fc{s_{14}}{z_{14}}+\fc{s_{24}}{z_{24}}\ri)\ ,\cr
F^{(4253)}&=-\int\limits_{0<z_2<z_3<z_4<1} dz_2 dz_3 dz_4\ 
\lf(\prod_{i<l} |z_{il}|^{s_{il}}\ri)\  \fc{1}{z_{31}}\ \fc{s_{14}}{z_{14}}\ \fc{s_{35}}{z_{53}}\ 
\lf(\fc{s_{12}}{z_{12}}+\fc{s_{24}}{z_{42}}\ri)\ ,\cr
F^{(5423)}&=-\int\limits_{0<z_2<z_3<z_4<1} dz_2 dz_3 dz_4\ 
\lf(\prod_{i<l} |z_{il}|^{s_{il}}\ri)\  \fc{1}{z_{31}}\ \fc{s_{15}}{z_{15}}\ \fc{s_{23}}{z_{23}}\ 
\lf(\fc{s_{14}}{z_{14}}+\fc{s_{45}}{z_{54}}\ri)\ ,\cr
F^{(4523)}&=-\int\limits_{0<z_2<z_3<z_4<1} dz_2 dz_3 dz_4\ 
\lf(\prod_{i<l} |z_{il}|^{s_{il}}\ri)\  \fc{1}{z_{31}}\ \fc{s_{14}}{z_{14}}\ \fc{s_{23}}{z_{23}}\ 
\lf(\fc{s_{15}}{z_{15}}+\fc{s_{45}}{z_{45}}\ri)\ ,\cr
F^{(5243)}&=-\int\limits_{0<z_2<z_3<z_4<1} dz_2 dz_3 dz_4\ 
\lf(\prod_{i<l} |z_{il}|^{s_{il}}\ri)\  \fc{1}{z_{31}}\ \fc{s_{15}}{z_{15}}\ \fc{s_{34}}{z_{43}}\ 
\lf(\fc{s_{12}}{z_{12}}+\fc{s_{25}}{z_{52}}\ri)\ ,\cr
F^{(2543)}&=-\int\limits_{0<z_2<z_3<z_4<1} dz_2 dz_3 dz_4\ 
\lf(\prod_{i<l} |z_{il}|^{s_{il}}\ri)\  \fc{1}{z_{31}}\ \fc{s_{12}}{z_{12}}\ \fc{s_{34}}{z_{43}}\ 
\lf(\fc{s_{15}}{z_{15}}+\fc{s_{25}}{z_{25}}\ri)\ ,}}

\eqn\Exvib{\eqalign{
F^{(3452)}&=-\int\limits_{0<z_2<z_3<z_4<1} dz_2 dz_3 dz_4\ 
\lf(\prod_{i<l} |z_{il}|^{s_{il}}\ri)\  \fc{1}{z_{21}}\ \fc{s_{13}}{z_{13}}\ \fc{s_{25}}{z_{52}}\ 
\lf(\fc{s_{14}}{z_{14}}+\fc{s_{34}}{z_{34}}\ri)\ ,\cr
F^{(4352)}&=-\int\limits_{0<z_2<z_3<z_4<1} dz_2 dz_3 dz_4\ 
\lf(\prod_{i<l} |z_{il}|^{s_{il}}\ri)\  \fc{1}{z_{21}}\ \fc{s_{14}}{z_{14}}\ \fc{s_{25}}{z_{52}}\ 
\lf(\fc{s_{13}}{z_{13}}+\fc{s_{34}}{z_{43}}\ri)\ ,\cr
F^{(5432)}&=-\int\limits_{0<z_2<z_3<z_4<1} dz_2 dz_3 dz_4\ 
\lf(\prod_{i<l} |z_{il}|^{s_{il}}\ri)\  \fc{1}{z_{21}}\ \fc{s_{15}}{z_{15}}\ \fc{s_{23}}{z_{32}}\ 
\lf(\fc{s_{14}}{z_{14}}+\fc{s_{45}}{z_{54}}\ri)\ ,\cr
F^{(4532)}&=-\int\limits_{0<z_2<z_3<z_4<1} dz_2 dz_3 dz_4\ 
\lf(\prod_{i<l} |z_{il}|^{s_{il}}\ri)\  \fc{1}{z_{21}}\ \fc{s_{14}}{z_{14}}\ \fc{s_{23}}{z_{32}}\ 
\lf(\fc{s_{15}}{z_{15}}+\fc{s_{45}}{z_{45}}\ri)\ ,\cr
F^{(5342)}&=-\int\limits_{0<z_2<z_3<z_4<1} dz_2 dz_3 dz_4\ 
\lf(\prod_{i<l} |z_{il}|^{s_{il}}\ri)\  \fc{1}{z_{21}}\ \fc{s_{15}}{z_{15}}\ \fc{s_{24}}{z_{42}}\ 
\lf(\fc{s_{13}}{z_{13}}+\fc{s_{35}}{z_{53}}\ri)\ ,\cr
F^{(3542)}&=-\int\limits_{0<z_2<z_3<z_4<1} dz_2 dz_3 dz_4\ 
\lf(\prod_{i<l} |z_{il}|^{s_{il}}\ri)\  \fc{1}{z_{21}}\ \fc{s_{13}}{z_{13}}\ \fc{s_{24}}{z_{42}}\ 
\lf(\fc{s_{15}}{z_{15}}+\fc{s_{35}}{z_{35}}\ri)\ ,}}
\smallskip
\bigskip

In \holdi\ we have displayed  the relation \FINDREL\ for one particular basis $\pi$.
Here, we want to present the relations \FINDREL\  for two other choices of basis.
For the new basis $\pi=\{(1,2,4,5,3,6),\ (1,4,2,5,3,6),\ (1,5,4,2,3,6),$
$(1,4,5,2,3,6),\ (1,5,2,4,3,6),\ (1,2,5,4,3,6)    \}$ we have
\eqn\KSIXb{\eqalign{
&K^\si_\pi=s_{36}^{-1}\cr
&\times\pmatrix{
\ss{t_1-s_1}& \ss{s_{13}} & \ss{0}& \ss{0} &\ss{0}&\ss{t_1-s_1+s_3}  \cr\crr
\ss{0}&\ss{0}&\ss{s_3+s_{13}}&\ss{s_{13}}&\ss{t_1-s_1+s_3}&\ss{0}\cr\crr
\ss{\fc{s_1 (t_3-s_4) d_{13}}{t_{145} s_{15}}}&\ss{\fc{(s_{36}-s_1) s_{13} d_{13}}{t_{145} s_{15}}}&\ss{\fc{-(s_3+s_{13}) s_{14} s_{25}}{t_{145} s_{15}}}&
\ss{\fc{-s_{13} s_{14} s_{25}}{s_{145} s_{15}}} &
\ss{\fc{d_8 s_{14} s_{35}}{t_{145} s_{15} }}&\ss{\fc{s_1 s_{35} d_{13}}{t_{145} s_{15}}} \cr\crr
\ss{\fc{s_1 (s_4-t_3)}{t_{145}}}&\ss{\fc{(s_1-s_{36}) s_{13}}{t_{145}}}&\ss{\fc{(s_3+s_{13}) d_5}{s_{145}}}&\ss{\fc{s_{13} d_5}{t_{145}}}&\ss{\fc{-(s_1+s_{24}) s_{35}}{t_{145}}}&
\ss{\fc{-s_1 s_{35}}{t_{145}}}\cr\crr
\ss{\fc{s_1 s_4 (s_1-t_1)}{t_{125} s_{15}}}&\ss{\fc{-s_1 s_4 s_{13}}{s_{125} s_{15}}}&\ss{\fc{ s_{14} (s_2+s_{35})d_3}{t_{125} s_{15}}} &\ss{\fc{s_{13}d_3d_7 }{t_{125} s_{15}}} &\ss{\fc{s_{14} s_{35}d_3 }{t_{125} s_{15}}}&\ss{\fc{s_1 (s_4-s_{36}) s_{35}}{s_{125} s_{15}}}\cr\crr
\ss{\fc{(t_1-s_1)d_6}{t_{125}}}&\ss{\fc{s_{13} d_6}{t_{125}}}&\ss{\fc{-s_{14} (s_2+s_{35})}{t_{125}}}&\ss{\fc{-d_7 s_{13}}{t_{125}}}&\ss{\fc{-s_{14} s_{35}}{t_{125} }}&\ss{\fc{d_1 s_{35}}{t_{125}}}\cr\crr}}}
and the following relation can be checked:
\eqn\holdii{
\lf(\matrix{
F^{(2453)}\cr
F^{(4253)}\cr
F^{(5423)}\cr
F^{(4523)}\cr
F^{(5243)}\cr
F^{(2543)}}\ri)=K^\ast\ 
\lf(\matrix{
F^{(2345)}\cr
F^{(3245)}\cr
F^{(4325)}\cr
F^{(3425)}\cr
F^{(4235)}\cr
F^{(2435)}}\ri)\ .}
On the other hand, for the third basis $\pi=\{(1,3,4,5,2,6),\ (1,4,3,5,2,6),\ (1,5,4,3,2,6)$, 
$(1,4,5,3,2,6),\ (1,5,3,4,2,6),\ (1,3,5,4,2,6)\}$ we have
\eqn\KSIXc{\eqalign{
&K^\si_\pi=s_{26}^{-1}\cr
&\times\pmatrix{
\ss{s_1}&\ss{s_1+s_2}&\ss{0}&\ss{s_1-s_3+t_2}&\ss{0}&\ss{0}\cr\crr
\ss{0}&\ss{0}&\ss{s_1-s_3+t_2}&\ss{0}&\ss{s_1+s_{24}}&\ss{s_1}\cr\crr
\ss{\fc{s_1 (s_{26}-s_{13}) d_{13}}{s_{145} s_{15}}}&\ss{\fc{-d_{9} s_{13} d_{13}}{s_{145} s_{15}}}&\ss{\fc{d_{10} s_{14} s_{25}}{s_{145} s_{15}}}&\ss{\fc{s_{13} s_{25} d_{13}}{s_{145} s_{15}}}&\ss{\fc{-s_{14} (s_1+s_{24}) s_{35}}{s_{145} s_{15}}}&\ss{\fc{-s_1 s_{14} s_{35}}{s_{145} s_{15}}}\cr\crr
\ss{\fc{s_1 (s_{13}-s_{26})}{s_{145}}}&\ss{\fc{d_{9} s_{13}}{s_{145}}}&\ss{\fc{-(s_3+s_{13}) s_{25}}{s_{145}}}&\ss{\fc{-s_{13} s_{25}}{s_{145}}}&\ss{\fc{-d_{12} (s_1+s_{24})}{s_{145}}}&\ss{\fc{-s_1d_{12}}{s_{145}}}\cr\crr
\ss{\fc{-s_1 s_4 s_{13}}{s_{246} s_{15}}}&\ss{\fc{-(s_1+s_2) s_4 s_{13}}{s_{246} s_{15}}}&\ss{\fc{s_{14} s_{25} d_0}{s_{246} s_{15}}}&\ss{\fc{s_{13} s_{25} (s_4-s_{26})}{s_{246} s_{15}}}&\ss{\fc{s_{14} (s_2+s_{25}) d_0}{s_{246} s_{15}}}&\ss{\fc{s_1 (s_{26}-s_{14})d_0}{s_{246} s_{15}}}\cr\crr
\ss{\fc{s_1d_{11}}{s_{246}}}&\ss{\fc{d_{11} (s_1+s_2)}{s_{246}}}&\ss{\fc{-s_{14} s_{25}}{s_{246}}}&\ss{\fc{-(s_3+s_{14}) s_{25}}{s_{246}}}&\ss{\fc{-s_{14} (s_2+s_{25})}{s_{246}}}&\ss{\fc{s_1 (s_{14}-s_{26})}{s_{246}}}\cr\crr}}}
and the following relation can be checked:
\eqn\holdiii{
\lf(\matrix{
F^{(3452)}\cr
F^{(4352)}\cr
F^{(5432)}\cr
F^{(4532)}\cr
F^{(5342)}\cr
F^{(3542)}}\ri)=K^\ast\ 
\lf(\matrix{
F^{(2345)}\cr
F^{(3245)}\cr
F^{(4325)}\cr
F^{(3425)}\cr
F^{(4235)}\cr
F^{(2435)}}\ri)\ .}
Hence, the relations \holdi, \holdii\ and \holdiii\ allow to express the additional set of $18$ 
functions \Exvi, \Exvia\  and \Exvib\ in terms of the minimal basis \exvi.

In the above matrices \KSIXb\ and \KSIXc\ we have introduced
$d_5=s_1+s_{24}-s_{36},\ d_6=-s_1+s_5+s_{35},\ 
d_7=s_1-s_5+s_{24}-s_{35}\ d_8=s_6-s_4+s_{13}-s_{24},\ d_{10}=s_1-s_3-s_4+s_6,\ 
d_{11}=s_3+s_{14}-s_{26},\ d_{12}=s_{26}-s_3-s_{13}$ and $d_{13}=s_{15}+s_{45}$.

\appendix{C}{Power series expansions in $\ap$ for $N\geq 7$}

In this appendix we give the $\ap$--expansions \LOW\ of the functions $F^\si$.
While for $N=4,5,6$ the latter can be found in subsection 2.5, here the cases $N\geq 7$ are dealt.
The strategy how to compute the power series expansion in $\ap$ for any generalized Euler 
integral is described in \refs{\StieOpr,\STii}.
Generically, this task amounts to evaluate generalized Euler--Zagier sums involving many
integer sums, which becomes quite tedious for $N\geq 6$.
A complementary approach to determine the $\ap$--expansion for the basis \revol\ can be set up
by imposing the factorization properties discussed in section 2.7. 

Specifically , in the following we display the first orders of the $24$ basis functions 
\revol\ specifying the $N=7$ amplitude:
\eqnn\SEVEN
$$\eqalignno{
F^{(2345)}&=1-\zeta(2)\ (s_5 s_6+s_1 s_7-t_1 t_4-s_5 t_5+t_4 t_5-s_1 t_7+t_1 t_7)\cr
&+\zeta(3)\ (-2 s_1 s_3 s_5+s_5^2 s_6+s_5 s_6^2+s_1^2 s_7+s_1 s_7^2+2 s_3 s_5 t_1+2 s_4 s_5 t_1+2 s_1 s_5 t_2\cr
&+2 s_1 s_5 t_3-2 s_5 t_1 t_3+2 s_1 s_2 t_4+2 s_1 s_3 t_4-2 s_3 t_1 t_4-t_1^2 t_4-2 s_1 t_2 t_4-t_1 t_4^2-2 s_4 s_5 t_5\cr
&-s_5^2 t_5+t_4^2 t_5-s_5 t_5^2+t_4 t_5^2-2 s_1 s_5 t_6-s_1^2 t_7-2 s_1 s_2 t_7+t_1^2 t_7-s_1 t_7^2+t_1 t_7^2)+\Oc(\ap^4)\ ,\cr
F^{(2354)}&=-\zeta(2)\ s_{46}\ (s_4-s_6+t_5)+\zeta(3)\ s_{46}\ (2 s_1 s_3+s_4^2+s_4 s_5-s_5 s_6-s_6^2-2 s_3 t_1-2 s_4 t_1\cr
&-2 s_1 t_2-2 s_1 t_3+2 t_1 t_3+s_4 t_4-s_6 t_4+2 s_4 t_5+s_5 t_5+t_4 t_5+t_5^2+2 s_1 t_6)+\Oc(\ap^4)\ ,\cr
F^{(2435)}&=\zeta(2)\ (s_3+t_1-t_5)\ (s_3+t_4-t_7)\cr
&+\zeta(3)\ (2 s_1 s_2 s_3+2 s_1 s_3^2-s_3^3+2 s_3^2 s_5+2 s_3 s_4 s_5-2 s_3^2 t_1+2 s_3 s_5 t_1+2 s_4 s_5 t_1-s_3 t_1^2\cr
&-2 s_1 s_3 t_2-2 s_3 s_5 t_3-2 s_5 t_1 t_3+2 s_1 s_2 t_4+2 s_1 s_3 t_4-2 s_3^2 t_4-3 s_3 t_1 t_4-t_1^2 t_4-2 s_1 t_2 t_4\cr
&-s_3 t_4^2-t_1 t_4^2-2 s_3 s_5 t_5-2 s_4 s_5 t_5+2 s_5 t_3 t_5+s_3 t_4 t_5+t_4^2 t_5+s_3 t_5^2+t_4 t_5^2-2 s_1 s_2 t_7\cr
&-2 s_1 s_3 t_7+s_3 t_1 t_7+t_1^2 t_7+2 s_1 t_2 t_7+s_3 t_5 t_7-t_5^2 t_7+s_3 t_7^2+t_1 t_7^2-t_5 t_7^2)+\Oc(\ap^4)\ ,\cr
F^{(2453)}&=-\zeta(2)\ s_{36}\ (s_3+t_1-t_5)\cr
&+\zeta(3)\ s_{36}\ (-2 s_1 s_2-2 s_1 s_3-s_3^2-2 s_3 s_4-2 s_4 t_1+t_1^2+2 s_1 t_2+2 s_3 t_3+2 t_1 t_3+s_3 t_4\cr
&+t_1 t_4+2 s_3 t_5+2 s_4 t_5-2 t_3 t_5-t_4 t_5-t_5^2+s_3 t_7+t_1 t_7-t_5 t_7)+\Oc(\ap^4)\ ,\cr
F^{(2534)}&=\zeta(2)\ s_{46}\ (s_3+s_6-t_3-t_5)+\zeta(3)\ s_{46}\ (2 s_1 s_3+2 s_3^2+s_3 s_4-s_3 s_5+s_3 s_6+s_4 s_6\cr
&-s_5 s_6-s_6^2-2 s_1 t_2-2 s_1 t_3-4 s_3 t_3-s_4 t_3+s_5 t_3-s_6 t_3+2 t_3^2-s_3 t_4-s_6 t_4+t_3 t_4\cr
&-3 s_3 t_5-s_4 t_5+s_5 t_5+3 t_3 t_5+t_4 t_5+t_5^2+2 s_1 t_6)+\Oc(\ap^4)\ ,\cr
F^{(2543)}&=-\zeta(2)\ s_{36}\ (s_3+s_6-t_3-t_5)+\zeta(3)\ s_{36}\ (-2 s_1 s_3-s_3^2-s_3 s_4-s_4 s_6+s_6^2\cr
&+2 s_1 t_2+2 s_1 t_3+2 s_3 t_3+s_4 t_3-t_3^2+s_3 t_4+s_6 t_4-t_3 t_4+2 s_3 t_5+s_4 t_5-2 t_3 t_5\cr
&-t_4 t_5-t_5^2-2 s_1 t_6+s_3 t_7+s_6 t_7-t_3 t_7-t_5 t_7)+\Oc(\ap^4)\ ,\cr
F^{(3245)}&=-\zeta(2)\ s_{13}\ (s_2-s_7+t_7)+\zeta(3)\ s_{13}\ (s_1 s_2+s_2^2+2 s_3 s_5-s_1 s_7-s_7^2+s_2 t_1-s_7 t_1\cr
&-2 s_5 t_2-2 s_5 t_3-2 s_2 t_4-2 s_3 t_4+2 t_2 t_4+2 s_5 t_6+s_1 t_7+2 s_2 t_7+t_1 t_7+t_7^2)+\Oc(\ap^4)\ ,\cr 
F^{(3254)}&=-2\ \zeta(3)\ s_{13}\ s_{25}\ s_{46}+\Oc(\ap^4)\ ,\cr
F^{(3425)}&=\zeta(2)\ s_{13}\ (s_3+s_7-t_2-t_7)+\zeta(3)\ s_{13}\ (-s_1 s_3+s_2 s_3+2 s_3^2+2 s_3 s_5-s_1 s_7\cr
&+s_2 s_7+s_3 s_7-s_7^2-s_3 t_1-s_7 t_1+s_1 t_2-s_2 t_2-4 s_3 t_2-2 s_5 t_2-s_7 t_2+t_1 t_2+2 t_2^2\cr
&-2 s_5 t_3+2 s_5 t_6+s_1 t_7-s_2 t_7-3 s_3 t_7+t_1 t_7+3 t_2 t_7+t_7^2)+\Oc(\ap^4)\ ,\cr
F^{(3452)}&=\zeta(2)\ s_{13}\ s_{26}+\zeta(3)\ s_{13}\ s_{26}\ (-s_1
        +s_2-s_7-t_1+2 t_3-2 t_6-t_7)+ \Oc(\ap^4) ,\cr
F^{(3524)}&=\fc{1}{4}\ \zeta(4)\ s_{13}\ s_{46}\ 
\lf(10\ s_{15}s_{24}+3\ s_{15}s_{26}+27\ s_{24}s_{35}+10\ s_{26}s_{35}\ri)+\Oc(\ap^5)\ ,\cr
F^{(3542)}&=\fc{1}{4}\ \zeta(4)\ s_{13}\ s_{26}\ 
\lf(-7\ s_{15}s_{24}-17\ s_{24}s_{35}+3\ s_{15}s_{46}+10\ s_{35}s_{46}\ri)+\Oc(\ap^5)\ ,\cr
F^{(4235)}&=-\zeta(2)\ s_{14}\ (s_3+t_4-t_7)\cr
&+\zeta(3)\ s_{14}\ (-2 s_2 s_3-s_3^2-2 s_3 s_5-2 s_4 s_5+s_3 t_1+2 s_3 t_2
+2 s_5 t_3-2 s_2 t_4+t_1 t_4\cr
&+2 t_2 t_4+t_4^2+s_3 t_5+t_4 t_5+2 s_2 t_7+2 s_3 t_7-t_1 t_7-2 t_2 t_7-t_5 t_7-t_7^2)+\Oc(\ap^4)\ ,\cr
F^{(4253)}&=\zeta(2)\ s_{14}\ s_{36}\cr
&+\zeta(3)\ s_{14}\ s_{36}\ (2 s_2+3 s_3+2 s_4-t_1-2 t_2-2 t_3-t_4-t_5-t_7)+\Oc(\ap^4)\ ,\cr
F^{(4325)}&=-\zeta(2)\ s_{14}\ (s_3+s_7-t_2-t_7)+\zeta(3)\ s_{14}\ (-s_2 s_3-s_3^2-2 s_3 s_5-s_2 s_7+s_7^2+s_3 t_1\cr
&+s_7 t_1+s_2 t_2+2 s_3 t_2+2 s_5 t_2-t_1 t_2
-t_2^2+2 s_5 t_3+s_3 t_5+s_7 t_5-t_2 t_5-2 s_5 t_6\cr
&+s_2 t_7+2 s_3 t_7-t_1 t_7-2 t_2 t_7-t_5 t_7-t_7^2)+\Oc(\ap^4)\ ,\cr
F^{(4352)}&=-\zeta(2)\ s_{14}\ s_{26}\cr
&+\zeta(3)\ s_{14}\ s_{26}\ (-s_2+s_3+s_7+t_1-t_2-2 t_3+t_5+2 t_6+t_7)+\Oc(\ap^4)\ ,\cr
F^{(4523)}&=\zeta(2)\ s_{14}\ s_{36}\cr
&+\zeta(3)\ s_{14}\ s_{36}\ (2 s_2+2 s_4-t_1+t_2+t_3-t_4-t_5-3 t_6-t_7)+\Oc(\ap^4)\ ,\cr
F^{(4532)}&=-\zeta(2)\ s_{14}\ s_{26}\cr
&+\zeta(3)\ s_{14}\ s_{26}\ (-s_2-s_4+s_7+t_1-t_2-t_3+t_5+2 t_6+t_7)+\Oc(\ap^4)\ ,\cr
F^{(5234)}&=\zeta(2)\ s_{15}\ s_{46}\cr
&+\zeta(3)\ s_{15}\ s_{46}\ (s_4-s_5-s_6+2 t_2-t_4-t_5-2 t_6)+\Oc(\ap^4)\ ,\cr
F^{(5243)}&=-\zeta(2)\ s_{15}\ s_{36}\cr
&+\zeta(3)\ s_{15}\ s_{36}\ (s_3-s_4+s_6-2 t_2-t_3+t_4+t_5+2 t_6+t_7)+\Oc(\ap^4)\ ,\cr
F^{(5324)}&=\fc{1}{4}\ \zeta(4)\ s_{15}s_{46}\ 
\lf(10\ s_{13}s_{24}+3\ s_{13}s_{26}-17\ s_{24}s_{35}-7\ s_{26}s_{35}\ri)+\Oc(\ap^5)\ ,\cr
F^{(5342)}&=\fc{1}{4}\ \zeta(4)\ s_{15}s_{26}\ 
\lf(-7\ s_{13}s_{24}+3\ s_{13}s_{46}+10\ s_{24}s_{35}-7\ s_{35}s_{46}\ri)+\Oc(\ap^5)\ ,\cr
F^{(5423)}&=-\zeta(2)\ s_{15}\ s_{36}\cr
&+\zeta(3)\ s_{15}\ s_{36}\ (-s_2-s_4+s_6-t_2-t_3+t_4+t_5+2 t_6+t_7)+\Oc(\ap^4)\ ,\cr
F^{(5432)}&=\zeta(2)\ s_{15}\ s_{26}+\zeta(3)\ s_{15}\ s_{26}\
(-s_6-s_7 +t_2+t_3-t_5-t_6-t_7)+\Oc(\ap^4)\ .&\SEVEN}$$
As  anticipated after \eqq \LOW\ there is one function starting only at $\zeta(3)\ap^3$ and 
a set of four functions starting not until  at $\zeta(4)\ap^4$.

We also have the expressions for $N\geq 8$. However, it is too elaborate
to present all expansions for $\geq 120$ basis functions \revol.
At any rate in \progress\ a detailed survey of the structure of the $\ap$--expansions \LOW\  is undertaken.

\listrefs

\end